\documentclass[useAMS,usenatbib,fleqn]{mn2e}
\usepackage{graphicx,color,pdfpages,amsmath,float,subfig}
\newcommand{\eagle}{\textsc{eagle}{ }}

\newcommand{\gadget}{\textsc{gadget}{ }}
\newcommand{\subfind}{\textsc{subfind}{ }}

\definecolor{coolblack}{rgb}{0.0, 0.2, 0.5}

\def\ltsima{$\; \buildrel < \over \sim \;$}
\def\simlt{\lower.5ex\hbox{\ltsima}}
\def\gtsima{$\; \buildrel > \over \sim \;$}
\def\simgt{\lower.5ex\hbox{\gtsima}}
\def\la{$\langle$}

\title[Numerical convergence of CDM haloes]{Numerical convergence of simulations of galaxy formation: the abundance and internal structure of cold dark matter haloes}

\author[Ludlow et al.] {\parbox{18cm}{
    Aaron D. Ludlow$^{1,\star}$,
    Joop Schaye$^{2}$ \&
    Richard Bower$^{3}$
  }\vspace{0.3cm}\\
  $^{1}${International Centre for Radio Astronomy Research, University of Western Australia, 35 Stirling Highway, Crawley,}\\
  {Western Australia, 6009, Australia}\\
  $^{2}${Leiden Observatory, Leiden University, PO Box 9513, 2300 RA Leiden, the Netherlands}\\
  $^{3}${Institute for Computational Cosmology, Department of Physics, Durham University, Durham DH1 3LE, U.K.}\\
}

\begin{document}

\maketitle 

\begin{abstract}
  We study the impact of numerical parameters on the properties of cold dark matter haloes formed in
  collisionless cosmological simulations.
  We quantify convergence in the median spherically-averaged circular velocity profiles for haloes
  of widely varying particle number, as well as in the statistics of their structural scaling
  relations and mass functions. In agreement with prior work focused on single haloes, our results suggest
  that cosmological simulations yield robust halo properties for a wide range of gravitational softening
  parameters, $\epsilon$, provided: 1) $\epsilon$ is not larger than a ``convergence radius'', $r_{\rm conv}$, 
  which is dictated by 2-body relaxation and determined by particle number, and
  2) a sufficient number of timesteps are taken to accurately resolve
  particle orbits with short dynamical times. Provided these conditions are met, median circular velocity 
  profiles converge to within $\approx 10$ per cent for radii beyond which the local 2-body relaxation 
  timescale exceeds the Hubble time by a factor $\kappa\equiv t_{\rm relax}/t_{\rm H}\simgt 0.177$, with
  better convergence attained for higher $\kappa$. We provide analytic estimates of $r_{\rm conv}$ that 
  build on previous attempts in two ways: first, by highlighting its explicit (but weak) softening-dependence and, 
  second, by providing a simpler criterion in which $r_{\rm conv}$ is determined entirely by 
  the mean inter-particle spacing, $l$; for example, better than $10$ per cent convergence in
  circular velocity for $r\simgt 0.05\,l$. We show how these analytic criteria can be used to assess 
  convergence in structural scaling relations for dark matter haloes as a function of their mass or 
  maximum circular speed.
\end{abstract}

\renewcommand{\thefootnote}{\fnsymbol{footnote}}
\footnotetext[1]{E-mail: aaron.ludlow@icrar.org} 

\section{Introduction}
\label{SecIntro}

Cosmological simulations have become an essential component of
astronomical science. Simulations of collisionless cold dark matter (CDM), in particular, have
matured to a point where both the statistical properties of large-scale structure,
as well as the highly non-linear structure of dark matter haloes are largely
agreed upon, even between groups employing widely varying simulation or analysis methods. Among these
are: the topology of large-scale structure \citep[e.g.][]{Gott1987,James2007,Blake2014};
the matter power spectrum \citep[e.g.][]{RESmith2003};
the clustering \citep[e.g.][]{Kaiser1984,White1987,Poole2015,Tinker2010}, mass function
\citep[e.g.][]{Jenkins2001,Reed2003,Tinker2008,Despali2016} and shapes
\citep[e.g.][]{Allgood2006,Despali2014,VeraCiro2014,VegaFerrero2017} of dark matter haloes;
their spherically averaged mass profiles \citep[e.g.][]{Navarro1996,Navarro1997,Bullock2001,Ludlow2013,Dutton2014};
mass assembly histories \citep[e.g.][]{vandenBosch2002,Zhao2009,Correa2015a,Correa2015b}
and the mass function and radial distribution of their substructure populations
\citep[e.g.][]{Ghigna1998,Stoehr2003,Gao2004,Springel2008b}.

The radial mass profile of dark matter haloes is a particularly important and robust prediction 
of N-body simulations. For relaxed haloes it can be approximated by the NFW profile 
\citep{Navarro1996,Navarro1997}, though slight deviations from this form have been reported 
extensively in the literature \citep[e.g.][]{Navarro2004,Ludlow2013,Dutton2014,LudlowAngulo2017}. 
The NFW profile has a central cusp where densities diverge as $\rho\propto r^{-1}$ and a steep 
outer profile where $\rho(r)$ tapers off as $r^{-3}$. In parametric form, the spherically
averaged density profiles can be well approximated by
\begin{equation}
  \rho(r)=\frac{\rho_s}{r/r_s(1+r/r_s)},
\end{equation}
where $\rho_s$ and $r_s$ are characteristic values of density and radius.

Agreement on these issues required pain-staking tests of numerical convergence
that demanded repeatability of simulation results, regardless of the numerical methods employed
or the numerical parameters adopted. A number of these studies led to the development of useful
``convergence criteria'' that can be used to disentangle aspects of simulations that
are reliably modelled from those that may be affected by numerical artifact. These studies differ
in the details, but uniformly agree that systematic convergence tests are necessary to
validate the robustness of a particular numerical result. Numerical requirements for
convergence in halo mass functions, for example, may differ substantially from those
required for convergence in shapes, dynamics or mass profiles.

For collisionless CDM, once a cosmological model has been specified, {\em only}
numerical parameters remain. The starting redshift and finite box size of
simulations, for example, affect halo mass functions and clustering, but leave the internal properties of
dark matter haloes largely unchanged \citep{Knebe2009,Power2006}.
Other parameters impose strict limits on spatial resolution, or otherwise alter the inner structure
of haloes in non-trivial ways. Of particular importance are the gravitational force 
softening, $\epsilon$ (which prevents divergent pairwise forces and suppresses large-angle
deflections), the integration timestep for the equations of motion, $\Delta t$, and the 
particle mass resolution, $m_p$. 

\citet[][hereafter P03]{Power2003} provided a comprehensive survey 
of how these numerical parameters affect the internal structure of a simulated CDM halo. They conclude
that convergence in mass profiles can be achieved for suitable choices of timestep, softening, 
particle number and force accuracy. For choices of softening that suppress discreteness effects, and
for timesteps substantially shorter than the local dynamical time, circular velocity profiles converge
to within $\simlt 10$ per cent roughly at the radius enclosing a sufficient number of particles to
ensure that the local 2-body relaxation time
exceeds a Hubble time. Their tests led to the development of what is now a standard choice for the 
``optimal'' softening for cosmological simulations, and to an empirical prescription for calculating 
the ``convergence radius'' of dark matter haloes. Their results -- which we put to the test in 
subsequent sections -- have been validated and extended by a number of follow-up studies 
\citep[e.g.][]{Diemand2004,Springel2008b,Navarro2010,Gao2012}. 

Strictly speaking, the criteria laid out by P03 mainly apply to convergence in the circular velocity profiles, 
$V_c(r)$, of {\em individual} haloes, and may not apply to convergence in other quantities of interest, 
such as their shapes \citep{VeraCiro2014}, mass functions \citep[e.g.][]{Reed2003}, to various aspects of 
their substructure distributions \citep[e.g.][]{Ghigna2000,Reed2005,Springel2008b}, or to population-averaged 
profiles of, for example, density or circular velocity. We focus on the latter in this paper.

P03 defined 
convergence empirically: the same simulation was repeated multiple times using different numerical parameters 
and the results were used to quantify the radial range over which $V_c(r)$ remained insensitive to those choices.
\citet{vandenBosch2018a} follow a different approach. Using a series of idealized numerical 
experiments, they argue that inappropriate choices for gravitational
softening are detrimental to the evolution of substructure haloes and that, as a result, many
state-of-the-art cosmological simulations are still subject to the classic ``over-merging''
problem \citep{Moore1996}. They additionally argue that discreteness-driven instabilities
in subhaloes with $\simlt 10^3$ particles forbids a proper assessment of their evolution
in strong tidal fields, limiting our ability to interpret convergence in their mass
functions or internal structure. 

In this paper, we interpret convergence in the median $V_c(r)$ profiles of halos in terms of
2-body relaxation, which varies from system-to-system depending on the number of particles
used to sample their mass profile. To do so, we carry out the same simulation several times
but with different numbers of particles, allowing us to compare the same sub-population
(i.e. those occupying a particular mass bin) at varying resolution. It is important to note,
however, that 2-body scattering may not be the sole driver of relaxation in collisionless
systems. Collective relaxation, for example, is a distinct process by which individual particle
trajectories are altered repeatedly in response to large-scale and potentially long-lived
fluctuations to the global potential \citep[see e.g.][]{Weinberg1993}. This is neglected in
simple analytic estimates of collisional relaxation rates, such as those presented in
Section~\ref{SSecRconv} \citep[see also][]{Chandrasekhar1943,Henon1961,Cohn1978}, but may
dominate when fluctuations are comparable to the size of the system under consideration. The
fluctuations may be physical, or form as a result of discreteness-driven noise in
collisionless $N$-body simulations. Indeed, it has been argued that this process may
give rise to artificial fragmentation that plagues traditional $N$-body simulations of warm dark
matter models, but appears to have measurable effects in CDM simulations as well
\citep[see, e.g.,][and references therein]{Power2016}.

Going beyond pure dark matter (DM), hydrodynamic simulations of galaxy formation are reaching new 
levels of maturity. The increase of computational resources and improved algorithms 
enable fully cosmological simulations of galaxy formation to be carried out;
simulations that often resolve tens of thousands of {\em individual} galaxies in volumes
approaching those required for cosmological studies. Notable among these are \eagle \citep{Schaye2015,Crain2015},
Illustris \citep{Vogelsberger2014,IllustrusTNG}, Horizon-AGN \citep{Dubois2014}, and
the Magneticum Pathfinder simulations \citep{Dolag2016}. Although sub-grid models in simulations
such as these must be calibrated to reproduce a desired set of observables (for {\textsc{eagle}},
the $z=0$ galaxy stellar
mass function and size-mass relation), many of their {\em predictions} have been ratified by
observations making them a useful tool for interpreting observational data and for illuminating
the complex physical processes that give rise to galaxy scaling relations. 

As discussed by \cite{Schaye2015}, the need to calibrate subgrid models in simulations
affects our ability to interpret numerical convergence, particularly when mass and spatial
resolution are improved. Clearly convergence is a requirement for predictive
power, but for a multi-scale process such as galaxy formation it should arguably be 
attained only {\em after} recalibration of the subgrid physics, thus allowing models to benefit 
from increased resolution by incorporating new, scale-dependent physical processes.

The convergence of hydrodynamic simulations is, in any event, poorly understood.
It remains unclear, for example, how robust predictions of galaxy formation models are to
small changes in the numerical parameters. To cite an example, \citet{Ludlow2019} recently
pointed out that simulated galaxy sizes, quantified by their projected half-mass radii, grow over
time as a result of spurious energy transfer between stellar and DM particles of unequal mass.
The effect can be suppressed by adopting a stellar to DM particle mass ratio that is close to
unity, even when other numerical and subgrid parameters are unchanged. We seek
to further address these issues using a suite of simulations drawn from the \eagle project. In this first
paper we study the sensitivity of our simulations to numerical parameters when only dark matter 
is present, seeking to illuminate and clarify shortcomings of prior work. In a follow-up paper,
we will address convergence in fully-hydrodynamical simulations using a well-calibrated galaxy 
formation model from the \eagle project. 

Our study is part of the \eagle Project. All of our runs were carried out using the same 
simulation code and adopt the same ``fiducial'' numerical parameters as \citet{Schaye2015}, which we 
vary systematically from run-to-run. We concentrate our discussion primarily on gravitational 
softening and the impact of 2-body collisions on the spatial resolution of N-body simulations, 
but consider mass resolution and timestepping when necessary. Softening has been studied in 
great detail in the past two decades, but this has not led to a clear consensus on what constitutes
an ``optimal'' softening length for a given simulation. 

The remainder of the paper is organized as follows. In Section~\ref{SSecSimSetup} we describe our simulation
suite and the variation of numerical parameters, as well as halo finding techniques (Section~\ref{SSecSubfind}).
We provide simple analytic estimates of plausible bounds on gravitational softening in Section~\ref{SecAnalytics};
we introduce the 2-body ``convergence radius'' in Section~\ref{SSecRconv}, highlighting its explicit dependence
on softening. We then present our main results: the convergence of median circular velocity profiles is discussed in 
Section~\ref{SSecConv}, followed by that of mass functions (Section~\ref{SSecMF}). We summarize and conclude 
in Section~\ref{SecConclusion}.

\begin{center}
 \begin{table*}
   \caption{Numerical aspects of our dark matter only simulations. The first column provides a run label, adopting the same
     nomenclature as \citet{Schaye2015}. $N_{\rm p}$ is the total number of simulation particles
     of mass $m_{\rm DM}$; $\epsilon_{\rm cm}$ and $\epsilon_{\rm phys}$ the co-moving and maximum physical softening
     lengths, respectively; $\epsilon_i/l$ are the softening lengths expressed in
     units of the mean inter-particle separation, $l={\rm L_b}/N_{\rm p}^{1/3}$; $f_\epsilon$ is the softening length in units of
     the ``fiducial'' \eagle value for a given $m_{\rm DM}$; {\tt ErrTolIntAcc} is {\sc gadget}'s 
     integration accuracy parameter; $z_{\rm phys}$ the redshift below which the softening
     remains fixed in physical (rather than comoving) units.}
   \begin{tabular}{c r c r r r r r r c r c c}\hline \hline
     & Name & $m_{\rm DM}$    & $N_{\rm p}$ & $\epsilon_{\rm phys}$  & $\epsilon_{\rm phys}/l$     & $\epsilon_{\rm cm}$ & $\epsilon_{\rm cm}/l$ & $f_\epsilon$                    & \tt{ErrTolIntAcc} & $z_{\rm phys}$ \\
     &      &[$10^5\, M_\odot$]&             & $[{\rm pc}]$       & $[10^{-2}]$             & $[{\rm pc}]$    & $[10^{-2}]$       & $[\epsilon/\epsilon_{\rm fid}]$ &                   & \\\hline
     & L0012N0752 & 1.8   & 752$^3$     &   175.0  &  1.05   & 665.0  &  4.00   & 1.0000      &  0.025         & 2.8  &\\\vspace{0.25cm}
     & L0012N0752 & 1.8   & 752$^3$     &   43.8   &  0.26   & 166.3 &  2.00   & 0.2500   &  0.010          & 2.8 &\\

     & L0012N0376 & 14.4  & 376$^3$     &   5600.0 &  16.84  & 2.1$\times10^3$  &  64.01 & 16.0000&  0.025         & 2.8 \\
     & L0012N0376 & 14.4  & 376$^3$     &   2800.0 &  8.42   & 10.6$\times10^3$ &  32.01 & 8.0000 &  0.025         & 2.8 \\
     & L0012N0376 & 14.4  & 376$^3$     &   1400.0 &  4.21   & 5320.0           &  16.00 & 4.0000 &  0.025         & 2.8 \\
     & L0012N0376 & 14.4  & 376$^3$     &   700.0  &  2.11   & 2660.0           &  8.00  & 2.0000 &  0.025         & 2.8 \\
     & L0012N0376 & 14.4  & 376$^3$     &   350.0  &  1.05   & 1330.0           &  4.00  & 1.0000 &  0.025, 0.0025 & 2.8 \\
     & L0012N0376 & 14.4  & 376$^3$     &   175.0  &  0.53   & 665.0            &  2.00  & 0.5000 &  0.025, 0.0025 & 2.8 \\
     & L0012N0376 & 14.4  & 376$^3$     &   87.5   &  0.26   & 332.5            &  1.00  & 0.2500 &  0.025, 0.0025 & 2.8 \\
     & L0012N0376 & 14.4  & 376$^3$     &   43.8   &  0.13   & 166.3            &  0.50  & 0.1250 &  0.025, 0.0025 & 2.8 \\
     & L0012N0376 & 14.4  & 376$^3$     &   21.9   &  0.07   & 83.13            &  0.25  & 0.0625 &  0.025, 0.0025 & 2.8 \\
     & L0012N0376 & 14.4  & 376$^3$     &   10.9   &  0.03   & 41.6             &  0.13  & 0.0313 &  0.025, 0.0025 & 2.8 \\\vspace{0.25cm}
     & L0012N0376 & 14.4  & 376$^3$     &   5.5    &  0.02   & 20.8             &  0.06  & 0.0156 &  0.025, 0.0025 & 2.8 \\

     & L0012N0376 & 14.4  & 376$^3$     &   350.0  &  1.05   & 350.0            &  1.05  & 1.0000 &  0.025         & 0.0 \\\vspace{0.25cm}
     & L0012N0376 & 14.4  & 376$^3$     &   350.0  &  1.05   & 2609.2           &  8.09  & 1.0000 &  0.025         & 10.0 \\

     & L0012N0188 & 115.0 & 188$^3$     &   700.0  &  1.05   & 2660.0           &  4.00   & 1.0000 &  0.025         & 2.8 \\
     & L0012N0188 & 115.0 & 188$^3$     &   350.0  &  0.53   & 1330.0           &  2.00   & 0.5000 &  0.025         & 2.8 \\
     & L0012N0188 & 115.0 & 188$^3$     &   175.0  &  0.26   & 665.0            &  1.00   & 0.2500 &  0.025         & 2.8 \\
     & L0012N0188 & 115.0 & 188$^3$     &   87.5   &  0.13   & 332.5            &  0.50   & 0.1250 &  0.025         & 2.8 \\
     & L0012N0188 & 115.0 & 188$^3$     &   43.8   &  0.07   & 166.3            &  0.20   & 0.0625 &  0.025         & 2.8 \\
     & L0012N0188 & 115.0 & 188$^3$     &   21.9   &  0.03   & 83.13            &  0.13   & 0.0313 &  0.025         & 2.8 \\
   \end{tabular}
   \label{TabSimParam}
 \end{table*}
\end{center}

\section{Simulations}
\label{SecSims}

\subsection{Simulation set-up}
\label{SSecSimSetup}

All runs were carried out in the same ${\rm L_b}=12.5\, {\rm Mpc}$ cubic periodic 
volume which was simulated repeatedly using different numbers of particles, $N_{\rm p}$, gravitational 
softening lengths, $\epsilon$, and timestep size, $\Delta t$. Each run adopted the set of ``Planck''
cosmological parameters used for \eagle \citep{Schaye2015,PlanckCollaboration2014}:
$\Omega_{\rm M}=\Omega_{\rm DM}+\Omega_{\rm bar}=1-\Omega_\Lambda=0.307$; 
$\Omega_{\rm bar}=0.04825$; $h=0.6777$; $\sigma_8=0.8288$; $n_s=0.9611$. Here $\Omega_i$ is the energy density of component $i$ 
expressed in units of the critical density, $\rho_{\rm crit}\equiv 3 H_0^2/(8\pi G)$; 
$h\equiv H_0/[100 \,{\rm km/}s/{\rm Mpc}]$ is Hubble's constant; $\sigma_8$ is the $z=0$ linear rms 
density fluctuation in 8 $h^{-1}{\rm Mpc}$; and $n_s$ is the primordial power spectral index. Initial 
conditions for each  simulation were generated using second-order Lagrangian perturbation theory at 
$z=127$ \citep{Jenkins2013}, which is sufficiently high to ensure that all resolved modes are initially 
well within the linear regime. We sample the linear density
field with $N_{\rm p}=188^3$, $376^3$ and $752^3$ equal-mass particles; the corresponding particle
masses are $m_{\rm DM}=1.15\times 10^7$, $1.44\times 10^6$ and $1.80\times 10^5  \, {\rm M}_\odot$.
All simulations sample resolved modes using the same initial phases.
For consistency we label each run in a way that encodes the box size and particle number, 
using the same nomenclature as \citet{Schaye2015}. For example, L0012N0376 corresponds to a
run with ${\rm L_b}=12.5\, {\rm Mpc}$ and $N_{\rm p}=376^3$ particles. The DM density
field is evolved using the same version of \textsc{P-gadget} \citep{Springel2005b} employed for the
\eagle project. 

It is common in the literature to refer to $\epsilon$ as the ``spatial resolution'' of a
simulation, not surprisingly given its dimensions. 
For cosmological simulations of uniform mass resolution it is customary to adopt a gravitational 
softening length that is a fixed fraction of the (comoving) mean inter-particle separation, 
\begin{equation}
  l\equiv {\rm L_b}/N_{\rm p}^{1/3},
\end{equation}
thus fixing the ratio $\epsilon/m_{\rm DM}^{1/3}$.
In \eagle the softening parameter, initially fixed in comoving coordinates, reaches a 
maximum {\em physical} value at redshift $z_{\rm phys}= 2.8$, after which it remains constant in physical 
coordinates. For ${\rm L_{b}=100}\, {\rm Mpc}$, $N_{\rm p}=1504^3$ and 
$\epsilon(z=0)=700\, {\rm pc}$, this implies, $\epsilon_{\rm phys}/l\approx 0.011$ for $z\leq 2.8$,
while $\epsilon_{\rm cm}/l\approx 0.04$ in co-moving coordinates for $z>2.8$. We will hereafter refer 
to $\epsilon_{\rm fid}(z=0)/l\approx 0.011$ as the ``fiducial'' softening length regardless of mass 
resolution, and will vary $\epsilon$ away from this value by 
successive factors of two. For ${\rm N_{\rm p}=752^3}$ the ``fiducial'' softening 
length is $\epsilon_{\rm fid}=175\, {\rm pc}$; $\epsilon_{\rm fid}=350\,{\rm pc}$ for $N_{\rm p}=376^3$,
and $700\, {\rm pc}$ for $N_{\rm p}=188^3$. We further note that $\epsilon$ 
refers to the Plummer-equivalent softening length, which is related to the ``spline'' softening length
used by \textsc{P-gadget} through
\begin{equation}
  \epsilon_{\rm sp}\equiv 2.8\times\epsilon. 
\end{equation}

To test the sensitivity of our results to changing $z_{\rm phys}$,
we have also carried out runs with $z_{\rm phys}=0$ (fixed co-moving $\epsilon$ at all $z$) and $\infty$
(fixed physical $\epsilon$). For convenience, we will sometimes reference softening parameters relative
to the fiducial value, hence defining the relative softening length $f_\epsilon\equiv \epsilon_i/\epsilon_{\rm fid}$.
Table~\ref{TabSimParam} lists all of the relevant numerical parameters for our simulations. 

\subsection{Halo identification}
\label{SSecSubfind}

We identify haloes in all of our simulations using the \subfind \citep{Springel2001b} algorithm.
\subfind first links dark matter particles into friends-of-friends (FoF) groups before separating them into a number of
self-bound ``subhaloes''. Each FoF group contains a central or ``main'' subhalo that contains most
of its mass, and a number of lower-mass substructures. For each FoF halo and its entire hierarchy
of subhaloes \subfind records a number of attributes, the most basic of which include its mass, ${\rm M_{FoF}}$ 
(for FoF haloes), position (defined as the location of the particle with the minimum potential energy)
and the magnitude and location of its peak circular speed, $V_{\rm max}$ and $r_{\rm max}$.
For FoF haloes (defined as ``main'' haloes in what follows), \subfind also records other common mass
definitions based on a variety of spherical overdensity (SO) boundaries: ${\rm M_{200}}$ is the mass
contained within a spherical region whose mean density is 200$\times\rho_{\rm crit}(z)$; 
${\rm M_\Delta}$ encloses a mean density of $\Delta\times\rho_{\rm crit}(z)$, where $\Delta$ is the
redshift-dependent virial overdensity of \citet{Bryan1998} ($\Delta=103.7$ for our adopted cosmology).
Using as a starting-point all FoF particles in a given halo, \subfind also computes its
gravitationally-bound mass, ${\rm M_{bound}}$, as well as the mass bound to each of its subhaloes. 
Each of these are common and sensible ways to define halo
masses, which we compare in Section~\ref{SSecMF_b}. Note that the virial mass of a halo implicitly
defines its virial radius, $r_\Delta$, and corresponding circular velocity, ${\rm V_\Delta}$: for an 
overdensity $\Delta$, for example, $r_\Delta=3M_\Delta/4\pi \Delta\rho_{\rm crit}$, and 
$V_\Delta=\sqrt{G\, M_\Delta/r_\Delta}$. 

In addition to halo properties, \subfind also records lists of particles belonging to each halo
(which include all bound as well as unbound particles within $r_{200}$), which we 
use to calculate their spherically-averaged enclosed mass profiles, $M(r)$.
Note that {\em all} particles contribute to $M(r)$, and not only those that \subfind
deems bound to the halo. This choice precludes any subtle dependence of our results on \textsc{subfind}'s
unbinding algorithm.
We construct these profiles for all main haloes in 50 equally-spaced logarithmic bins spanning
$-5\leq \log r/[{\rm Mpc}]\leq 0$ which we then use to build {\em median} circular velocity 
profiles, ${\rm V_c}(r)=\sqrt{G\, M(r)/r}$, as a function of halo mass, and various other structural 
scaling relations. Note that {\em all} particles are used to calculate $M(r)$, and not just those 
deemed bound to the main halo by \textsc{subfind}. The results presented in the following sections 
are obtained using these spherically-averaged profiles, and the halo masses returned by \textsc{subfind}.

Assessing convergence in properties of substructure is challenging
\citep[see][for a recent in depth analysis]{vandenBosch2018a,vandenBosch2018b}. For that
reason, we will focus our analysis on isolated haloes expected to host central galaxies in
hydrodynamic runs.

\section{Analytic expectations}
\label{SecAnalytics}

\subsection{Preliminaries: limits on gravitational softening}
\label{SSecPrelim}

Softening gravitational forces in N-body simulations has distinct advantages. In particular, it
suppresses large-angle deflections due to (artificial) 2-body scattering, thereby
permitting the use of low-order orbit integration schemes. This substantially decreases
wall-clock times required for a given N-body problem. There are, however, drawbacks: when 
$\epsilon$ is small shot noise in the particle load can result in large near-neighbor forces,
or when large, to systematic suppression of inter-particle forces. Both effects 
can jeopardize the results of N-body simulations and the optimal choice of gravitational softening 
should provide a compromise between the two. 

The finite particle mass and limited force resolution inherent to particle-based simulations
can give rise to adverse discreteness effects if $\epsilon$ is not properly chosen, and the debate
over what constitutes a wise choice persists. Some studies,
designed specifically to annotate discreteness-driven noise in simulations, suggest 
a safe {\em lower} limit to softening of $\epsilon/l\simgt 1-2$ \citep[e.g.][]{Melott1997,Power2016}.
This is supported by others who argue that various two-point statistics of the DM density
field are not converged on scales $\simlt l$ \citep{Splinter1998}. It is important to note,
however, that discreteness noise does not propagate from the small scales where it is introduced
to larger ones, being typically confined to scales of order $\sim \epsilon$ to $\sim 2\,l$
\citep{Romeo2008}.

Whether simulations are trustworthy below scales $\sim l$ remains a matter of debate. 
Claims above to the contrary are often based on particular cases: cold, plane-symmetric
collapse and simulations with truncated linear power spectra give rise to spurious clustering
as a result of discreteness-driven relaxation. The effect is prominent in simulations of
hot and warm DM models \citep[e.g.][]{AvilaReese2001,Bode2001,Knebe2003},
where counterfeit haloes form as a result of anisotropic
force errors occurring during the initial phases of filamentary collapse \citep{Hahn2013}. These haloes follow
a spatial pattern that reflects the Lagrangian inter-particle spacing, but remain prominent
in warm DM simulations that adopt isotropic softening lengths of order $l$ \citep{Power2016}.
Better success at stifling these haloes has been achieved using adaptive, anisotropic
softening lengths \citep{Hobbs2016} or alternatives to traditional N-body methods \citep{Angulo2013}.

\citet{Power2016} argue that spurious haloes are also present in CDM models, but their
prevalence has not been quantified due to difficulties disentangling them from
``genuine'' ones. Using simulations
of hot DM models, \citet{WangWhite2007} showed that spurious structures dominate the halo mass function on
scales below a limiting mass given by
\begin{equation}
  M_{\rm lim}\approx 10.1\, \overline{\rho}\, l\, k_{\rm peak}^2,
  \label{Wang}
\end{equation}
where $\overline{\rho}=\Omega_{\rm DM}\,\rho_{\rm crit}$ is the mean matter density
and $k_{\rm peak}$ is the wavenumber at which the dimensionless matter power spectrum,
$k^3\,P(k)$, reaches a maximum. Assuming
$M_{\rm lim}=(4/3)\,\pi\,r_{\rm lim}^3\,\Omega_{\rm DM}\Delta\rho_{\rm crit}$, we can 
write eq.~\ref{Wang} in terms of the maximum comoving size of spurious haloes:
\begin{equation}                          
  \begin{split}
    r_{\rm lim} & \approx 1.3 \, l\, \biggr(\frac{\Omega_{\rm DM}}{\Delta}\biggl)^{1/3}\biggr(\frac{k_{\rm Ny}}{k_{\rm peak}}\biggl)^{2/3},\\
    &\approx 0.07\, l\,\biggr(\frac{k_{\rm Ny}}{k_{\rm peak}}\biggl)^{2/3},
    \label{WangSize}
  \end{split}
\end{equation}
where $k_{\rm Ny}\equiv \pi/l$ is the Nyquist frequency, and in the second expression
we have used $\Omega_{\rm DM}=0.307$ and $\Delta=200$. For CDM models, finite resolution
suggests that $k_{\rm peak}\approx k_{\rm Ny}$, implying $r_{\rm lim}\approx 0.07\, l$
(assuming eq.~\ref{Wang} remains valid in this regime).

It is useful to compare $r_{\rm lim}$ to the comoving virial radius of the lowest-mass
haloes resolved in cosmological CDM simulations.
Noting that $M_{200}=N_{200}\, m_{\rm DM}$ and $m_{\rm DM}=\rho_{\rm crit}\, \Omega_{\rm DM}\, l^3$,
this can be written
\begin{equation}
  \begin{split}
    r_{200} =& \biggl(\frac{3\,\Omega_{\rm DM}}{8\,\pi}\biggr)^{1/3}\,\biggl(\frac{N_{200}}{100}\biggr)^{1/3}\, l \\ 
    \approx&\,\, 0.332 \, l \, \biggl(\frac{N_{200}}{100}\biggr)^{1/3}.
  \end{split}
  \label{eq:eps-rvir-limhi}
\end{equation}
Note, for CDM, $r_{200}=r_{\rm lim}\approx 0.07\, l$ when $N_{200}\approx 1$, suggesting spurious
haloes are negligible for models in which $k^3\,P(k)$ increases to arbitrarily small scales. For
models with suppressed small-scale power, on the other hand, spurious haloes may dominate at low
masses. For example, $r_{\rm lim}\simgt l$ provided the Power spectrum ``turn over'' is resolved with
$\simgt 54$ resolution elements.

To avoid biasing gravitational collapse at the resolution
limit of CDM simulations, the comoving softening length should therefore remain {\em smaller} than
the comoving virial radius of the lowest-mass haloes that can be resolved. Indeed, \citet{Lukic2007}
and \citet{Power2016} showed that CDM halo mass functions are substantially suppressed on scales
below which the softening length exceeds the halo virial radius.
For a conservative resolution limit of $N_{200}= 100$, eq.~\ref{eq:eps-rvir-limhi}
suggests that $\epsilon$ should remain {\em smaller} than about one third of the mean 
inter-particle spacing; for $N_{200}=20$, $\epsilon/l\simlt 0.2$ is required.

Most recent large-scale simulations adopt softening lengths substantially smaller than
these limits, but large enough to ensure that the lowest-mass haloes resolved by the simulation
remain approximately collisionless\footnote{No haloes are truly collisionless in particle-based
    simulations. We use the phrase here to refer to haloes in which collisional dynamics driven by
    2-body interactions remain small compared to those dictated by the global potential.}
at all times. One such criterion demands that the specific binding energy of low-mass haloes,
$V_{200}^2\simeq (10\, G\, H\, N_{200}\, m_{\rm DM})^{2/3}$, remains larger than the binding
energy of two DM particles separated by $\epsilon$: $v_\epsilon^2=G\,m_{\rm DM}/\epsilon$. The condition 
$v_\epsilon^2\ll V_{200}^2$ imposes a {\em lower} limit on $\epsilon$ of
\begin{equation}
  \begin{split}
    \epsilon_v &\gg l\, \biggl(\frac{3\, \Omega_{\rm DM}}{800\, \pi}\frac{1}{N_{200}^2}\biggr)^{1/3} \\
    &\approx 3.32\times 10^{-3}\, l\, \biggr(\frac{N_{200}}{100}\biggl)^{-2/3}.
  \end{split}
  \label{eq:eps-rvir-limlow}
\end{equation}
Eq.~\ref{eq:eps-rvir-limlow} can also be expressed $\epsilon_v > r_{200}/N_{200}$, where we have
used the fact that $M_{200}=\Omega_{\rm DM}\,\rho_{\rm crit}\,N_{200}\,l^3$. 
This is comparable to the softening length required to suppress large-angle deflections during close 
encounters, given by $\epsilon_{90} \sim b_{90}=2Gm_{\rm DM}/v^2$, where $b_{90}$  is the impact parameter
giving rise to $\sim 90^\circ$ deflections \citep{BinneyTremaine87} and $v^2 \approx GM/r$ is the typical
speed of particles at distance $r$. This condition therefore requires 
$\epsilon_{90}\simgt 2\, r_{200}/N_{200}$, a factor of 2 larger than $\epsilon_v$. 
Suppressing large {\em accelerations} during to close encounters results in stricter limits on softening 
(see P03). For example, requiring that the maximum stochastic acceleration due to close
encounters, $a_\epsilon=Gm_{\rm DM}/\epsilon^2$, remains smaller than the {\em minimum}
mean-field acceleration across the entire halo, $a_{\rm min}=G\, M_{200}/r_{200}^2$,
imposes a lower limit of $\epsilon_{\rm acc}\simgt r_{200}/\sqrt{N_{200}}$. 

For $N_{vir}\sim 20$ eq.~\ref{eq:eps-rvir-limlow} implies a lower limit of $\epsilon/l \simgt 0.01$,
comparable to values adopted for essentially all recent large-scale simulations projects.
The Bolshoi simulation \citep{Klypin2011}, for example, had a force resolution of $\approx 0.016\, l$, 
while the Multi-dark simulations \citep{Klypin2016} adopted $\epsilon/l=0.014-0.026$ (Plummer equivalent);
the Millennium \citep{Springel2005a}, Millennium-II \citep{Boylan-Kolchin2009}
and Millennium-XXL \citep{Angulo2012} each used $\epsilon/l\approx 0.022$; $\epsilon/l=0.016$
for DUES FUR \citep{Alimi2012}; $\epsilon/l=0.020$ for Copernicus Complexio \citep{Hellwing2016}. 
Cosmological hydrodynamical simulations use comparable values of softening:
as mentioned above, \eagle adopted a physical softening length of $\epsilon/l=0.011$ ($\epsilon/b=0.04$
in co-moving coordinates at $z>2.8$), while Illustris-TNG used maximum physical value of
$\epsilon/l\approx 0.012$ \citep{Springel2018}. All values above are quoted as physical softening lengths
at $z=0$, unless stated otherwise. We will see in Section~\ref{SSecRconv} that 
the median circular velocity profiles of haloes in collisionless cosmological simulations
are affected by 2-body scattering at radii that generally exceed these softening lengths. For example, $V_c(r)$
converges to better than 10 per cent a radii $r\simgt 0.055\times l$, and to better than
3 per cent for $r\simgt 0.10\times l$ (section~\ref{SSecRconvSim}). In all of these runs, softening is unlikely
to compromise the mass profiles of haloes on scales not already influenced by 2-body relaxation.

As mentioned above, the ``optimal'' softening, $\epsilon_{\rm opt}$, for a given simulation is the one that balances large 
force errors due to shot noise with biases resulting from large departures from the Newtonian force 
law. \citet{Merritt1996} suggested that $\epsilon_{\rm opt}$ should be chosen to minimize the average
square error in force evaluations relative to what is expected from an equivalent smooth matter 
distribution. A drawback of this approach is that $\epsilon_{\rm opt}$ depends not only on the
mass distribution under consideration -- which is not generally known a priori -- but also on the number of
particles in the system, $N$. \citet{Dehnen2001}, for example, found that the optimal softening for a
\citet{Hernquist1990} halo is roughly $\epsilon_{\rm opt}/a\simeq 0.017\, (N/10^5)^{-0.23}$, where $a$ is
its scale radius. \citet{vandenBosch2018a} find that
$\epsilon_{\rm opt}/r_{200}=0.005\times (N_{200}/10^5)^{-1/3}$, a result obtained by studying
the stability of the central cusp of an isolated NFW halo
($N_{200}=10^5$ and $c=r_{200}/r_s=10$) to long-term secular evolution ($t\approx 60\,{\rm Gyr}$); the scaling with
$N_{200}$ is motivated by the work of \citet{vankampen2000}. We will see in Section~\ref{SSecRconvz} that this 
is sufficiently small to avoid compromising the spatial resolution in halo centres in cosmological simulations 
using equal-mass particles. In our study, we consider a broad range
of softening lengths, spanning  $\epsilon/l\approx 0.17$ to $\approx 1.6\times 10^{-4}$. 

\begin{figure}
  \includegraphics[width=0.47\textwidth]{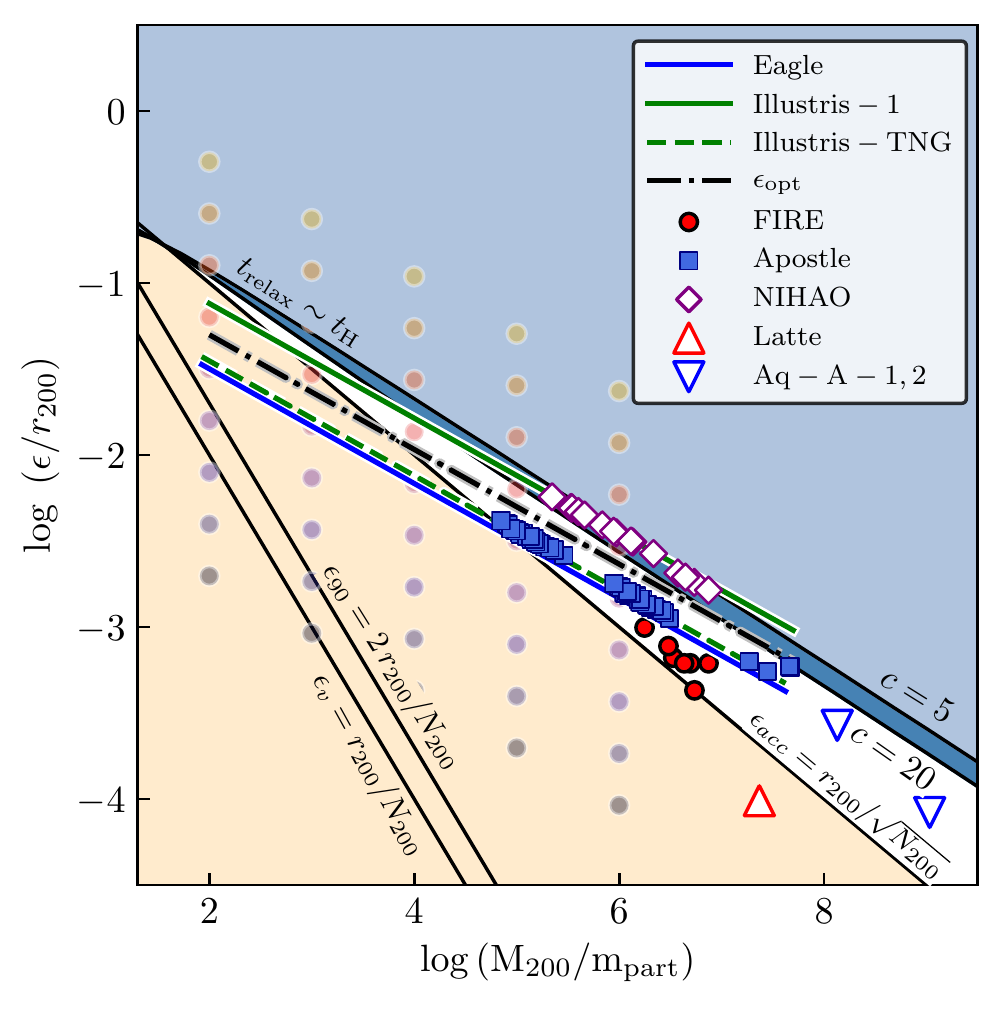}
  \caption{Limits on the gravitational softening and spatial resolution of collisionless N-body simulations as
    a function of particle number $N_{200}=M_{200}/m_{\rm DM}$. Thin black lines show the minimum softening lengths,
    $\epsilon_v$ and $\epsilon_{90}$, required for collisionless dynamics, and length scale $\epsilon_{acc}$
    needed to suppress large stochastic accelerations during close encounters between particles. The beige
    shaded region thus highlights softening lengths that may result in large force errors due to shot
    noise in the particle distribution. The thick black lines show the radius at which the 
    collisional relaxation time for NFW haloes (with concentrations $c=5$ and 20) is equal to the
    Hubble time, $t_{\rm relax}=t_{\rm H}$, and provides an estimate of the {\em minimum} spatial scale that can
    be resolved due to 2-body scattering. The dot-dashed black line shows the ``optimal'' softening, $\epsilon_{\rm opt}$,
    advocated by \citet{vandenBosch2018a}. The blue shaded region indicates $t_{\rm relax}\simgt t_{\rm H}$ for
    a given $N_{200}$; softening lengths chosen below this region guarantee that the dynamics are not governed by
    softened forces. Coloured lines and symbols indicate several recent N-body and hydrodynamical simulations:
    Blue lines, square and triangles show, respectively, the \textsc{eagle}, Apostle and Aquarius (level-1 and
    2) simulations
    \citep[][respectively]{Schaye2015,Sawala2016,Springel2008b}; the red circles and triangle show the FIRE and Latte simulations,
    respectively \citep{Fire,Latte}; the Illustris and Illustris-TNG simulations are highlighted using solid and dashed green lines,
    respectively \citep{IllustrusI,IllustrusTNG}, and purple squares indicate NIHAO \citep{Wang2015,Buck2019}. The grid of coloured circles
    show the values of $\epsilon$ used in our study, and approximately span the range of $N_{200}$ resolved by
    our simulations. Values of $\epsilon$ plotted here correspond to the $z=0$ values quoted by each author for
    DM particles.}
  \label{fig:soft}
\end{figure}

Figure~\ref{fig:soft} compares the softening lengths used in a number of recent state-of-the-art
simulations, and shows for comparison the characteristic values $\epsilon_v$, $\epsilon_{90}$ and $\epsilon_{acc}$
mentioned above. Note that, in all cases, $\epsilon$ is expressed in units of the virial radius, $r_{200}$,
and plotted as a function of $N_{200}$.

In summary, a number of previous studies suggest that there can be considerable errors affecting the
results of N-body simulations on scales $\simlt l$ \citep[e.g.][]{Peebles1989,Melott1997,Splinter1998,Romeo2008}.
These may be driven by discreteness-noise \citep[e.g.][]{Power2016} or errors that affect phases of the Fourier
component on such scales \citep[see, e.g.,][]{Melott2007}.
Their severity, however, depends on the linear power spectrum adopted, or the
type of simulation or statistic studied. Regardless, this underscores the need to establish
independent convergence criteria for {\em all results} obtained from simulations, and to abstain from the common view
that the softening length, or particle mass, represent meaningful ``resolution limits''. 

\subsection{2-Body relaxation and the convergence radius}
\label{SSecRconv}

\subsubsection{The relaxation timescale}
\label{SSSecRelT}

It is important to emphasize that softening pair-wise forces {\em does not} necessarily
increase 2-body relaxation times, which generally impose stricter constraints on the spatial resolution
of N-body simulations. To see why, consider the cumulative effect of 2-body interactions incurred by a test
particle as it crosses an $N$-particle system with surface mass density $\Sigma \approx N/\pi R^2$. Following
\citet{BinneyTremaine87} \citep[see also][]{Huang1993,FaroukiSalpeter1982}, we assume that any one encounter
induces a small velocity perturbation $|\delta v_\perp| \ll v$ perpendicular to the particle's direction
of motion, but leaves its trajectory unchanged. The perturbation due to a single encounter can be expressed
\begin{equation}
  |\delta v_\perp | \approx \frac{2\, G\, m_{\rm DM}\, b}{(b^2+\epsilon^2)\,v},
  \label{eq:vpert}
\end{equation}
where $b$ is the impact parameter and $\epsilon$ the (Plummer) softening length. In a single crossing, the test
particle will experience $\delta n\approx 2\pi \Sigma \,b\, db$ such collisions with impact parameters spanning $b$ to $b+db$. 
Integrating the square velocity change over all such encounters yields
\begin{equation}
  \begin{split}
    \Delta v_\perp^2&=8\,N\biggr(\frac{G m_{\rm DM}}{R v}\biggl )^2\int_{b_{\rm min}}^{b_{\rm max}} b^3 (b^2 + \epsilon^2)^{-2} db \\
                    &=\frac{4 v^2}{N}\biggr[\ln(\epsilon^2 + b^2) + \frac{\epsilon^2}{\epsilon^2+b^2}\biggl]_{b_{\rm min}}^{b_{\rm max}},
  \end{split}
  \label{eq:delv}
\end{equation}
where $b_{\rm min}$ and $b_{\rm max}$ are, respectively, the minimum and maximum impact parameters, and 
we have assumed a typical velocity $v^2=G\, m_{\rm DM}\, N / R$. 

The {\em relaxation time} can be defined in terms of the number of orbits a test particle must execute 
in order to lose memory of its initial trajectory, i.e. $n_{\rm orb}\sim v^2/\Delta v_\perp^2\sim t_{\rm relax}/t_{\rm orb}$.
In the limit $b\gg \epsilon$, eq.~\ref{eq:delv} implies 
\begin{equation}
  \frac{t_{\rm relax}}{t_{\rm orb}}=\frac{N}{8\ln(b_{\rm max}/b_{\rm min})}\simeq \frac{N}{8 \ln (N/2)},
  \label{eq:trel_BT}
\end{equation}
where $t_{\rm orb}=2\, \pi\,r/v$ is the local orbital time, and we have assumed $b_{\rm max}=R$ and 
$b_{\rm min}=b_{90}\equiv 2G\, m_{\rm DM}/v^2$. Eq.~\ref{eq:trel_BT} is comparable to the
classic relaxation time
calculated by \citet{BinneyTremaine87} for Keplerian forces, and depends only on the
enclosed number of particles. Note that for a narrow
range of impact parameter, $b_{\rm max}/b_{\rm min}=(1+\Delta)$, eq.~\ref{eq:trel_BT} implies
$t_{\rm relax}/t_{\rm orb}\propto \ln(1+\Delta)^{-1}$: fixed intervals of $b$ therefore contribute
equally to relaxation, which is sensitive to both {\em close and distant} encounters. As a result,
softening forces on small scales does little to prolong relaxation, which is primarily driven by
large numbers of distant perturbers. 

What is the appropriate choice of $b_{\rm min}$ for haloes in cosmological simulations? If $\epsilon$
is chosen in order to suppress large-angle deflections, then the assumption $b_{\rm min}=b_{90}$ may
need revision. Indeed, for a Plummer potential there is a well-defined impact parameter\footnote{This can be seen by
  maximizing the radial gradient of a Plummer force law and solving for $\epsilon$. The minimum
  impact parameter obtained this way is slightly smaller than that which maximizes the velocity perturbation
  in eq.~\ref{eq:vpert}, which occurs at $b=\epsilon$.}, $b_{\rm min}=\epsilon/\sqrt{2}$, 
for which the perpendicular pair-wise force between particles is {\em maximized}, tending to zero for both larger and
smaller $b$. Inserting this into eq.~\ref{eq:delv}, and assuming $b_{\rm max}=R$, we obtain
\begin{align}
  \frac{t_{\rm relax}}{t_{\rm orb}}&=\frac{N}{4}\biggr[\ln\biggr(\frac{R^2}{\epsilon^2} + 1\biggl) + 
    \frac{\epsilon^2-2R^2}{3(\epsilon^2+R^2)}-\ln\biggr(\frac{3}{2}\biggl)\biggl]^{-1} \label{eq:delke1}\\    
  &\approx \frac{N}{4}\biggr[\ln\biggr(\frac{R^2}{\epsilon^2}\biggl)-\ln\biggr(\frac{3}{2}\biggl)-\frac{2}{3}\biggl]^{-1}\label{eq:delke2}\\    
  &\approx \frac{N}{8 \ln (R/\epsilon)}\label{eq:delke3},
\end{align}
where the last two steps are valid provided $R\gg \epsilon$. Note that eq.~\ref{eq:delke3} depends logarithmically
on $\epsilon$, and is equivalent to eq.~\ref{eq:trel_BT} if $b_{\rm max}=R$ and $b_{\min}=\epsilon$.
This confirms our expectation that softened forces give rise to modest changes in $t_{\rm relax}$, even for
very small values of the softening length \citep[see also, e.g.,][]{Huang1993,Theis1998,Dehnen2001,Diemand2004b}.
We will test this explicitly in Section~\ref{SSecRconvSim}.

\subsubsection{The convergence radius}
\label{SSSecRconvP03}

Collisional relaxation imposes a lower limit on the spatial resolution of N-body simulations that
is typically {\em larger} than the limits on gravitational softening outlined in the previous
section. Based on an extensive suite of simulations of {\em individual} Milky Way-mass haloes,
P03 concluded that convergence 
is obtained at radii that enclose a sufficient number of particles so that the local 
two-body relaxation timescale is comparable to or longer than a Hubble time, $t_H(z)\sim t_{200}(z)$,
where $t_{200}(z)=2\,\pi\,r_{200}(z)/V_{200}(z)$ is the circular orbital time at $r_{200}$.
The ``convergence radius'', $r_{\rm conv}$, implied by eq.~\ref{eq:trel_BT} can thus be approximated by the 
solution to 
\begin{equation}                          
  \begin{split}
    \kappa_{\rm P03}(r,z)&\equiv \frac{t_{\rm relax}(r)}{t_{\rm 200}(z)}\\
    &=\frac{N}{8\ln N}\frac{r/V_c(r)}{r_{200}(z)/V_{200}(z)}\\
    &=\frac{\sqrt{200}}{8}\frac{N}{\ln N}\biggr(\frac{\overline{\rho}(r)}{\rho_{\rm crit}(z)}\biggl)^{-1/2}
  \label{eqP03}
  \end{split}
\end{equation}
where $\overline{\rho}(r)$ is the mean internal density enclosing $N\equiv N(r)$ particles.
We can rewrite eq.~\ref{eqP03}  in order to incorporate its explicit dependence on softening implied by 
eq.~\ref{eq:delke1}, resulting in
\begin{multline}                         
  \kappa_\epsilon(r,z)=\frac{\sqrt{200}\, N}{4}\, \biggr[\ln\biggr(\frac{R^2}{\epsilon^2} + 1\biggl) + \\
    \frac{\epsilon^2-2R^2}{3(\epsilon^2+R^2)}-\ln\biggr(\frac{2}{3}\biggl)\biggl]^{-1}\biggr(
    \frac{\overline{\rho}(r)}{\rho_{\rm crit}(z)}\biggl)^{-1/2}.
  \label{eqKapLud}
\end{multline}

Note that we have used subscripts on eqs.~\ref{eqP03} and \ref{eqKapLud} in order to distinguish
the P03 convergence radius from our softening-dependent estimate above, a convention we retain throughout 
the paper. Either equation, however, can be used to 
estimate $r_{\rm conv}$ once an empirical relationship between $\kappa\equiv t_{\rm relax}/t_{\rm H}$ 
and ``convergence'' has been determined from simulations. 
P03 found that circular velocity profiles converge to better than $10$ per cent provided
$\kappa_{\rm P03}\simgt 0.6$. \citet[][hereafter N10]{Navarro2010} showed that better
convergence can be obtained for larger values of $\kappa_{\rm P03}$; at $\kappa_{\rm P03}=7$, for example,
${\rm V_c}(r)$ profiles converge to $\approx 2.5$ per cent. 
We follow P03 and N10 and quantify convergence in circular velocity profiles using 
$\Delta V_c\equiv (V_c^{\rm high}-V_c^{\rm low})/V_c^{\rm low}$,
where $V_c^{\rm low}$ and $V_c^{\rm high}$ are the {\em median} profiles for haloes of fixed mass
in our low- and highest-resolution simulations, respectively. 

\subsubsection{A simpler convergence criterion}
\label{SSSecRconvApprox}

As a useful approximation, we can rewrite the convergence radius implied by eq.~\ref{eqP03} as
\begin{equation}
  \frac{r_{\rm conv}}{r_{200}}=\frac{\kappa_{\rm P03}^{2/3}\, C}{N_{200}^{1/3}},
  \label{eqxconv}
\end{equation}
where $C\equiv 4\,(\ln N_c/\sqrt{N_c})^{2/3}$ and $N_c\equiv N(<r_{\rm conv})$; these quantities depend 
weakly on concentration and $N_{200}$, but also on 
$\kappa_{\rm P03}$. To illustrate, we set $\kappa_{\rm P03}=1$ and solve eq.~\ref{eqP03} assuming
an NFW mass profile to determine $N_c$ for a range of concentrations and ${\rm N_{200}}$. 
We find that, for $10^2 < N_{200} < 10^8$, $C$ varies by a 
factor $\simlt 2.4$ for $c=20$ and $\simlt 1.9$ for $c=5$.
Neglecting this weak dependence, eq.~\ref{eqxconv} implies that the ratio $r_{\rm conv}/r_{200}$ is
approximately {\em independent of mass} and that, to first order, $r_{\rm conv}/r_{200}\propto N_{200}^{-1/3}$. 
As a result, haloes of a given $N_{200}$ will be ``converged'' roughly to a fixed fraction of their virial radii 
regardless of mass. 

We can use these finding to cast eq.~\ref{eqxconv} into convenient forms that depend only on particle mass, or 
mean inter-particle spacing:
\begin{align}
  r_{\rm conv}(z)&=C\,\kappa_{\rm P03}^{2/3} \, \biggr(\frac{3\,m_{\rm DM}}{800\,\pi\,\rho_{\rm crit}(z)}\biggl)^{1/3}\label{eqxconvII}\\
                 &=C\,\kappa_{\rm P03}^{2/3} \, \biggr(\frac{3\, \Omega_{\rm DM}}{800\,\pi}\biggl)^{1/3}\,l(z)\label{eqxconvIII},
\end{align}
where we have used the fact that $m_{\rm DM}=\Omega_{\rm DM}\, \rho_{\rm crit}(z)\,l(z)^3$;
$l(z)\equiv l/(1+z)$ is the mean inter-particle spacing in physical units. The latter result, eq.~\ref{eqxconvIII}, 
implies that the ratio $r_{\rm conv}(z)/l(z)$ should be largely independent of redshift, halo mass and particle 
mass; {\em the convergence radius is simply a fixed fraction of the mean inter-particle spacing.} 
We will test these scalings explicitly in later sections.  

Figure~\ref{fig:soft} plots $r_{\rm conv}$ (for $\kappa_{\rm P03}=1$; thick black lines)
for NFW profiles with $c=5$ and $c=20$, representative of the vast majority of DM haloes
that form in typical cosmological simulations. Due to the weak
dependence of $N_c$ on $N_{200}$, these curves are only slightly steeper than $r_{\rm conv}/r_{200}\propto N_{200}^{1/3}$.
As a result, adopting a fixed softening parameter for cosmological simulations with uniform mass resolution
will not necessarily compromise spatial resolution at {\em any} mass scale. Take for example the solid blue
line in Figure~\ref{fig:soft}, which plots $\epsilon/r_{200}\propto N_{200}^{1/3}$ adopted for the \eagle
simulation. Here, $\epsilon$ is smaller than $r_{\rm conv}$ by more than a factor of $\sim$2 at all relevant
halo masses (but will eventually exceed $r_{\rm conv}$ at very large ${\rm N_{200}}$).

\section{Median mass profiles}
\label{SSecConv}

\begin{figure*}
  \centering
  \footnotesize
  \subfloat{}{}{\includegraphics[width=.4\textwidth]{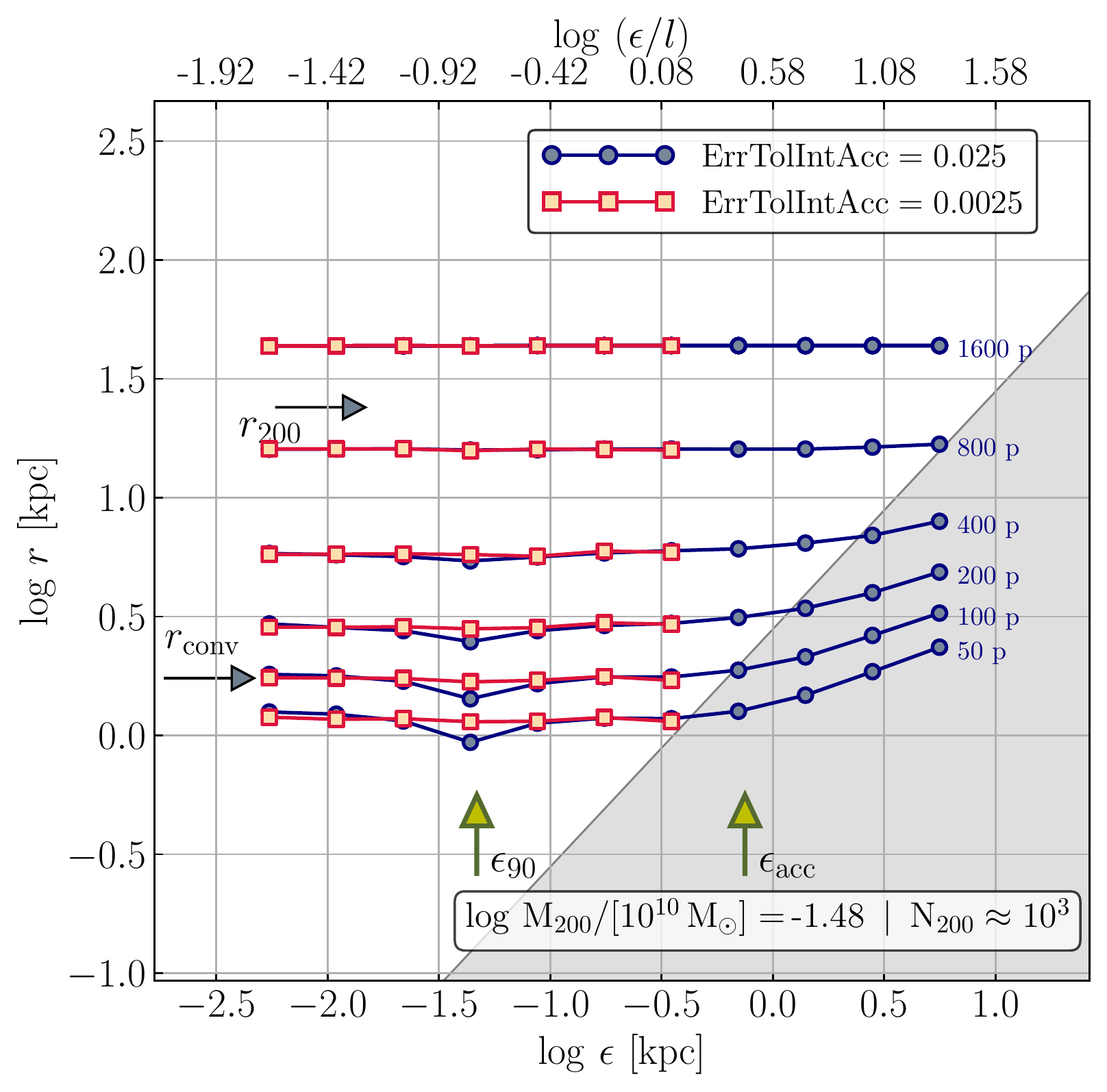}}\quad
  \subfloat{}{}{\includegraphics[width=.4\textwidth]{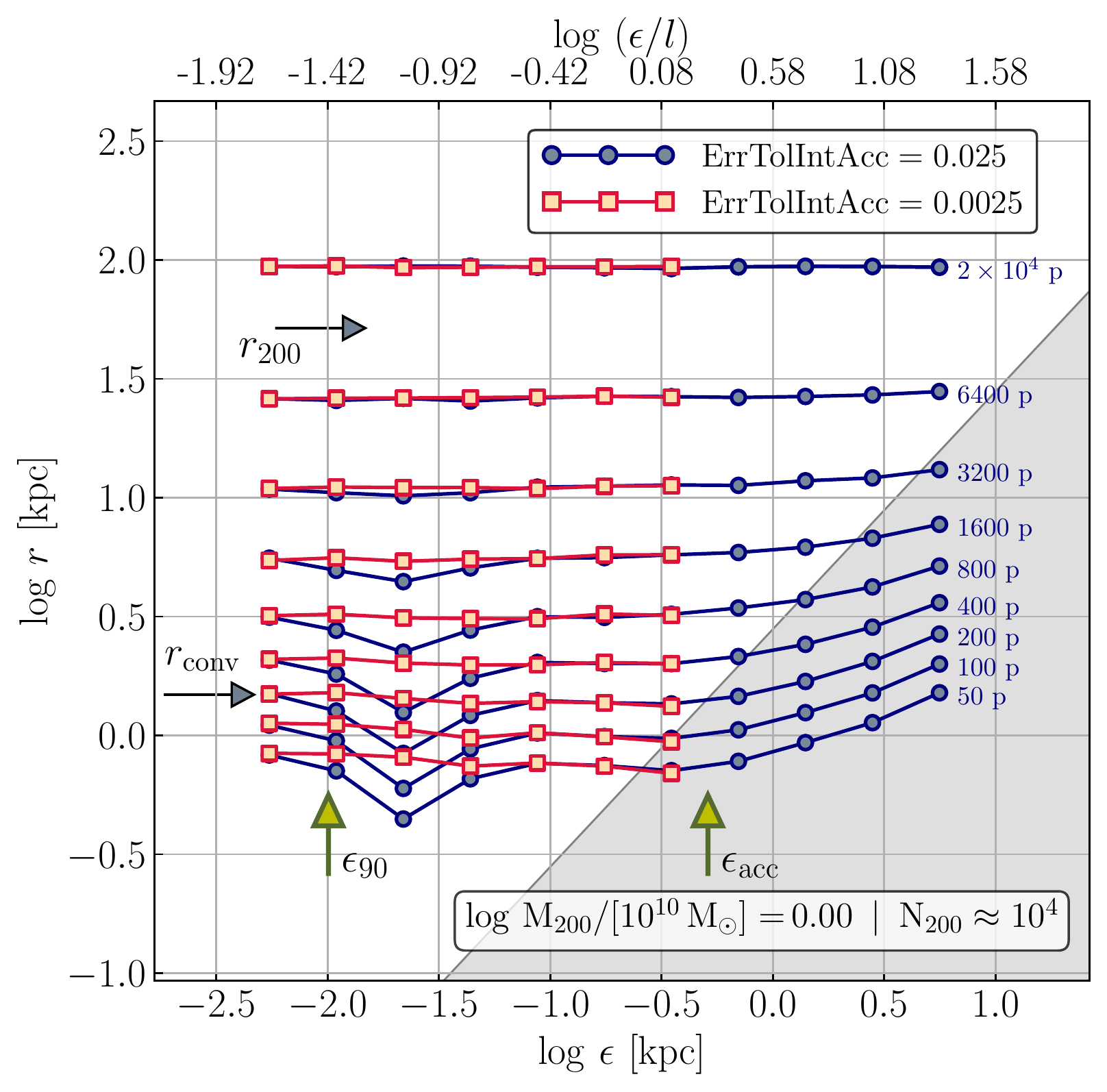}}\\
  \subfloat{}{}{\includegraphics[width=.4\textwidth]{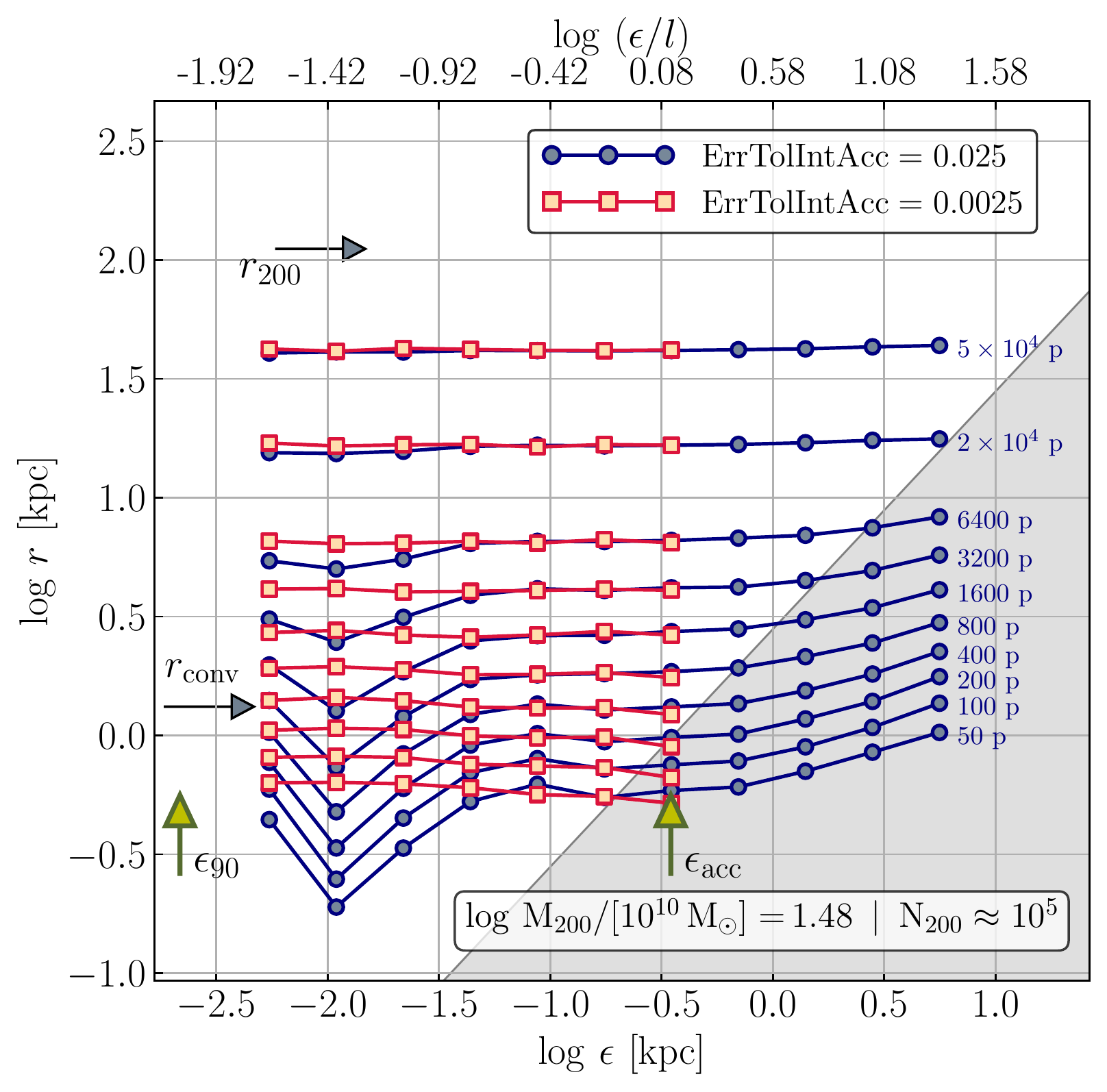}}\quad
  \subfloat{}{}{\includegraphics[width=.4\textwidth]{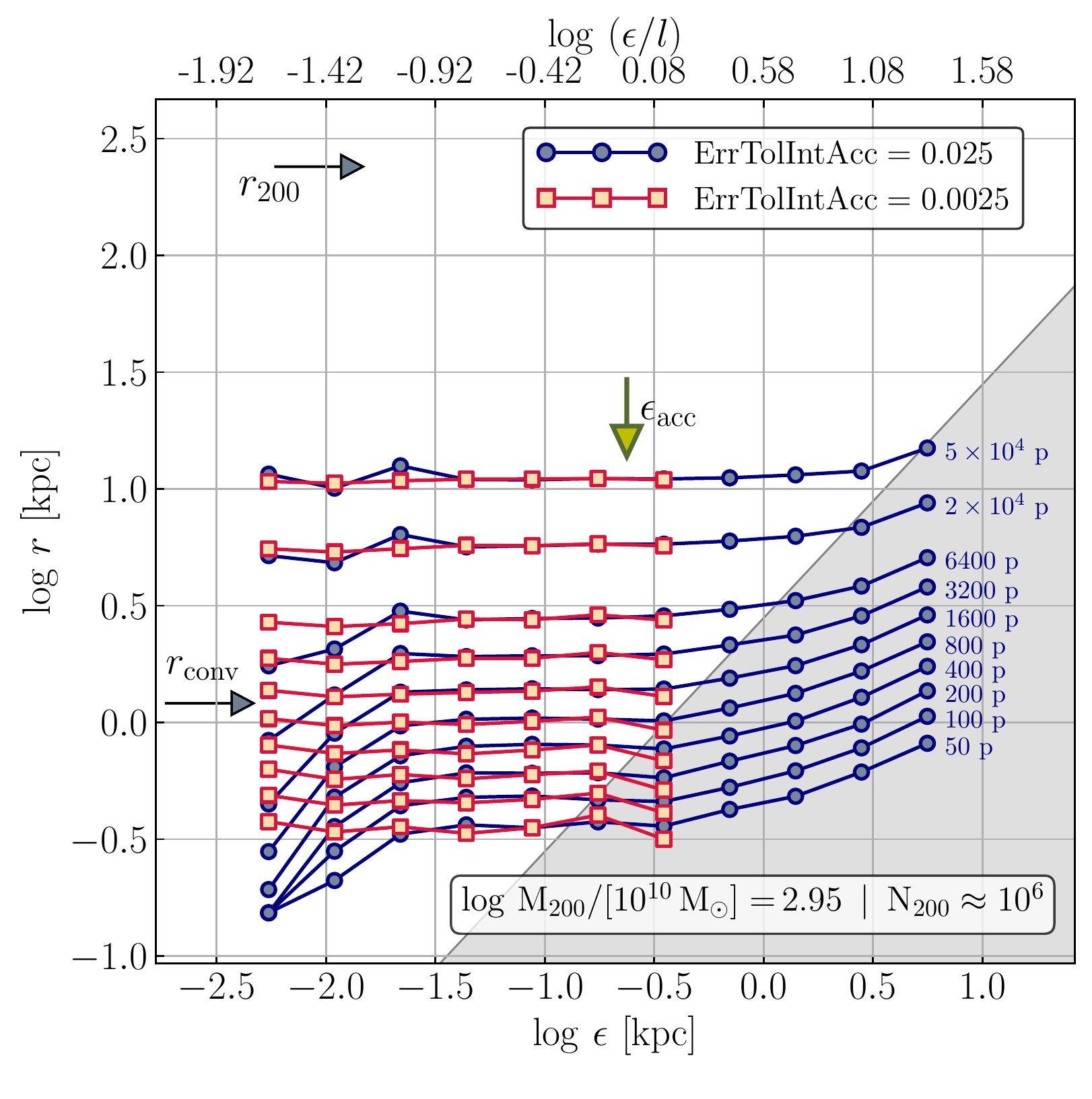}}
  \caption{Average radii enclosing fixed numbers of particles as a function of the 
    softening length, $\epsilon$, for haloes in four separate mass bins. All runs used 
    $N_{\rm p}=376^3$ particles. Connected blue circles
    correspond to runs carried out with {\sc gadget}'s default integration accuracy parameter, 
    {\tt ErrTolIntAcc}=0.025, and red squares to runs with {\tt ErrTolIntAcc}=0.0025. From top
    left to bottom right, panels correspond to halo masses equivalent to $N_{200}\sim 10^3$,
    $10^4$, $10^5$ and $10^6$ DM particles. The grey shaded regions highlight
    $r\le\epsilon_{\rm sp}=2.8\times\epsilon$, the inter-particle length scale
    beyond which forces become exactly Newtonian. Green arrows mark two estimates 
    of softening required to suppress discreteness effects, $\epsilon_{\rm acc}$ and $\epsilon_{90}$
    (see section~\ref{SSecPrelim} for details). The right-pointing arrows mark the virial
    radius, $r_{200}$, and P03's convergence radius, $r_{\rm conv}$, for each mass bin.}
  \label{fig:eps_int}
\end{figure*}

The most comprehensive attempt to establish the impact
of numerical parameters on halo mass profiles has been the work of P03.
Their results imply that, provided timestep size, softening and 
starting redshift are wisely chosen, particle number is the primary factor determining 
convergence. 

One limitation of the work carried out by P03 and N10
is that they were based on simulations of a
{\em single} $\sim 10^{12}\,{\rm M_\odot}$ dark matter halo that was
resolved with {\em at least} $\sim 10^4$ particles, over 300 times that of the poorest resolved
systems in typical cosmological runs. Can the conclusions of P03 be extrapolated to haloes with
$\sim 10^2$ particles, or fewer? Is the P03 radial convergence criterion valid for {\em stacked}
mass profiles, or for haloes whose masses differ substantially from $\sim 10^{12}\,{\rm M_\odot}$?
We devote this section to addressing these questions.

\subsection{Integration accuracy}
\label{SSecIntAcc}

The central regions of dark matter haloes are difficult to simulate due to their high
densities, which reach many orders of magnitude above the cosmic mean.
Crossing times there are $\sim 10^{-3}$ to $10^{-4}\times t_H(z)$ implying that thousands of
orbits per particle must be integrated to ensure accuracy and reliability. 
Adding to this, haloes are not smooth and integration errors
may accumulate during close encounters between particles. To prevent this, integration
must be carried out with a minimum (but a priori unknown) level of precision.
In \textsc{gadget}, particles take adaptive timesteps of length $\Delta t=\sqrt{2\eta\epsilon/|\mathbf{a}|}$,
where $|\mathbf{a}|$ is the magnitude of the local acceleration, and $\eta$ is a parameter that
allows some additional control over the step size. Clearly smaller timesteps are required
when $\epsilon$ is small. In the \gadget parameter file, $\eta$ is referred to as ${\tt ErrTolIntAccuracy}$
and takes on a default value of 0.025.

Figure~\ref{fig:eps_int} plots the average enclosed mass profiles of haloes drawn from our $N_{\rm p}=376^3$ run,
and highlights the importance of accurate integration. Panels correspond to different
mass bins, which were selected so that (from top left to bottom right) $N_{200}\approx 10^3$,
$10^4$, $10^5$ and $10^6$. The curves show the radii, $r_{\rm N_p}$, enclosing a given number of particles
(indicated to the right of each curve) and are plotted as a function of the gravitational softening, 
$\epsilon$. As a guide, the virial radius for each mass bin is
indicated by the arrow on the left side of each panel. Connected (blue) circles show results for \gadget's
default integration accuracy parameter, $\tt{ErrTolIntAcc}=0.025$. 

For comparison, the vertical green arrows in each panel mark two estimates of minimum softening 
needed to suppress discreteness effects. The first, $\epsilon_{\rm acc}=r_{200}/\sqrt{{\rm N}_{200}}$,
ensures that the {\em maximum} stochastic acceleration felt by a particle ($\sim Gm/\epsilon^2$) remains
smaller than the {\em minimum} mean field acceleration of the halo ($\sim GM_{200}/r_{200}^2$). The other,
$\epsilon_{90}=2\,r_{200}/{\rm N}_{200}$, is less restrictive and is required to prevent large-angle deflections
during two-body encounters. 

There are a couple points to note in this figure. First, central densities are noticeably suppressed 
for radii smaller than the ``spline'' softening length, i.e. $r \simlt \epsilon_{\rm sp}=2.8\times \epsilon$
(in \gadget, pairwise forces become exactly Newtonian for separations larger than $\epsilon_{\rm sp}$).
This can be seen by noting that all radii appear to increase slightly once they cross into the grey shaded region, 
which delineates $r=\epsilon_{\rm sp}$. Spatial resolution is clearly compromised at {\em all}
radii $\simlt \epsilon_{\rm sp}$.
More interestingly, however, the central regions of haloes show a clear increase in central density 
(i.e. the blue lines curve downward) when the softening parameter is reduced below a particular value. The 
value of $\epsilon$ at which this occurs depends on particle number, with more massive systems exhibiting 
symptoms at smaller $\epsilon$. Note too that further reducing $\epsilon$ does not result in a
denser centre: for very small $\epsilon$, densities are again reduced. The asymptotic central 
density therefore depends non-monotonically on $\epsilon$, suggesting a numerical origin. 

The connected (red) squares show, for a subset of $\epsilon$, the same results but for a
series of runs in which $\eta$ was reduced from 0.025 to 0.0025 (this
increases the total number of timesteps by a factor of roughly $\sqrt{10}\approx 3.16$). These curves
are clearly flatter, suggesting that halo mass profiles are robust to changes in softening across a
wide range of masses provided: a) $\epsilon$ is sufficiently small so that $r\simgt \epsilon_{\rm sp}$,
and b) care is taken to ensure particle orbits are resolved with a sufficient number of timesteps. For
the remainder of the paper, all results from runs for which $\epsilon < \epsilon_{\rm fid}$ were carried
out with {\tt ErrTolIntAcc}=0.0025, unless stated otherwise.

These results are qualitatively consistent with those of P03, but differ in the details. These
authors report that \gadget runs with fixed timestep developed artificially dense ``cuspy'' 
centres, where resolution is poor. Based on this, they argue for an $\epsilon$-dependent
adaptive timestepping criterion, $\Delta t_i=\sqrt{\eta\, \epsilon_i/a_i}$, where $\eta=0.04$.
This is the same criterion adopted for our runs, but with a considerably larger value
of $\eta$, perhaps due to the much smaller softening parameters tested in our study. This may
indicate the need for a timestepping criterion with a stronger dependence on $\epsilon$. We 
defer this task to future work. 

\begin{figure}
  \includegraphics[width=0.47\textwidth]{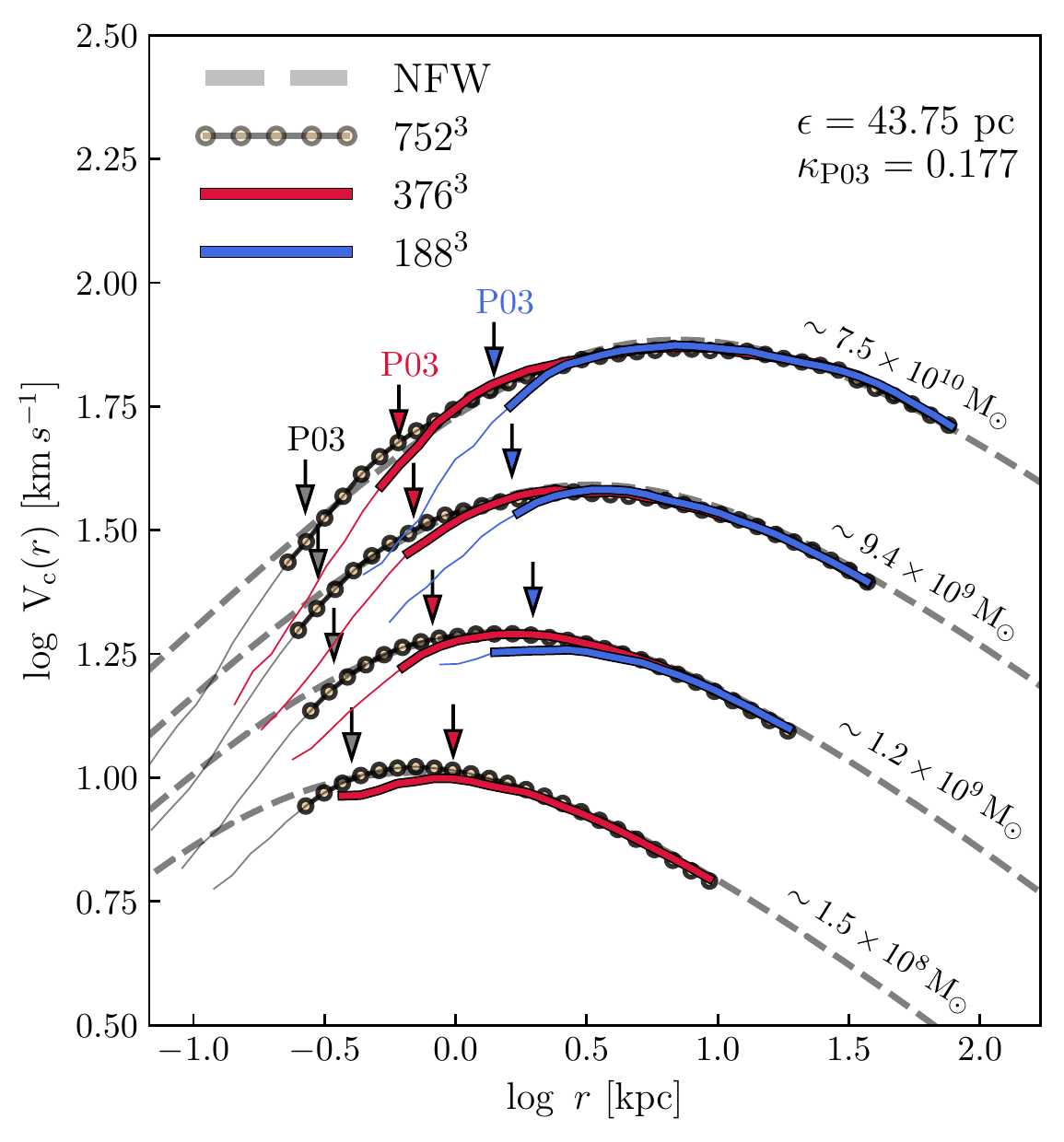}
  \caption{Median circular velocity profiles for haloes in four distinct mass bins.
    Each run used the same softening length, 
    $\epsilon=43.75\, {\rm pc}$, but a different total number of particles:
    $N_{\rm p}=752^3$ (grey circles), $N_{\rm p}=376^3$ (red lines) and $N_{\rm p}=188^3$
    (blue lines), corresponding to a factor of 64 in mass resolution. 
    Dashed lines show NFW profiles with a
    concentration parameter equal to the median values for haloes in our $N_{\rm p}=752^3$ run
    in the same mass bins. Thick lines (or points in the case of $N_{\rm p}=752^3$) extend down to
    the radius beyond which the circular velocity profiles agree with the theoretical ones
    to within $\sim$10 per cent; thin lines extend the curves to radii enclosing $\sim 20$ 
    particles. Downward pointing arrows mark the P03 ``convergence radius'' for
    $\kappa_{\rm P03}=0.177$ (eq.~\ref{eqP03}).}
  \label{fig:Vcirc}
\end{figure}

\begin{figure}
  \includegraphics[width=0.47\textwidth]{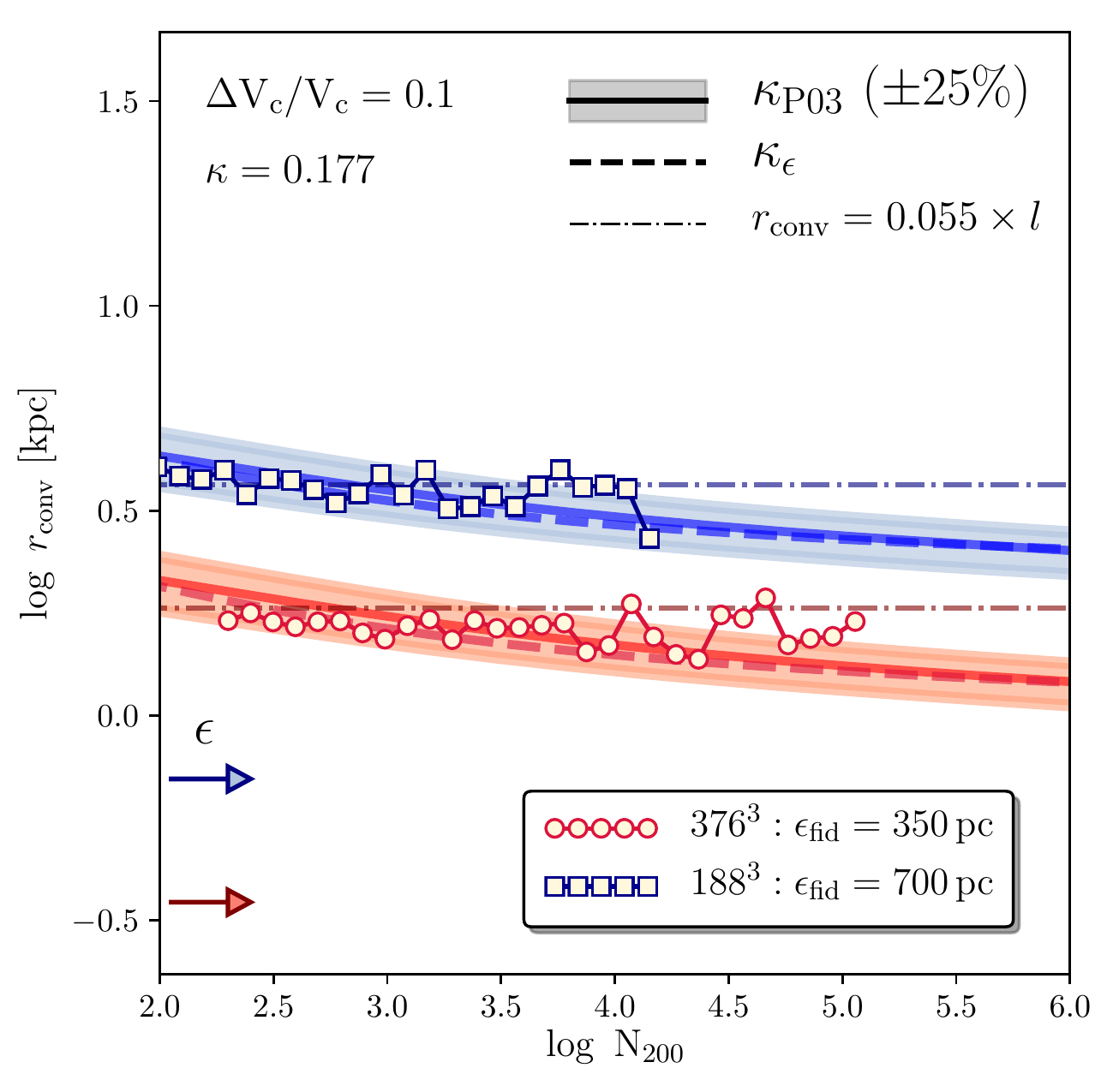}
  \caption{``Convergence radius'' beyond which the median circular velocity profiles
    converge to within 10 per cent of those in our $N_{\rm p}=752^3$ run
    plotted as a function of particle number, $N_{200}$. Red circles correspond 
    to results from $N_{\rm p}=376^3$ runs, blue squares to $N_{\rm p}=188^3$. Both runs 
    adopted the fiducial softening length for their particle mass. Only bins containing
    at least 8 haloes have been included. Solid lines show
    $r_{\rm conv}$ computed using the P03 criterion (eq.~\ref{eqP03}) for each resolution,
    assuming $\kappa_{\rm P03}=0.177$ (shaded regions show variations brought about
    by a $\pm$25 per cent change in $\kappa_{\rm P03}$). Dashed lines show the convergence radius 
    implied by eq.~\ref{eqKapLud} for $\epsilon_{\rm fid}$ and $\kappa_\epsilon=0.177$. Horizontal
    dot-dashed lines show eq.~\ref{eqp03ApproxII}, which approximates $r_{\rm conv}$ as a 
    fixed fraction of the mean inter-particle spacing.}
  \label{fig:VcircConv}
\end{figure}

Provided a sufficient number of timesteps are taken, none of our runs appear to
be adversely affected by $\epsilon$ being {\em too small}. Indeed, enclosed mass profiles
appear stable at all $r\simgt \epsilon$ even when $\epsilon\simlt \epsilon_{90}$, suggesting that hard scattering
does not unduly influence the innermost structure of DM haloes, at least for the values of
$\epsilon$ tested here. This may be because such scattering events are
sufficiently rare (occurring with a cross-section proportional to $\epsilon^2$)
that their cumulative effects are dynamically unimportant across a Hubble time.
Averaged over much {\em longer} timescales, scattering-induced cores {\em do}
develop in idealized simulations of isolated NFW haloes (see, for example,
Figure 6 of \citealt{vandenBosch2018a}, who study the secular evolution of an
NFW halo for $\approx 60$ Gyr).

These results suggest that the median mass profiles of dark matter haloes
are remarkably robust to changes in gravitational softening provided it is not
so {\em large} that it compromises enclosed masses. This is true
even for haloes containing as few as $N_{200}=10^3$ particles, and for radii
enclosing as few as $50$. This does not, however,
imply numerical convergence. Indeed, provided other numerical parameters
are carefully chosen, 2-body relaxation places much stronger constraints 
on convergence (P03), and dense haloes centres -- which occupy only a small fraction of 
their total mass and volume -- are highly susceptible to low particle number.
The systematic effects of 2-body scattering on the innermost mass profiles of dark
matter haloes must therefore be carefully considered when seeking to quantify numerical
convergence. As discussed above, a useful tactic for separating converged from unconverged parts of a halo
is to identify the radius at which the collisional relaxation time is some multiple 
$\kappa\equiv t_{\rm rel}/t_H(z)$ of a Hubble time (P03). We turn our attention to this
in the following subsections.

\subsection{Median circular velocity profiles} 

\begin{figure*}
  \includegraphics[width=1\textwidth]{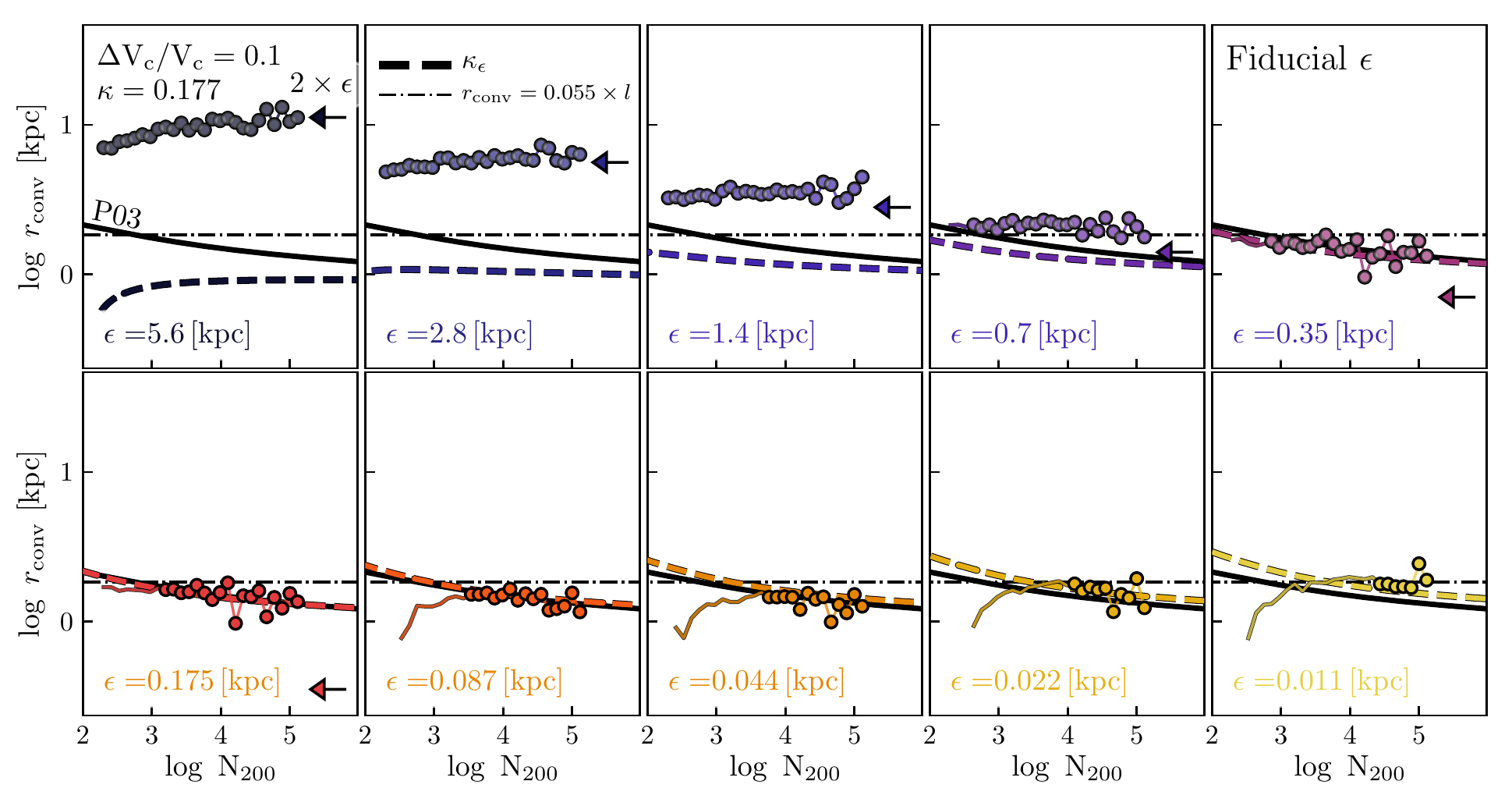}
  \caption{Radii beyond which median circular velocities converge to within 10 per cent as a function of $N_{200}$
    for $N_{\rm p}=376^3$ runs carried out with different gravitational softening lengths.
    Only bins containing at least 8 haloes are shown.
    The solid black curve shows the predicted convergence radius using the P03 criterion
    for $\kappa_{\rm P03}=0.177$; dashed lines show the predictions of eq.~\ref{eqKapLud} for each
    value of $\epsilon$, again for $\kappa_\epsilon=0.177$. Both sets of curves were constructed using
    the mass-concentration relation predicted by the model of \citet{Ludlow2016}, assuming a
    Planck cosmology. The horizontal dot-dashed lines show eq.~\ref{eqp03ApproxII}, which approximates 
    $r_{\rm conv}$ as a fixed fraction of the mean inter-particle spacing. For consistency with several subsequent 
    figures, points and lines have been colour-coded by 
    softening length, which decreases from top-to-bottom, left-to-right. Note that softening lengths
    {\em larger} than the expected $r_{\rm conv}$ compromise spatial resolution and result in {\em measured}
    convergence radii of $\approx 2\times\epsilon$ (indicated using coloured
    arrows). For $\epsilon$ smaller than this value the minimum spatial resolution is set by 2-body
    relaxation, and is essentially independent of $\epsilon$ over the range of values studied. }
  \label{fig:VcircConvB}
\end{figure*}

Figure~\ref{fig:Vcirc} shows the median circular velocity profiles
of haloes in four separate mass bins and at three different
resolutions. Grey dashed lines show NFW fits to the profiles of haloes in our 
$N_{\rm p}=752^3$ run (grey lines and circles). Because these curves agree well with the simulated
profiles over a large radial range, we can use them to estimate the convergence radius by identifying
the point at which the simulated profiles first dip below the theoretical 
ones by a certain amount. Thick lines (or points in the case of ${\rm N_p}=752^3$)
cover radii for which $V_c$ departs from dashed lines by less than 10 per cent;
thin lines extend to radii enclosing ${\rm N}\geq 20$ particles. 

Note that measured convergence radii are, for a given resolution, only weakly
dependent on halo mass (the thick segments or points end at similar
radii for a given set of curves). Haloes in our $N_{\rm p}=752^3$ run, for example,
have convergence radii that differ by at most $\sim$30 per cent across the entire
mass range plotted (which corresponds to a factor of $512$ in $N_{200}$ and $8$ 
in $r_{200}$), and the difference is similar for the other resolutions. Two haloes with 
the same {\em total} number of particles therefore have very different convergence radii 
depending on their mass. For example, a halo of mass $1.2\times 10^9\,{\rm M_\odot}$
in our 752$^3$ run has a measured convergence radius of $\approx 0.60\, {\rm kpc}$,
whereas $7.5\times 10^{10}\,{\rm M_\odot}$ haloes in our 188$^3$ run--resolved with the same
number of particles--$r_{\rm conv}\approx 3.5\, {\rm kpc}$, about a factor of 6 larger.

This scaling is indeed expected from eq.~\ref{eqxconv}. 
Neglecting the weak dependence of concentration on halo mass, two haloes with the same number of
particles identified in simulations of different mass resolution, will have
convergence radii that resolve comparable fractions of their virial radii, but differ, on
average, by a factor
$(r_{200}^{\rm A}/r_{200}^{\rm B})\propto (m_{\rm p}^{\rm A}/m_{\rm p}^{\rm B})^{1/3}=({\rm M_{200}^{\rm A}}/{\rm M_{200}^{\rm B}})^{1/3}$,
indicating a smaller convergence radius in the higher-resolution run.
The downward pointing arrows in Figure~\ref{fig:Vcirc} mark Power's convergence radii
for each simulation volume and mass bin, which agree well with these empirical estimates.
Note that these convergence radii have been approximated using the NFW profiles assuming
$\kappa_{\rm P03}=0.177$, smaller than the value advocated by P03. We discuss possible
reasons for this difference in Section~\ref{SecConclusion}.

\subsection{The convergence radius of collisionless cold dark matter haloes}
\label{SSecRconvSim}

\subsubsection{Dependence on halo mass}

The convergence radius can also be calculated explicitly by comparing the circular velocity
profiles of haloes to those of the same mass but in a higher-resolution simulation. 
Figures~\ref{fig:VcircConv} shows, as a function of $N_{200}$, the radius at which the
median profiles in our $N_{\rm p}=376^3$ (red circles) and 188$^3$ (blue squares) first
depart from those in the $N_{\rm p}=752^3$ run by more than 10 per cent. Note that
only bins containing at least 8 haloes have been included.

For comparison, we also plot the convergence radii expected from eq.~\ref{eqP03}
as solid lines of corresponding colour (assuming $\kappa_{\rm P03}=0.177$; 
smaller than the value of $\approx 0.6$ advocated by P03 for $\Delta V_c/V_c = 0.1$),
which agree well with the measured values of $r_{\rm conv}$. Our measurements are also 
recovered well by eq.~\ref{eqKapLud}, which is shown using dashed lines in Figure
\ref{fig:VcircConv} for $\kappa_\epsilon=0.177$. Both sets of curves were constructed
using NFW haloes that follow the mass-concentration relation of \citet{Ludlow2016}, assuming a Planck cosmology,
and, for the case of eq.~\ref{eqKapLud}, the softening length adopted for each particular 
run. The shaded region highlights the expected change in $r_{\rm conv}$ brought about by 
increasing or decreasing $\kappa_{P03}$ by 25 per cent. 

Note, however, that eqs.~\ref{eqP03} and ~\ref{eqKapLud} predict that $r_{\rm conv}$ should depend
weakly on mass. While not inconsistent with our numerical results, a much simpler 
approximation for $r_{\rm conv}$ can be obtained from eq.~\ref{eqxconvIII} after we specify a 
value of $C$. We find $C\approx 2.44$ provides a conservative upper-limit on $r_{\rm conv}$, leading to the 
following approximation for the P03 convergence radius:
\begin{align}
  r_{\rm conv} & \approx 0.77\, \biggr(\frac{3\, \Omega_{\rm DM}}{800\,\pi}\biggl)^{1/3}\,l(z)\label{eqp03ApproxI} \\
               & \approx 5.5\times 10^{-2}\,l(z)\label{eqp03ApproxII},
\end{align}
which is valid for $\Delta V_c/V_c \approx 0.1$. The dot-dashed horizontal lines in Figure~\ref{fig:VcircConv}
show eq.~\ref{eqp03ApproxII} for our $N_p=188^3$ (blue) and $376^3$ runs (red). 

\subsubsection{Dependence on gravitational softening}

Based on the discussion in Section~\ref{SSecRconv}, we expect $r_{\rm conv}$ to depend slightly but
systematically on $\epsilon$, particularly for haloes resolved with small numbers of particles.
Figure~\ref{fig:VcircConvB} plots $r_{\rm conv}$ versus $N_{200}$ for a series of
$N_{\rm p}=376^3$ runs carried out with different softening lengths. As before, $r_{\rm conv}$
is estimated by comparing the point at which these profiles first depart from
those in our highest resolution run ($N_{\rm p}=752^3$, $\epsilon=43.75 \, {\rm pc}$)
by a certain amount. Only bins containing at least 8 haloes are shown. All panels
shows results for $\Delta V_c/V_c= 0.1$, with $\epsilon$ decreasing from top to bottom, 
left to right by a factor of two between consecutive panels. For each value of $\epsilon$ 
we use filled circles to indicate $N_{200}>2\, r_{200}/r_{\rm conv}$
(or equivalently, $r_{\rm conv}\simgt \epsilon_{90}$),
which, based on our analytic estimates, should be unaffected by
large-angle scattering of particles during close encounters (see Section~\ref{SSecPrelim}).
The solid black line in each panel shows the convergence radii expected from
eq.~\ref{eqP03}; dashed lines show eq.~\ref{eqKapLud}, which depend explicitly on $\epsilon$
(both assume $\kappa=0.177$). The dot-dashed horizontal lines show eq.~\ref{eqp03ApproxII}.

When $\epsilon$ is {\em large}, the measured convergence radii scale roughly 
as $r_{\rm conv}\sim 2\times \epsilon$ (shown as arrows on the right side of each panel),
approximately independent of halo mass. Once $\epsilon$ becomes smaller than
the analytic estimates of $r_{\rm conv}$ (solid, dashed or dot-dashed lines), the
measured values bottom-out and exhibit, at most, a weak dependence on $\epsilon$ and $N_{200}$
thereafter.
The weak mass-dependence is described reasonably well by eqs.~\ref{eqP03} and ~\ref{eqKapLud},
but may also be approximated by the much simpler relation, eq.~\ref{eqp03ApproxII}, in which
$r_{\rm conv}$ is a fixed fraction of the mean inter-particle spacing, regardless of 
mass. We conclude that, provided $\epsilon$ is negligibly small, 
eqs.~\ref{eqP03} and~\ref{eqKapLud} provide a reasonable upper limit to the 
values of $r_{\rm conv}$ measured from the median $V_c(r)$ profiles of haloes composed of
as few as $N_{200}\approx 100$ particles, provided $\kappa_{\rm P03}\approx 0.177$ (for $\Delta V_c/V_c=0.1$). 
The dependence of $t_{\rm relax}$ on $\epsilon$ anticipated from eq.~\ref{eqKapLud} adds only a minor
correction, but may become increasingly important as $\epsilon$ becomes arbitrarily small. 

\begin{figure}
  \includegraphics[width=0.47\textwidth]{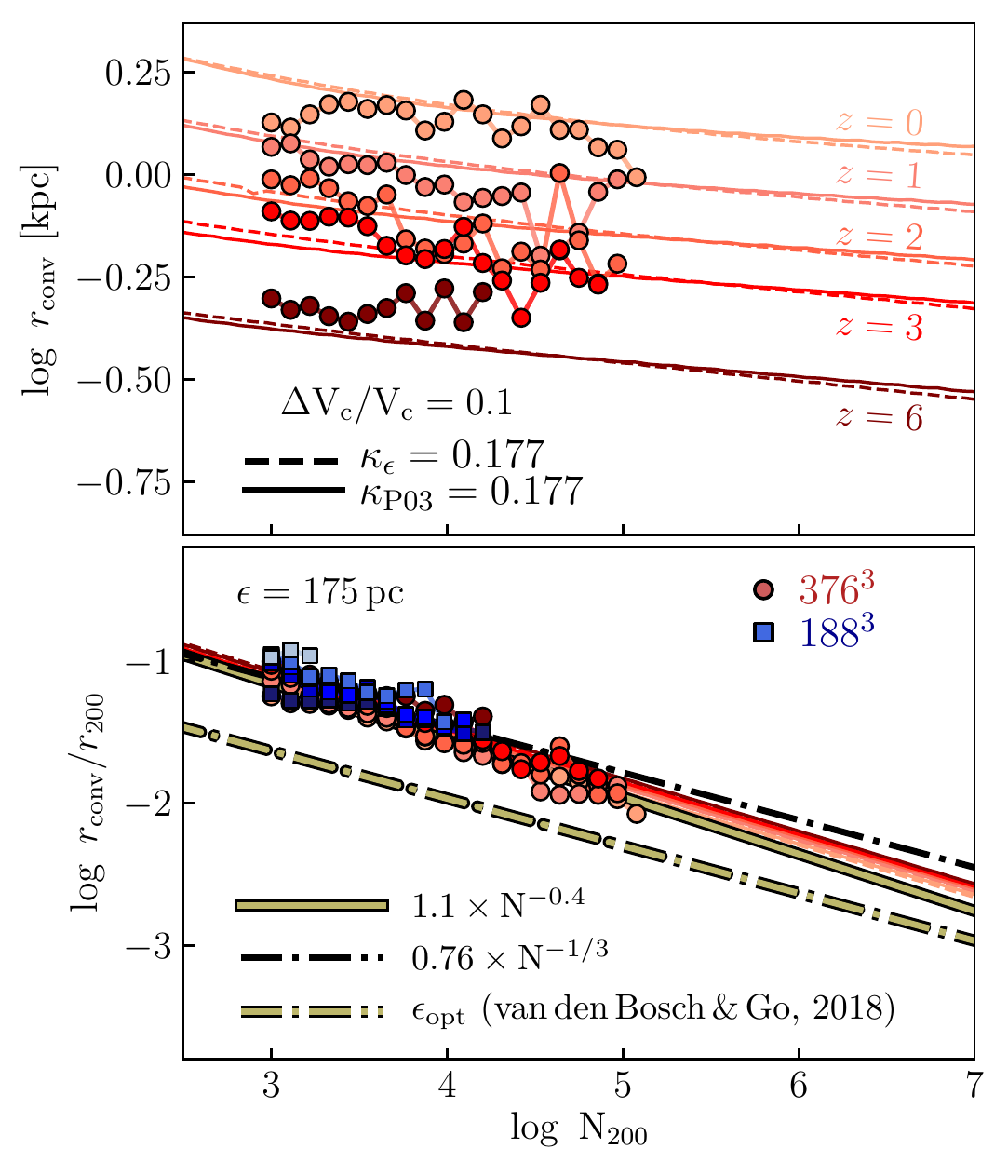}
  \caption{Physical convergence radii for $\Delta V_c/V_c = 0.1$, plotted as a
    function of $N_{200}$ for haloes identified at different redshifts.
    The upper panel shows results from our $N_{\rm p}=376^3$ run carried out with
    $\epsilon=175\,{\rm pc}$ ($\epsilon_{\rm fid}/2$).
    These results are again plotted in the lower panel, but after rescaling
    $r_{\rm conv}$ by $r_{200}(z)$. Blue squares in the lower panel show results from
    our $N_{\rm p}=188^3$ runs carried out with the same softening length. The thick solid line
    in the lower panel shows the scaling $r_{\rm conv}/r_{200}=1.1\times N_{200}^{-0.4}$, which
    approximates our numerical results reasonably well; the dashed line shows the ``optimal''
    softening for ($c=10$) NFW haloes advocated by \citet{vandenBosch2018a}. The thick dot-dashed
    line corresponds to $r_{\rm conv}/l=0.055$ (eq.~\ref{eqp03ApproxII}).}
  \label{fig:rc_z}
\end{figure}

\subsubsection{Dependence on redshift}

Convergence radii anticipated from eqs.~\ref{eqP03} and \ref{eqKapLud} depend not only on enclosed 
particle number, $N(r)$, and gravitational softening, but also on redshift. We show this explicitly 
in the upper panel of Figure~\ref{fig:rc_z}, where we plot the convergence radii of haloes
identified in one of our $N_{\rm p}=376^3$ runs at several different redshifts (this run used $z_{\rm phys}=2.8$ and a 
maximum physical softening length of $\epsilon_{\rm fid}/2=175\, {\rm pc}$). As in previous Figures, the solid and dashed
lines show the analytic estimates of $r_{\rm conv}$ expected from eqs.~\ref{eqP03} and \ref{eqKapLud}, which describe
the numerical results reasonably well. Convergence radii clearly depend on redshift in a way that
is well captured by these simple analytic prescriptions.

Note too the strong redshift-dependence:
from $z=0$ to $z=6$, for example, physical convergence radii vary by as much as a factor
of $(1+z)=7$ at essentially all mass scales probed by our simulations,
consistent with the redshift dependence of $r_{200}(z)$ or $l(z)$. This result
may at first seem puzzling, but is indeed expected from eq.~\ref{eqP03}: haloes that follow a {\em universal}
NFW mass profile whose concentration depends only weakly on mass and redshift should have
convergence radii that scale approximately as $r_{\rm conv}(z)\propto r_{200}(z)/N_{200}^{1/3}$.
We show this explicitly in the lower panel of Figure~\ref{fig:rc_z}, where we have rescaled the
convergence radii above by $r_{200}(z)$, and included one of our $N_{\rm p}=188^3$ runs (blue points).
All curves now follow a similar scaling, implying that the {\em co-moving} convergence radius
is largely independent of redshift. The dot-dashed black line in the lower panels shows
eq.~\ref{eqp03ApproxII}, for which $r_{\rm conv}$ is a fixed fraction of the mean inter-particle
spacing: $r_{\rm conv}/l=0.055$. This simple approximation describes our numerical results well, but can be 
improved slightly using $r_{\rm conv}/r_{200}=1.1\times N_{200}^{-0.4}$ (heavy
solid line in the lower panels), which is slightly steeper than $N^{-1/3}$. These convergence radii are 
similar to, but slightly larger than, the ```optimal'' softening for NFW haloes advocated by \citet{vandenBosch2018a}.
For $N_{200}$ spanning $\sim 10^2$ to $\sim 10^7$, $\epsilon_{\rm opt}$
is a factor of 2 to 3 {\em smaller} than these estimates of $r_{\rm conv}$, and should therefore not compromise
spatial resolution. In addition, since $\epsilon_{\rm opt}\propto N_{200}^{-1/3}$, the ratio
$\epsilon_{\rm opt}/m_{\rm DM}^{1/3}$ remains fixed for all $N_{200}$: the softening is
optimal at {\em all masses}, regardless of $N_{200}$, making it a potentially desirable choice for
cosmological simulations of equal mass particles. 

\begin{figure}
  \includegraphics[width=0.47\textwidth]{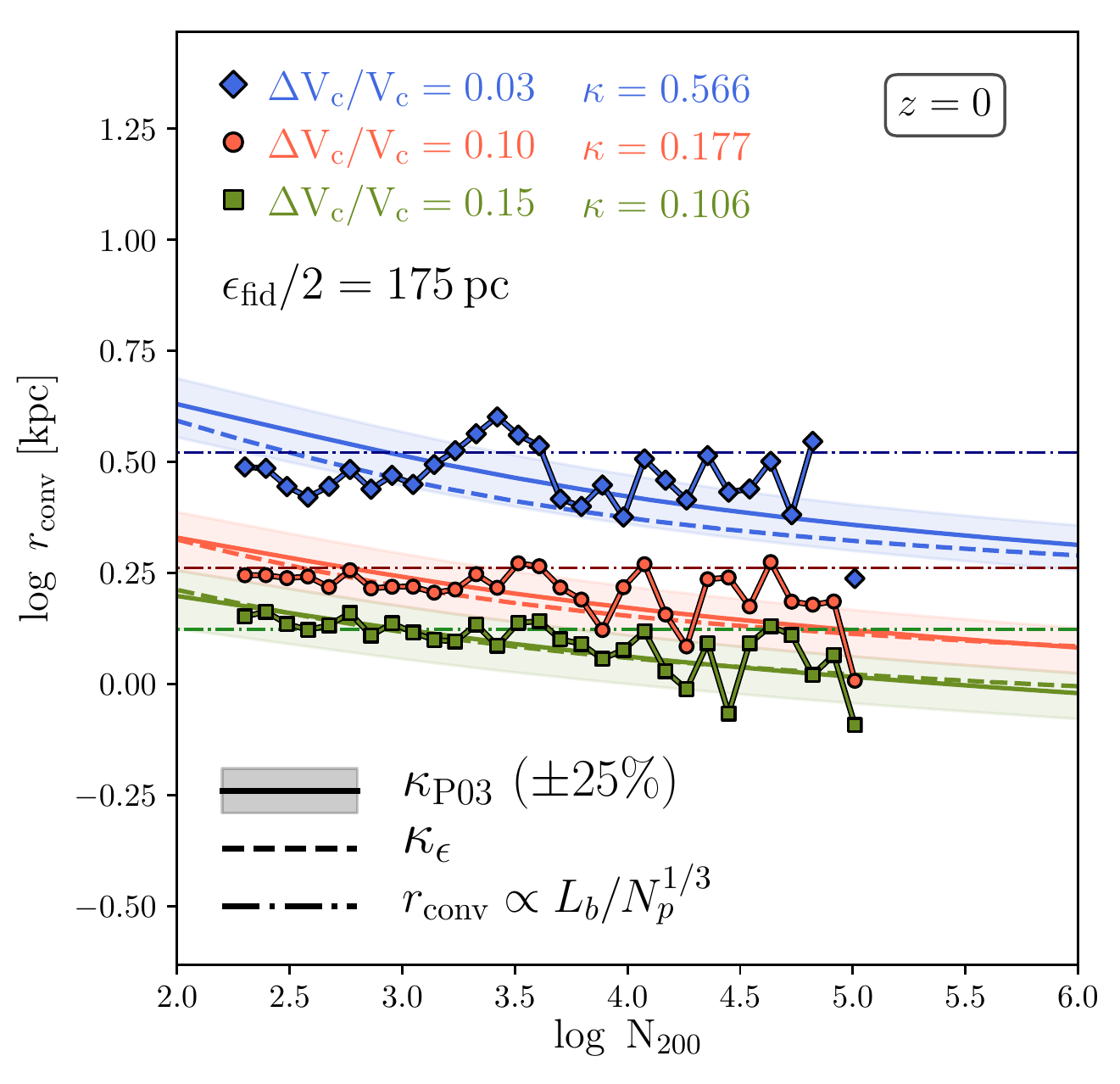}
  \caption{Convergence radii for $\Delta{\rm V_c/V_c}=0.03$. $0.1$ and $0.15$ as a function of
    $N_{200}$ for $z=0$ haloes in our $N_{\rm p}=376^3$ run. Solid lines show the
    scaling expected from eq.~\ref{eqP03} for different values of $\kappa_{\rm P03}$, with shading
    indicating the deviation expected for $\Delta\kappa_{\rm P03}/\kappa_{\rm P03}=\pm 0.25$. Dashed lines
    show eq.~\ref{eqKapLud} for the same values of $\kappa_\epsilon$. Note that better convergence requires
    higher values of $\kappa$, in agreement with N10. Dot-dashed horizontal lines show 
    constant fractions of the mean inter-particle spacing corresponding to 0.040 (green), 0.055 (red) 
    and 0.10 (blue).}
  \label{fig:rc_kappa}
\end{figure}

\begin{figure*}
  \centering
  \footnotesize
  \subfloat{}{}{\includegraphics[width=.75\textwidth]{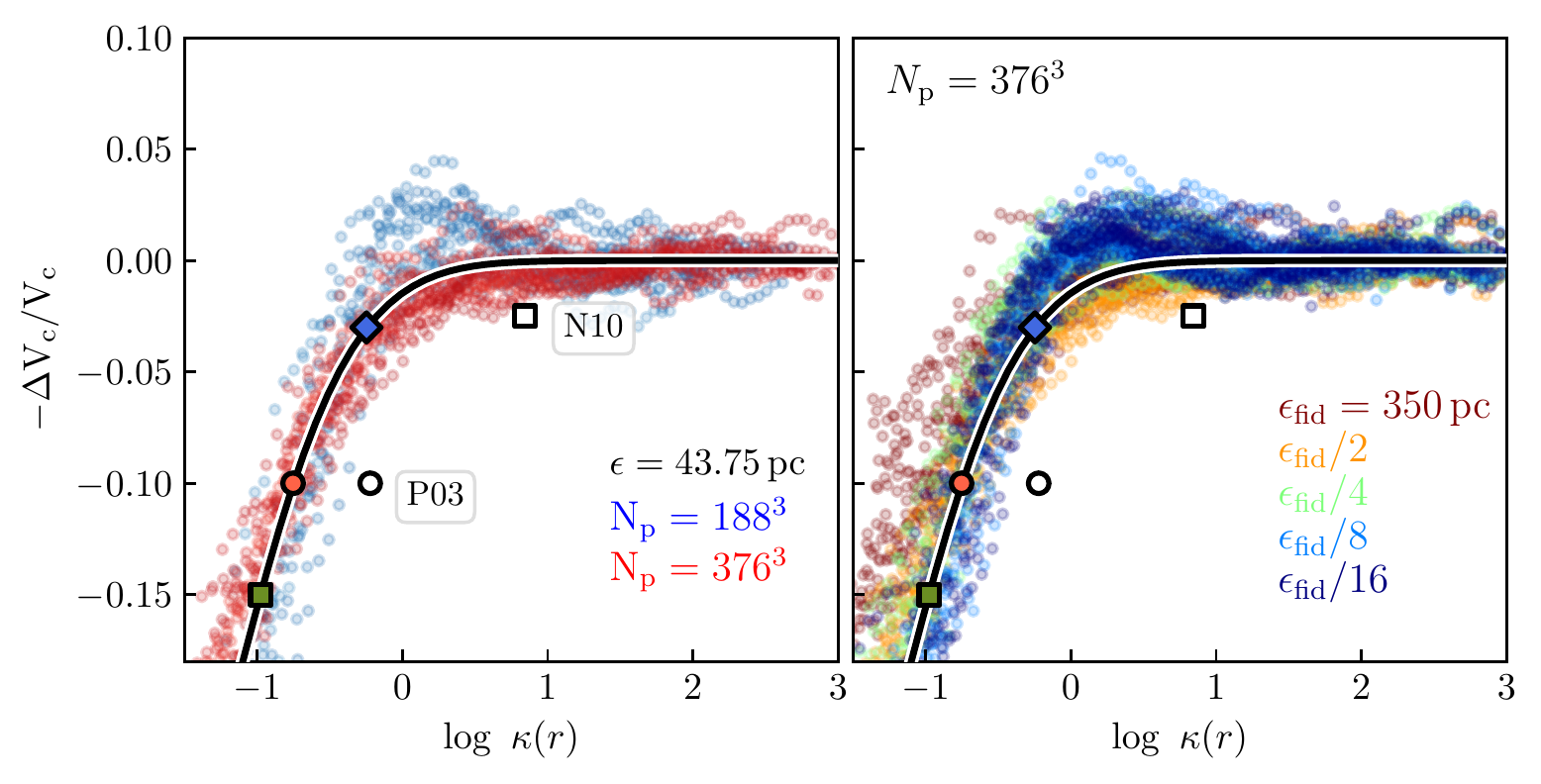}}\\
  \subfloat{}{}{\includegraphics[width=.75\textwidth]{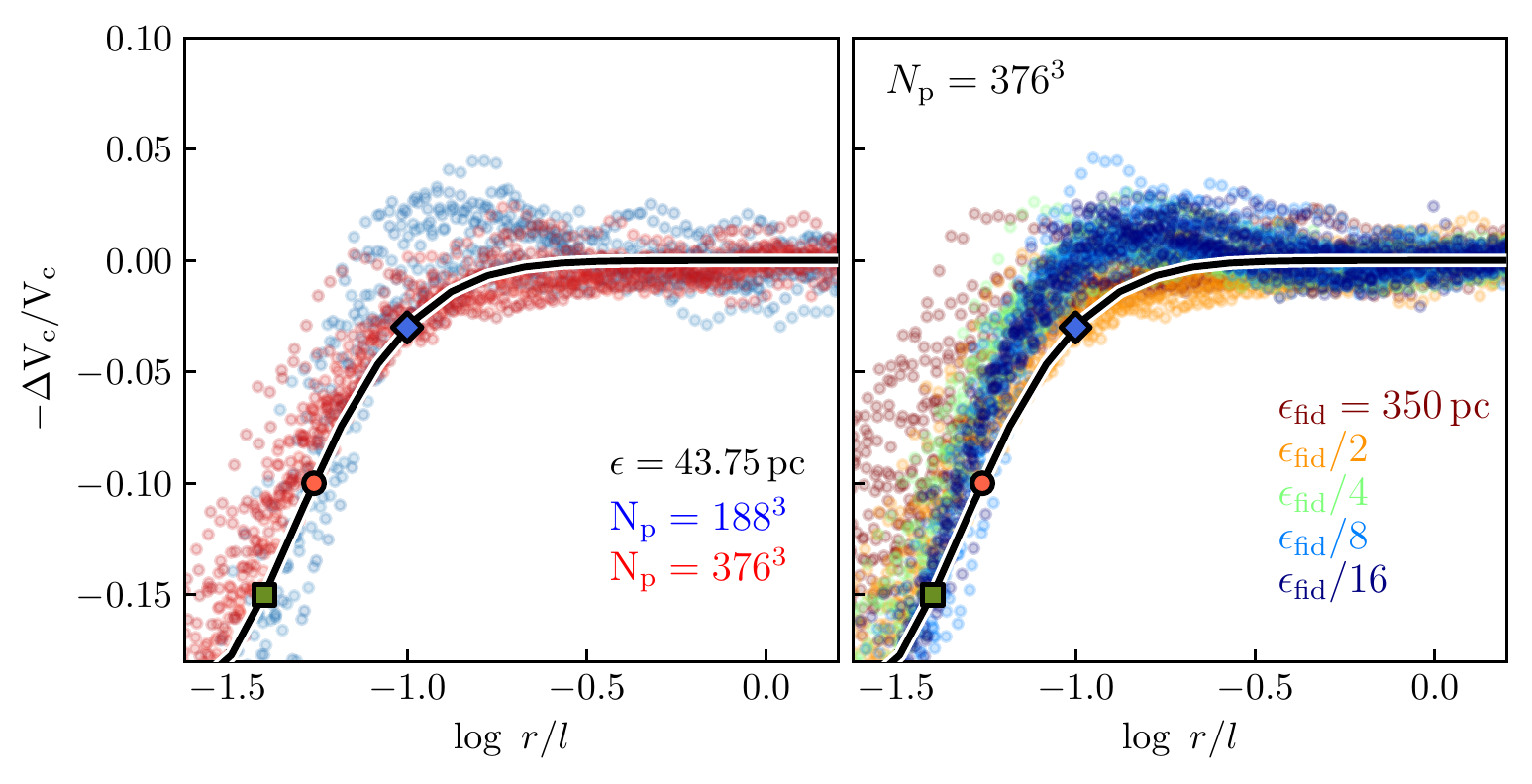}}
  \caption{Deviation in circular velocity profiles between low and high-resolution runs plotted as a function
    of $\kappa(r)\equiv t_{\rm rel}(r)/t_H(z)$ (upper panels) and $r/l$ (lower panels). The left panels show results for our $N_{\rm p}=376^3$
    (red) and $188^3$ (blue) runs for a fixed softening length of $\epsilon=43.75\, {\rm pc}$;
    the right-hand panels show $N_{\rm p}=376^3$ runs for a variety of $\epsilon$ (indicated in the
    legend). In all cases, $\Delta {\rm V_c/V_c}$ is calculated with respect to our $N_{\rm p}=752^3$,
    $\epsilon=43.75\,{\rm pc}$ run. (Note the sign convention: positive values of $\Delta{\rm V_c}$ indicate
    suppressed densities in our low-resolution runs.) Median profiles are shown in 40 
    equally-spaced bins of $\log\,M_{200}$ that span a lower limit corresponding to 
    ${\rm M_{200}}=100\times m_{\rm DM}$ up to $M_{200}\approx 10^{12}\,{\rm M_\odot}$. The white square and 
    circle in each upper panel show the empirical measurements of P03 and N10, 
    respectively, both of whom simulated a single DM halo within varying particle number; coloured points 
    correspond to values used in Figure~\ref{fig:rc_kappa}. The solid black line shows approximate empirical 
    fits to the simulation results (eq.~\ref{eq:poly}; see text for details). These may be used to estimate 
    $\kappa$ for a desired $\Delta V_c/V_c$, or to obtain directly the corresponding radius, $r/l$. }
  \label{fig:delVc_kappa}
\end{figure*}

\subsubsection{Dependence on $\kappa$}

In previous sections we estimated convergence radii in our simulations by comparing the median circular
velocity profiles of haloes in our $N_{\rm p}=376^3$ simulation with those of the same virial mass but in a higher-resolution 
simulation ($N_{\rm p}=752^3$). The radius within which profiles deviate by more than 10 per cent marked $r_{\rm conv}$. 
For analytic estimates of $r_{\rm conv}$ based on eqs.~\ref{eqP03}
and \ref{eqKapLud} this corresponds to particular values of $\kappa$, about $0.177$. However, as discussed
in detail in N10, better convergence can be obtained for higher values of $\kappa$, which occurs at larger
radii where relaxation times are substantially longer.

In Figure~\ref{fig:rc_kappa} we plot three separate estimates of convergence radii, corresponding to
fractional departures between median $V_c(r)$ profiles in our low and high-resolution runs of
$\Delta V_c/V_c=0.03$, 0.1 and 0.15. Better convergence is obtained at larger radii,
and requires larger values of $\kappa$: the solid and dashed lines show convergence radii estimated
from eqs.~\ref{eqP03} and \ref{eqKapLud}, respectively; blue, orange and green denote 
$\kappa=0.566$, 0.177 and 0.106, respectively (as before, the shaded region indicates $\pm 25$ per cent 
in $\kappa_{\rm P03}$). Horizontal dot-dashed lines indicate fixed fractions of the mean inter-particle
distance, corresponding to $r_{\rm conv}/l=0.040$, 0.055 and 0.10 for $\Delta V_c/V_c=0.03$, 0.1 and 0.15, respectively.

Figure~\ref{fig:delVc_kappa} summarizes a number of previous results. Here we plot, in the top panels, the residual difference in the
circular velocity profiles between haloes in our low- and highest-resolution runs as a function of $\kappa(r)=t_{\rm rel}(r)/t_H(z)$.
The left hand panel shows results for $N_{\rm p}=376^3$ (red) and 188$^3$ (blue); both runs used a ($z=0$) softening length of
$\epsilon=43.75\,{\rm pc}$. The panel on the right compares several $N_{\rm p}=376^3$ runs using different
softening. In all cases, residuals are calculated with respect to our $N_{\rm p}=752^3$ 
($\epsilon=43.75\,{\rm pc}$) in 40 equally-spaced bins of $\log\, M_{200}$ spanning the range $100\, m_{\rm DM}$
to $10^{12}\, {\rm M}_\odot$. White circles and squares mark the empirical results of P03 and N10, respectively;
both are more conservative than ours \citep[which supports the conclusions of][]{Zhang2018}.

As expected from previous plots, deviations in $V_c$ are largely independent of both mass and force
resolution, but correlate strongly with the enclosed relaxation timescale $\kappa(r)$.
Overall, the convergence of the {\em median circular velocity profiles} can be approximated reasonably
well by
\begin{equation}
  \frac{\Delta V_c}{V_c}(\psi)=-\log (1-10^{a\, \psi^2 + b\,\psi + c}),
  \label{eq:poly}
\end{equation}
where we have defined $\psi\equiv\log\kappa+d$, and $(a,b,c,d)=(-0.4,-0.6,-0.55,0.95)$ (heavy black line
in the upper panels of Figure~\ref{fig:delVc_kappa}).

In the lower panels of Figure~\ref{fig:delVc_kappa} we plot $\Delta {\rm V_c/V_c}$ for the same runs,
but now as a function of halo-centric distance normalized by the mean inter-particle separation, $l$.
Convergence in circular velocity is achieved at spatial scales that roughly correspond to fixed fractions
of $l$, which provides a much simpler convergence criterion than that advocated
by P03. A conservative upper limit to the convergence radius as a function of $\Delta {\rm V_c/V_c}$ can again 
be approximated by eq.~\ref{eq:poly}, but with modified parameters: $\psi\equiv\log (r/l)+d$ and 
$(a,b,c,d)=(-1.96,-0.76,-0.52,1.49)$ (thick black line in the lower panels). The outsized color points in 
Figure~\ref{fig:delVc_kappa} correspond to the curves in Figure~\ref{fig:rc_kappa}.

\subsection{Scaling relations}
\label{SSecScalings}

\begin{figure}
  \includegraphics[width=0.47\textwidth]{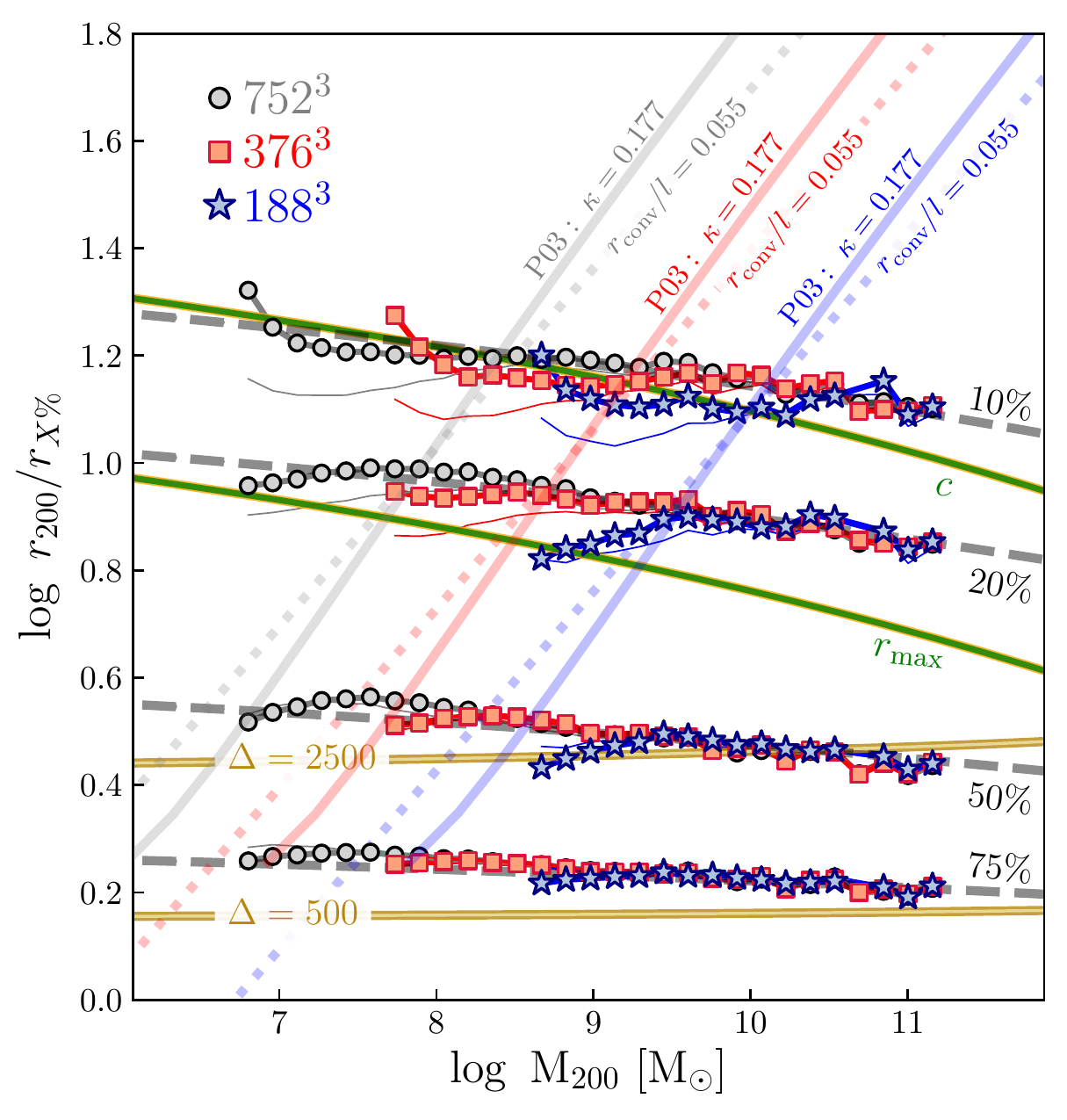}
  \caption{Ratio of the virial radius $r_{200}$ to that enclosing a fixed fraction of the 
    virial mass plotted versus $\rm{ M_{200}}$. Different coloured curves correspond
    to simulations of different mass resolution: our $N_{\rm p}=752^3$ run is shown in grey, the $376^3$ 
    run in red, and the $188^3$ in blue. Connected points correspond to runs carried out with $\epsilon=43.75\,{\rm pc}$;
    thin solid lines to our fiducial runs. From bottom to top each set of curves shows 
    the ratio of $r_{200}$ to the radius $r_{X\%}$ enclosing 75, 50, 20 and 10 per cent of the virial mass 
    ${\rm M_{200}}$, as indicated by the labels. Curves extend down to the halo masses
    corresponding to $N_{200}=32$ particles. Faint diagonal lines of corresponding colour show 
    the convergence radius anticipated from eq.~\ref{eqP03} for each mass resolution 
    (shown as solid lines and assuming $\kappa_{\rm P03}=0.177$), or from $r_{\rm conv}/l=0.055$ (dotted lines).     
    Good convergence in these scaling relations is achieved at mass scales above 
    those corresponding to $r_{X\%}\simgt r_{\rm conv}$ (i.e. to the right of the faint lines marking $r_{\rm conv}$). For comparison, heavy
    dashed grey lines show the predictions of the analytic model of \citet{Ludlow2016}, approximating
    haloes as NFW profiles. Using the same model, we also show $c=r_{200}/r_s$, $r_{200}/r_{\rm max}$, and
    two radii enclosing fixed overdensities of $\Delta=500$ and $\Delta=2500$.}
  \label{fig:scaling_converged}
\end{figure}

These results help clarify why structural scaling relations for dark matter
haloes converge even for systems resolved with only a few hundred particles, a result
we show explicitly in Figure~\ref{fig:scaling_converged}. Here we plot, as a function of
$M_{200}$, the ratio of the virial radius to the radius $r_{X\%}$ enclosing X per cent of the 
halo mass (the curve labelled ``50\%'', for example, 
is the half-mass radius--virial mass relation), for all haloes resolved with
$N_{200}\geq 32$ particles. As in previous figures, different colours
correspond to different resolutions.  Points
connected by thick lines in Figure~\ref{fig:scaling_converged} correspond to runs
carried out with fixed $\epsilon = 43.75\,{\rm pc}$, and the thin lines to the
``fiducial'' softening, which is a factor of 4 to 16 times larger, depending on
resolution. The faint diagonal lines show the P03 convergence 
radius for $\kappa_{\rm P03}=0.177$ (solid) and assuming $r_{\rm conv}/l=0.055$ (dotted). These provide a good indication
of the mass scale above which the scaling relations are converged.
Note too that, as expected, the {\em converged} relations are largely independent of $\epsilon$. 
Dashed lines show the expected trends assuming NFW profiles and the $c(M)$ relation of \citet{Ludlow2016}.
For comparison we also plot the expected concentration, $c=r_{200}/r_{-2}$, and the ratio
$r_{200}/r_{\rm max}$ using the same model, as well as radii that enclose fixed overdensities of $\Delta=500$ and
2500.

Resolving the innermost structure of
haloes is clearly challenging. Resolving $r_{-2}$ with a {\em systematic} bias in ${\rm V_c}(r_{-2})$ of 
less than $10$ per cent (assuming eq.~\ref{eqP03} with $\kappa_{\rm P03}=0.177$), for example, requires $N_{200}\approx 844,\, 1083$, and $1353$ particles for
our $N_{\rm p}=188^3$, $376^3$ and $752^3$ runs, respectively. Resolving $r_{\rm max}\approx 2.2\times r_{-2}$ is
less demanding, requiring only $N_{200}\approx 170,\, 214,$ and $266$ for the same respective $N_{\rm p}$.

\begin{figure*}
  \centering
  \footnotesize
  \subfloat{}{}{\includegraphics[width=.48\textwidth]{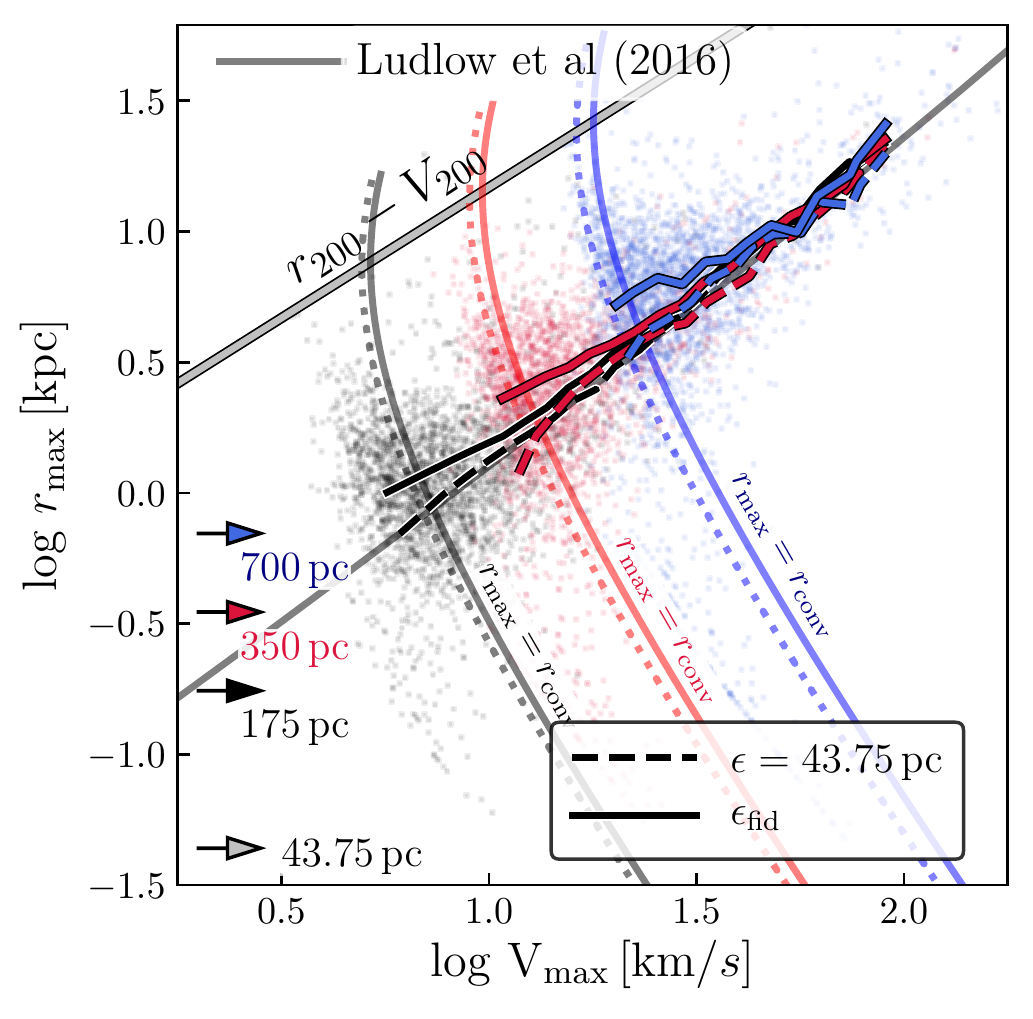}}\quad
  \subfloat{}{}{\includegraphics[width=.48\textwidth]{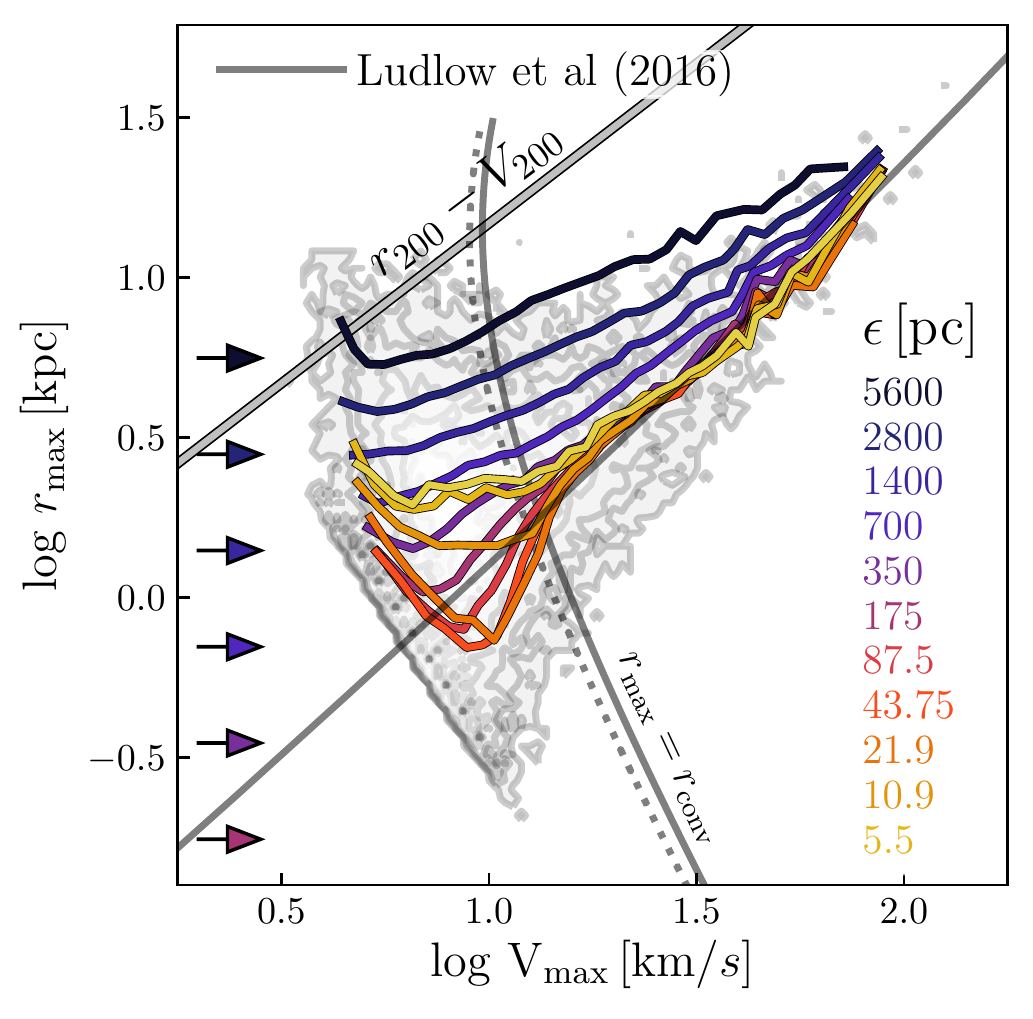}}\\
  \caption{The $r_{\rm max}-{\rm V_{\rm max}}$ relations for runs carried out with
    different mass (left) and force (right) resolution. In the left-hand panel blue, red and black lines
    correspond to runs with $N_{\rm p}=188^3,\, 376^3,\,$ and $752^3$; thick solid lines show
    median trends in bins of $M_{200}$ for runs adopting our fiducial softening length; thick 
    dashed lines correspond to $\epsilon=43.75\,{\rm pc}$.
    Only haloes with $N_{200}\geq 100$ have been used. In the right-hand panel we plot
    results from our $N_{\rm p}=376^3$ runs for a variety of softening lengths, indicated using
    different coloured lines (contours highlight the scatter among individual haloes in the 
    fiducial run). In this case all haloes, regardless of $N_{200}$, have been
    plotted. In both panels, thick grey lines show the $r_{200}-V_{200}$ relation and the
    $r_{\rm max}-{\rm V_{\rm max}}$ relation expected for pure CDM haloes \citep[see][]{Ludlow2016};
    arrows on the left indicate the values of $\epsilon$ used for the various runs.
    The $r_{\rm max}-{\rm V_{\rm max}}$ relations for haloes whose mass is fixed by
    the constraint $r_{\rm max}=r_{\rm conv}$ are shown using solid colored curves (for eq.~\ref{eqP03}, 
    with $\kappa_{\rm P03}=0.177$) or dotted curves (using eq.~\ref{eqp03ApproxII}).}
  \label{fig:VmaxRmax}
\end{figure*}

We show this explicitly in the left-hand panel
of Figure~\ref{fig:VmaxRmax}, where we plot the $V_{\rm max}-r_{\rm max}$ relations for haloes in 
runs of different resolution. Thick curves correspond to the median $r_{\rm max}$ and $V_{\rm max}$ in bins of
$M_{200}$, and extend down to a lower mass limit corresponding to $N_{200}=100$.
We adopt the same colour scheme as in previous figures and use
solid lines for our fiducial runs, and dashed lines for runs where $\epsilon$ was kept fixed
at $43.75\,{\rm pc}$. Note that curves corresponding to
a given mass resolution begin to converge when $N_{200}$ exceeds the lower limits provided above.
Convergence is not perfect, however, as systematic differences in $V_{\rm max}$ and $r_{\rm max}$
of order 10 per cent are expected at these mass scales.
Solid (faint) lines of corresponding colour, for example, show the $r_{\rm max}-V_{\rm max}$ relation for haloes
whose mass is kept fixed at those lower limits (the relevant curves are labelled $r_{\rm max}-r_{\rm conv}$).
Dotted lines show the analogous relations for convergence radii equal to $r_{\rm conv}=0.055\, l$.

Fixing $\epsilon$ at $43.75\,{\rm pc}$, well below the fiducial value,
appears to improve convergence between runs of different mass resolution even slightly {\em below }
these mass scales. Indeed, at first glance, convergence even seems better than expected
from limits imposed by 2-body relaxation. This result, however, is fortuitous, as shown in the right-hand of Figure~\ref{fig:VmaxRmax}
where $r_{\rm max}-V_{\rm max}$ relations are plotted for our $N_{\rm p}=376^3$ runs for a variety of $\epsilon$.
It is clear from this figure that a $r_{\rm max}\simgt r_{\rm conv}$ is a {\em requirement} for
convergence in the median relations (i.e. convergence is only achieved to the right of the solid or dotted 
grey lines labelled $r_{\rm max}=r_{\rm conv}$).

\subsection{Convergence Requirements for cosmological simulations}
\label{SSecReq}

\begin{figure*}
  \includegraphics[width=0.99\textwidth]{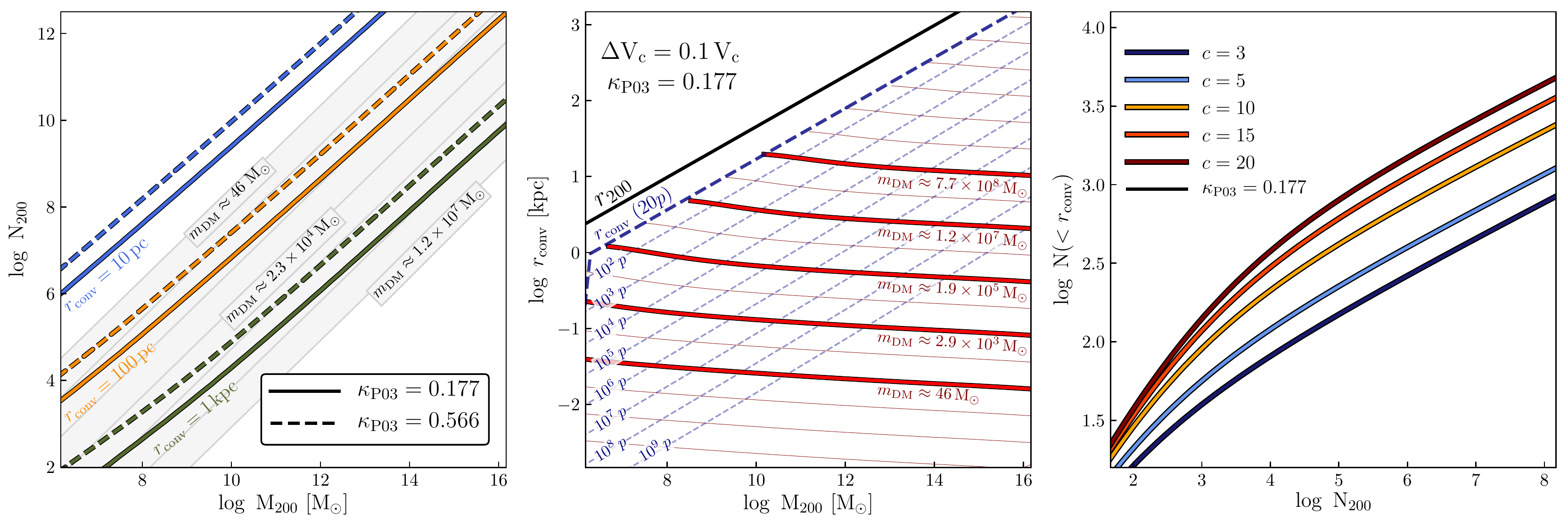}
  \caption{{\em Left panel:} Total number of particles, $N_{200}$, required
    to resolve the innermost 10$^3$, 100 and 10 pc of NFW
    haloes as a function of their mass (obtained from eq.~\ref{eqP03}). Dashed lines assume
    convergence of $V_c$ to within 3 per cent ($\kappa_{\rm P03}=0.566$) and
    solid curves to within $10$ per cent ($\kappa_{\rm P03}=0.177$). As a guide, the thin grey
    lines show $N_{200}$ versus virial mass for several values of particle mass,
    increasing from $m_{\rm DM}=46\, {\rm M}_\odot$ to $1.2\times 10^7\ {\rm M}_\odot$ by factors
    of 8. {\em Middle panel:} Power's convergence radius (for $\kappa_{\rm P03}=0.177$) as a function of
    halo mass for different mass resolutions. Thick
    burgundy lines correspond to particle masses decreasing from $7.7\times 10^8\, {\rm M}_\odot$
    by successive factors of 64; thin lines
    correspond to additional values of $m_{\rm DM}$ that differ from these by factors of 8.
    The heavy black line shows $r_{200}$ and the dashed blue line 
    the convergence radius expected for haloes of ${\rm N_{200}=20}$ particles, where all other
    curves terminate. The dashed lines show convergence radii for haloes of mass $M_{200}$ resolved
    with different numbers of particles (labelled along each line). {\em Right panel:} Number of particles
    contained within the convergence radius, $N(<r_{\rm conv})$, as a function of $N_{200}\equiv N(<r_{200})$
    for several values of the concentration parameter. All curves assume that
    haloes follow an NFW profile and a mass-concentration relation consistent with the model of 
    \citet{Ludlow2016} for the Planck cosmology.}
  \label{fig:scaling_convergedII}
\end{figure*}

The results of previous sections suggest that we can use eq.~\ref{eqP03} or ~\ref{eqKapLud} to
place restrictions on the {\em minimum} number of particles required to reach a desired spatial 
resolution in halo centres. The left panel of Figure~\ref{fig:scaling_convergedII} plots the 
total number of particles, $N_{200}$, required to resolve the innermost
$r_{\rm conv}=1\, {\rm kpc}$ (green), 0.1 ${\rm kpc}$ (orange) and 0.01 ${\rm kpc}$ (blue)
of an NFW halo as a function of its virial mass $M_{200}$. Solid and dashed lines 
correspond to $\kappa_{\rm P03}=0.177$ and 0.566 (using eq.~\ref{eqP03}), respectively.
All curves assume the mass-concentration relation predicted by the model of \citet{Ludlow2016}.
This plot highlights the difficulty
of resolving the inner regions of dark matter haloes. For example, to resolve the central
$1\,{\rm kpc}$ of a $10^{15}\, {\rm M_\odot}$ galaxy cluster with $10$ per cent accuracy
(corresponding to $\approx 0.06$ per cent of $r_{200}$) requires sampling its virial mass
with $> 10^{8.4}$ particles, comparable to the particle number required to resolve the
innermost $100\, {\rm pc}$ of Milky Way mass haloes. Note too that achieving a spatial
resolution of $\sim 10\,{\rm pc}$ for a simulated halo
of mass comparable to that of the Milky Way would require $N_{200}\sim 10^{11}$
particles, far higher than any cosmological simulation published to date. Note, however, that
these scales are baryon-dominated in hydrodynamical simulations, and the relevance of these
criteria are not obvious in that case. 

The middle panel of Figure~\ref{fig:scaling_convergedII} shows the convergence radius
(computed using eq.~\ref{eqP03}
assuming an NFW profile and $\kappa_{\rm P03}=0.177$)
as a function of halo mass expected for simulations of different uniform mass resolution
(burgundy lines), for which $m_{\rm DM}$ varies by successive factors of 8.
Heavy lines highlight several particular values of $m_{\rm DM}$. Note that for 
$m_{\rm }=1.2\times 10^7\,M_\odot$ (comparable to the particle mass in the 
$100 \, {\rm Mpc}$, $N_{\rm p}=1504^3$ \eagle simulation), convergence radii
vary from $\approx 3 \,{\rm kpc}$ for $10^{10}\, {\rm M}_\odot$ haloes, 
to $\approx 2 \,{\rm kpc}$ at ${\rm M_{200}}=10^{15}\, {\rm M}_\odot$,
equivalent to roughly 3 to 4 fiducial softening lengths. Targeting convergence radii below
$\simlt 100 \, {\rm pc}$ for haloes with virial masses $M_{200}\simgt 10^8\, M_\odot$
requires dark matter particle masses of only a few thousand ${\rm M}_\odot$; 
achieving $\simlt 10\,{\rm pc}$ resolution requires $m_{\rm DM}\approx {\rm M}_\odot$.

Dashed blue lines show, for comparison, the convergence radii as a function of mass
for fixed values of $N_{200}$, ranging from 20 particles (thick dashed line) to $N_{200}=10^9$.
This is comparable to the highest-resolution dark matter-only simulations carried
out to date -- the Aquarius \citep{Springel2008b} and Ghalo \citep{Stadel2008} simulations --
which achieve a maximum spatial resolution of order $100\,{\rm pc}$ in a Milky Way mass
dark matter halo. 
These results imply that the current state-of-the-art in cosmological hydrodynamical
simulations (e.g. \textsc{eagle}, Illustris, Apostle, FIRE, Auriga) are {\em not} fully 
converged on scales relevant for galaxy formation. 

The right-most panel of Figure~\ref{fig:scaling_convergedII} plots the number of particles
within $r_{\rm conv}$ as a function of $N_{200}$ for several different values of the (NFW)
concentration (we assumed $\kappa_{\rm P03}=0.177$ and chose values of $c$ that span the extremes
of concentrations measured in typical cosmological simulations). Once a mass profile and concentration
have been chosen, $N(<r_{\rm conv})$
depends {\em only} on $N_{200}$. Haloes with $N_{200}\approx 10^3$ will typically have between
$\approx 30$ and 100 particles within $r_{\rm conv}$ for $c=3$ and 20, respectively; haloes with
$N_{200}\approx 10^6$ will have between $\approx 260$ and 1500, respectively.

\subsection{Section summary}
\label{Sec1Summ}

Before moving on, we first summarize the results of this section:
\begin{itemize}
  
\item Provided a sufficient number of timesteps are taken, the median
  mass profiles of simulated haloes are independent of softening at virtually all radii
  similar to or greater than the spline softening
  length, $r\simgt \epsilon_{\rm sp}$. This confirms the findings of P03, but extends their results 
  to a broader range of halo mass, and to more extreme values of $\epsilon$.
  This holds true even for runs with very small softening and for haloes
  resolved with only $\sim 10^3$ particles, despite the fact that strong discreteness
  effects were naively expected for $\epsilon\simlt r_{200}/\sqrt{N_{200}}$. 
  Indeed, the enclosed mass profiles of dark matter haloes are remarkably robust to changes
  in $\epsilon$, even for much smaller values of $\epsilon$. 

\item Nevertheless, the fact that the median mass profiles of haloes are largely insensitive to $\epsilon$
  says little about the radial range over which they can be considered reliably resolved.
  For a given DM halo the {\em minimum resolved} radius, $r_{\rm conv}$, 
  depends primarily on total particle number, $N_{200}$, and roughly coincides with the 
  radius within which the enclosed 2-body relaxation time first exceeds the Hubble time by some factor 
  $\kappa$. The precise value of $\kappa$ dictates the level of convergence: we find that for $\kappa\approx 0.177$,
  median circular velocity profiles have converged to within 10 per cent; 3 per cent convergence in 
  $V_c(r)$ requires $\kappa\approx 0.566$. This is true regardless of $m_{\rm DM}$ and of $\epsilon$, provided 
  the condition $\epsilon \simlt r_{\rm conv}$ is met. Note that these values of $\kappa$ are
    smaller than those advocated by N10 and P03 for similar levels of convergence (e.g. $\kappa=7$
    for 2.5 per cent convergence, and $\kappa=0.6$ for 10 per cent), suggesting that median
    halo mass profiles may converge more easily than those of individual systems.

\item The convergence radius scales roughly as $r_{\rm conv}/r_{200}\propto N_{200}^{-1/3}$
  (see eq.~\ref{eqxconv}), implying $r_{\rm conv}/l\approx {\rm constant}$, independent of 
  redshift, halo mass and particle mass. Indeed, for $\Delta V_c/V_c=0.1$ ($\kappa_{\rm P03}=0.177$)
  we find that $r_{\rm conv}(z)=0.055\times l(z)$ describes our numerical data about as well as the
  P03 criterion, typically to within 20 per cent.
  A better approximation can be obtained with a slightly steeper dependence on $N_{200}$: 
  $r_{\rm conv}/r_{200}=1.1\times N_{200}^{-0.4}$ (see Figure~\ref{fig:rc_z}). 

\item Cosmological simulations should adopt softening lengths at least a factor of 2 smaller than
  $r_{\rm conv}$ in order to ensure that biased force estimates do not compromise their spatial
  resolution. Recently, \citet{vandenBosch2018a} suggested that the ``optimal'' softening for
  NFW haloes scales with particle number approximately as $\epsilon_{\rm opt}/r_{200}=0.005\times (N_{200}/10^5)^{-1/3}$,
  which is a factor of 2 to 4 smaller than $r_{\rm conv}$ over virtually all values of
  $N_{200}$ relevant for cosmological simulations ($10^2\leq N_{200}\leq 10^8$), and is ``optimal''
  regardless of halo mass. Indeed, their results imply that $\epsilon_{\rm opt}/l\approx 0.017$,
  comparable to values adopted by the majority of recent large-scale cosmological simulations.

\end{itemize}

\section{Halo mass functions}
\label{SSecMF}

Accurately characterizing the dark matter halo mass function, $n(M)$, either from simulation or theory,
is necessary for several reasons. At high mass, for example, the shape and time evolution of $n(M)$ encodes 
clues that provide important  constraints on cosmological models \citep[e.g.][]{Eke1996}.
The mass function is also an important probe of
dark matter, since many particle candidates predict strong, scale-dependent deviations from the
expectations of the canonical cold dark matter model \citep[see, e.g.][]{Schneider2015,Angulo2013}.
Accurate predictions for halo mass functions are also required for galaxy formation
theory, since galaxies are assumed to form in halo centres via dissipative collapse of primordial and
recycled gas \citep[see, e.g.,][and a number of subsequent works]{White1991}. The abundance of dark
matter haloes is therefore closely related to the observed luminosity function of galaxies, and is
a central aspect of theoretical models that attempt to reproduce observed galaxy number densities. 

The latter point is particularly important for hydrodynamical simulations that possess star-forming
haloes ($\simgt$ a few $\times 10^9\,{\rm M_\odot}$) close their resolution limit. \eagle is an example
of a simulation in which haloes of mass $\sim 10^9\, {\rm M}_\odot$ are resolved with only a few
hundred dark matter particles. The abundance and structure of these haloes, and consequently
the statistics of the first generation of low-mass galaxies, are likely subject to numerical
artifact. In this section we quantify the robustness of the halo mass function to changes in particle mass
and in gravitational softening length, focusing particularly on the lowest mass haloes. 

\begin{figure}
  \includegraphics[width=0.47\textwidth]{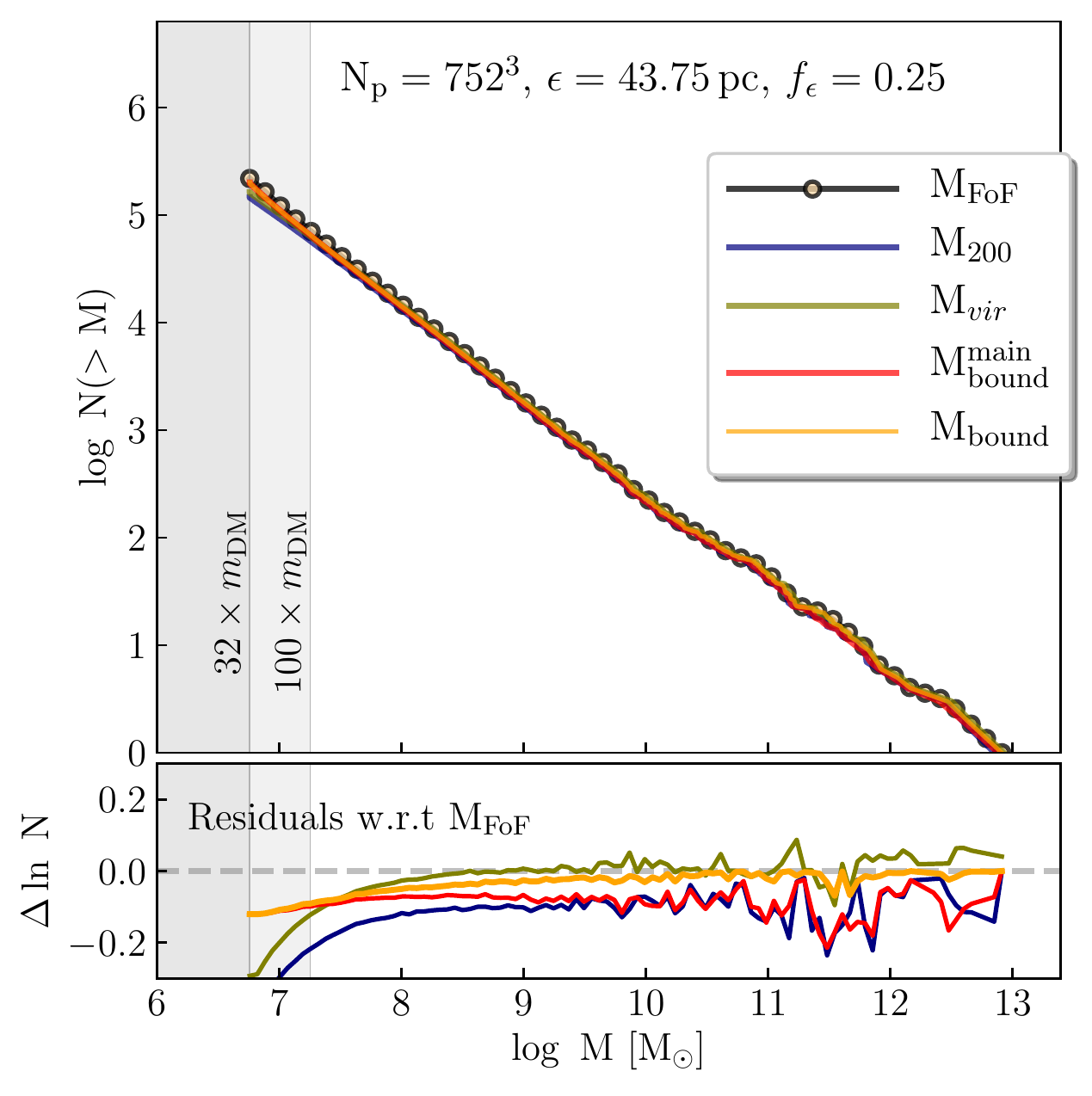}
  \caption{Cumulative halo mass functions in our $N_{\rm p}=752^3$ run
    for different mass definitions: ${\rm M_{FoF}}$ is the 
    friends-of-friends mass; ${\rm M_{200}}$ and ${\rm M}_{vir}$ are the virial masses enclosing a 
    mean density of $200\times \rho_{\rm crit}$ and $\Delta\times\rho_{\rm crit}$, respectively
    ($\Delta$ is the virial overdensity of \citet{Bryan1998}); 
    ${\rm M_{bound}}$  and ${\rm M_{bound}^{main}}$ are, respectively, the {\em bound} halo mass 
    including and excluding contributions from substructure. Residuals in the lower panel show the 
    departure of each curve from the FoF mass function. The 100 and 32-particle limits are indicated
    by the shaded regions. This run used a ($z=0$) softening length 
    $\epsilon=43.75\,{\rm pc}$, a factor $f_\epsilon=0.25$ smaller than our fiducial
    choice for this particle mass.}
  \label{fig:MF_def}
\end{figure}

\begin{figure*}
  \centering
  \footnotesize
  \subfloat{}{}{\includegraphics[width=.4\textwidth]{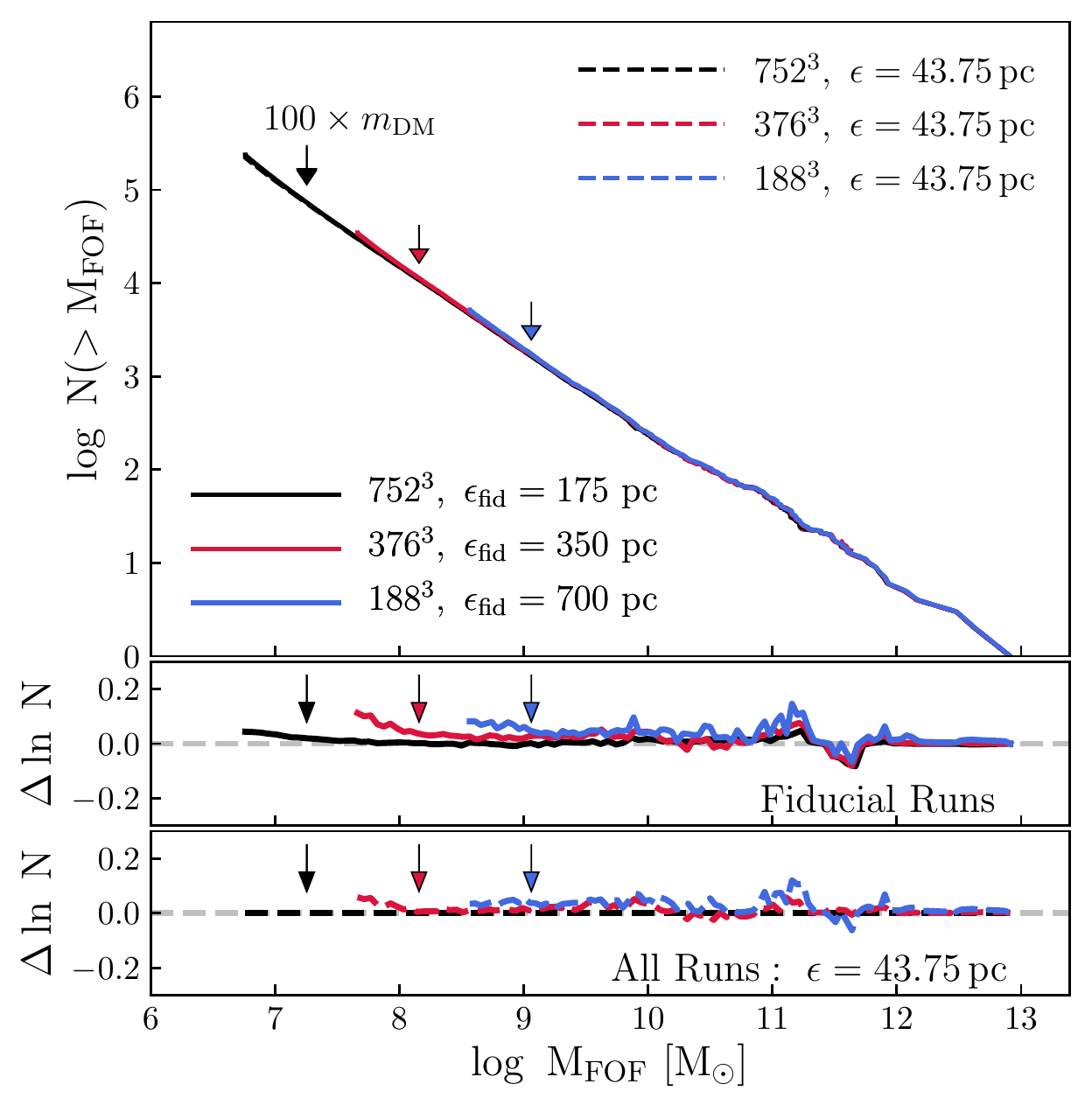}}\quad
  \subfloat{}{}{\includegraphics[width=.4\textwidth]{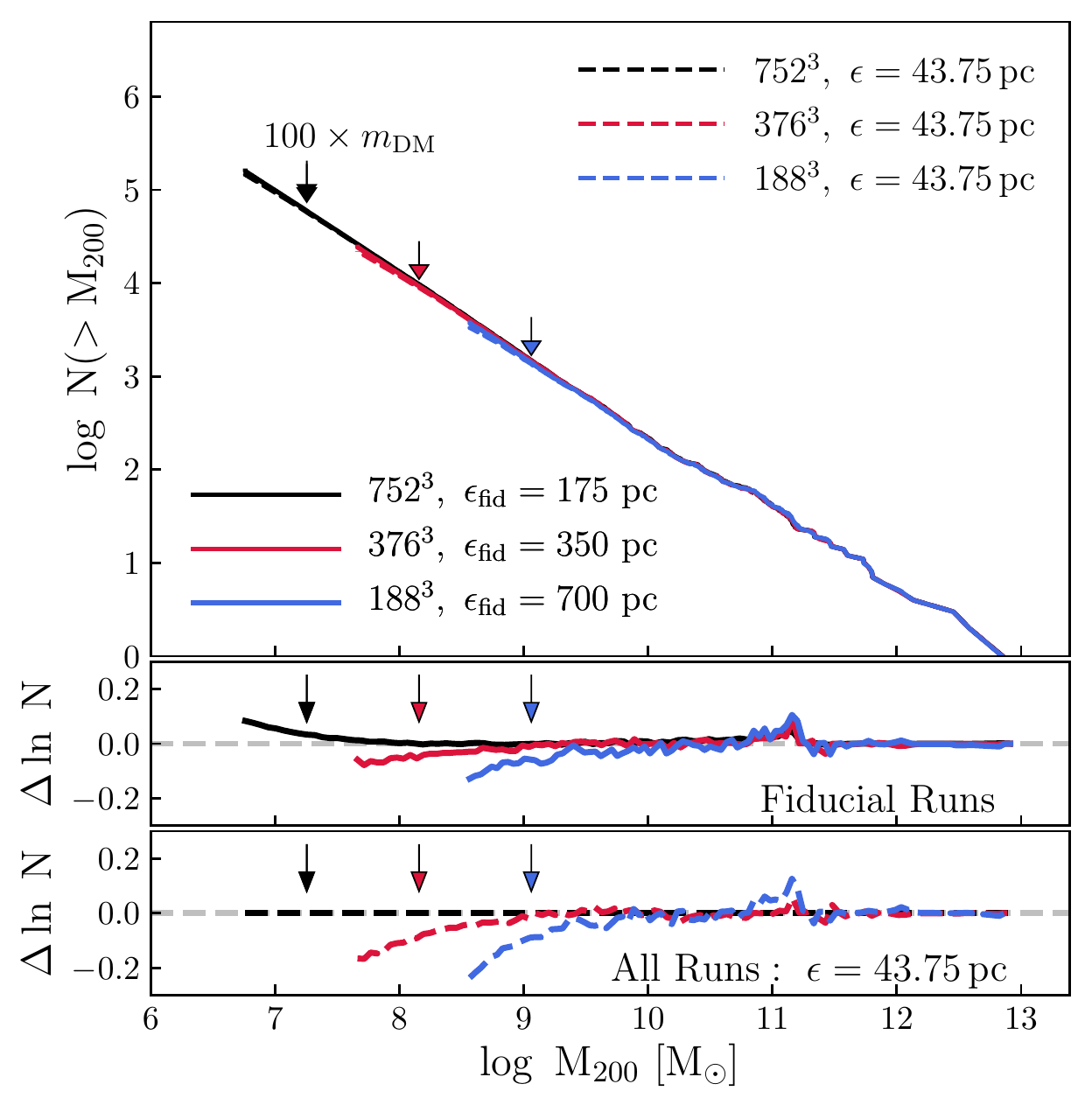}}\\
  \subfloat{}{}{\includegraphics[width=.4\textwidth]{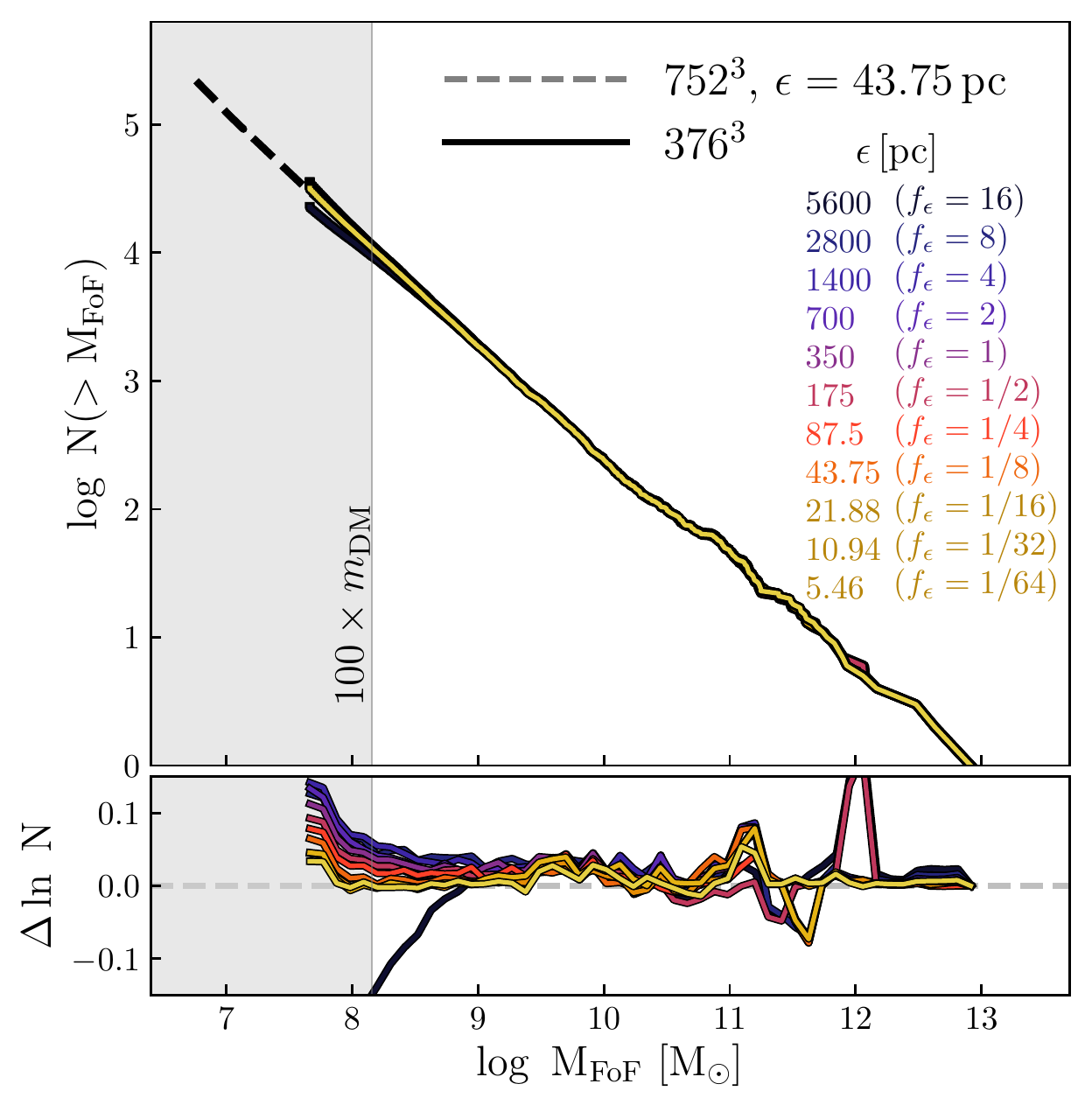}}\quad
  \subfloat{}{}{\includegraphics[width=.4\textwidth]{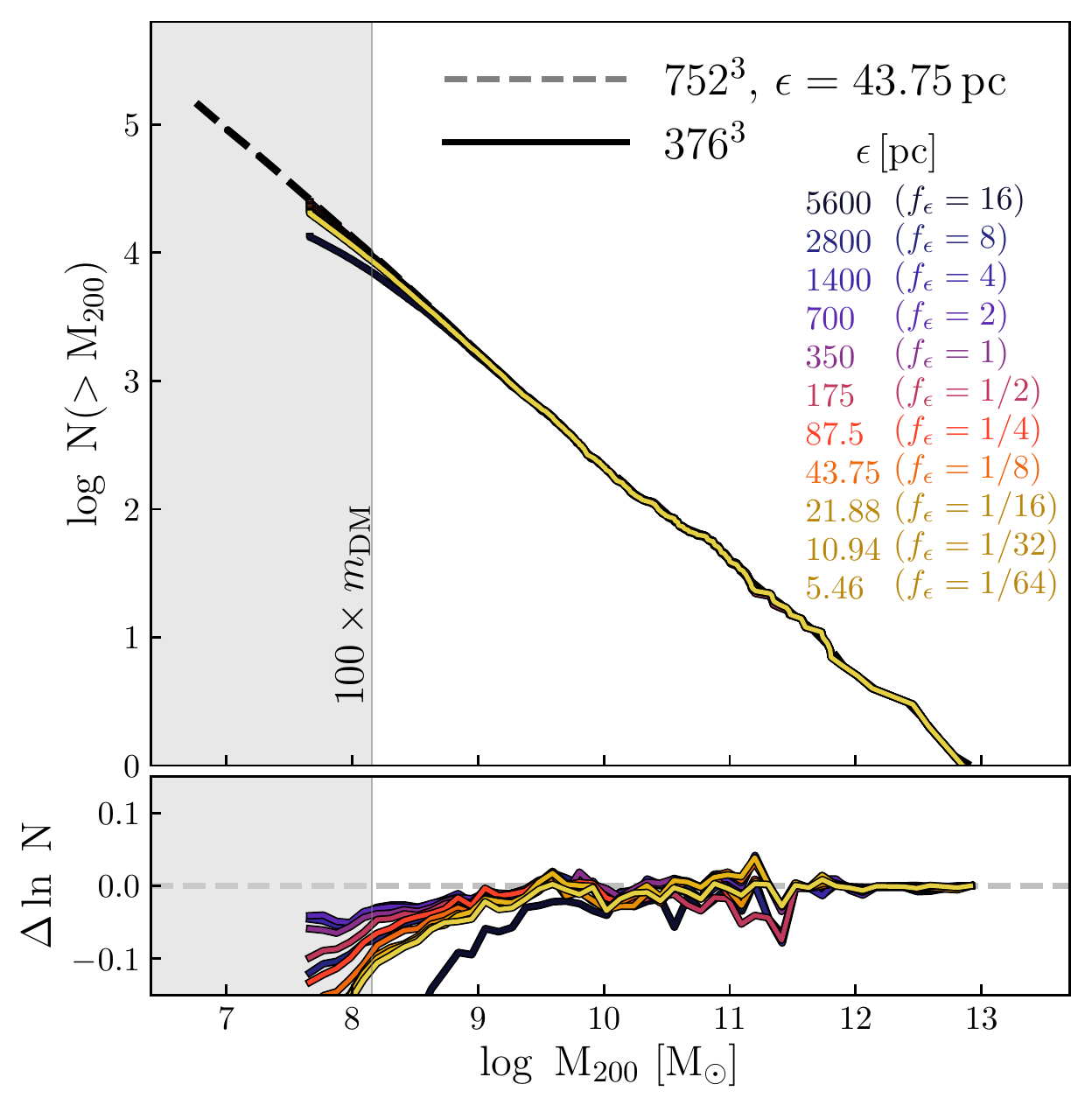}}
  \caption{{\em Upper panels}: Resolution dependence of cumulative halo mass functions for FoF (left) and
    ${\rm M_{200}}$ (right) mass definitions. In both panels, results are shown for six separate runs.
    Three adopt the ``fiducial'' softening parameter, corresponding (at $z=0$) to a fixed fraction $f\approx 0.011$
    of the Lagrangian mean inter-particle separation, but vary the total particle number, $N_p$ (solid lines). These
    are compared to three additional runs carried out with the same three $N_p$, but with $\epsilon$
    held fixed at a value of $43.75\,{\rm pc}$ (dashed lines). Downward arrows mark the mass of 100 DM particles
    for each $N_{\rm p}$, above which the mass functions agree with the higher resolution runs to within $\simlt 10$
    per cent. Upper residual panels show the departure of our fiducial runs from the $N_{\rm p}=752^3$ run with
    the smallest softening, $\epsilon=43.75\,{\rm pc}$ (i.e. from the dashed black line); lower residual panels
    compare runs at fixed softening, $\epsilon=43.75\,{\rm pc}$, regardless of resolution. {\em Lower panels}:
    Dependence of cumulative mass functions on softening. As on top, left and right hand panels correspond to
    FoF and ${\rm M_{200}}$ masses, respectively. Dashed lines show results for our
    $N_{\rm p}=752^3$, $\epsilon=43.75\,{\rm pc}$ run; coloured lines correspond to $N_{\rm p}=376^3$ runs 
    for a variety of softening lengths, as indicated in the legend. Note that, despite $\epsilon$ differing by
    over a factor of $\sim 10^3$, the mass functions have typically converged at the level of 
    $\approx 5$ per cent for ${\rm N_{FoF}\simgt 100}$, and to $\approx 10$ per cent for ${\rm N_{200}\simgt 100}$. 
    The lower panels plot the residuals with respect to the $N_{\rm p}=752^3$ run with $\epsilon =43.75\,{\rm pc}$.}
  \label{fig:MF}
\end{figure*}


\subsection{Dependence on halo mass definition}
\label{SSecMF_b}

Before studying the sensitivity of halo mass functions to numerical parameters,
it is useful to examine their dependence on how mass is defined, in order
to have a useful gauge for later on. 
Figure~\ref{fig:MF_def} shows the cumulative mass functions of haloes in our $N_{\rm p}=752^3$
($\epsilon=43.75\, \rm{pc}$) run for several possibilities. The connected points correspond
to FoF masses, solid blue and green lines to the overdensity masses ${\rm M}_{200}$ and 
${\rm M}_{vir}$, respectively. Red and yellow curves use the {\em bound} halo 
mass either excluding (red) or including (yellow) the contribution of its substructure
population (see Section~\ref{SSecSubfind} for details). 
Residuals in the lower panel are computed with respect to the FoF mass definition.

All mass functions have a similar shape, which can be approximated by a power-law over the mass
range covered by the run. The residuals, however, betray more clearly the differences. Based on mass
definition alone, $n(M)$ may differ by of order a few to $\approx$20 per cent for haloes resolved with
$\simgt 100$ particles,
with differences becoming even larger towards lower mass. FoF masses generally result
in the highest number densities at fixed $\rm M_{\rm FoF}$, and those based on $\rm M_{200}$ the lowest. Note that even
relatively well-resolved haloes have number densities that differ by roughly 10 per cent between
$\rm M_{\rm FoF}$ and $\rm M_{200}$ (corresponding to an average systematic mass difference of
roughly 15 to 20 per cent). 

We next consider the sensitivity of halo mass functions to numerical parameters, focusing on
mass and force resolution. As we will see, systematic differences brought about by varying numerical
parameters are virtually always smaller than those associated with mass definition. 

\subsection{Dependence on particle number and softening}
\label{SSecMF_b}

The upper panels of Figure~\ref{fig:MF} show the cumulative mass functions for ${\rm M_{FoF}}$ (left) and 
${\rm M_{200}}$ (right) in our $N_{\rm p}=752^3$, 376$^3$ and 188$^3$ boxes. 
Solid lines correspond to runs carried out with the fiducial softening parameter, 
$\epsilon/b\approx 0.011$ (at $z=0$); dashed lines are used for runs in which $\epsilon$ was kept
fixed at $43.75\, {\rm pc}$ (physical, also at $z=0$), regardless of $m_{\rm DM}$. The upper residual panel plots the departure of each
fiducial run from the black dashed line (i.e. from the run with the highest mass resolution and smallest
force softening). Differences are small, typically $\simlt 5$ per cent 
for haloes containing ${\rm N}\simgt 100$ particles (indicated using downward-pointing arrows),
and $\simlt 10$ per cent at {\em all} masses resolved by our simulation. The lower residual panels compare results for
fixed-softening runs (we use $\epsilon=43.75\, {\rm pc}$,  corresponding to
$\epsilon/\epsilon_{\rm fid}=1/4$, $1/8$ and $1/16$ for $N_{\rm p}=752^3$, 376$^3$ and 188$^3$,
respectively). Albeit slightly, this choice of softening improves agreement at low FoF masses, but has the
opposite effect for $\rm M_{200}$. Overall, the $\rm{ M_{FoF}}$ mass functions appear more stable to changes in 
$N_{\rm p}$ than those using ${\rm M_{200}}$. Nevertheless, it is worth emphasizing that, in both cases,
departures remain small compared to variations arising from different mass definitions. 

The lower two panels of Figure~\ref{fig:MF} show, for $N_{\rm p}=376^3$, the $\rm M_{\rm FoF}$ (left) and
$\rm M_{200}$ (right) mass functions for a large range of $\epsilon$ ($5.46\simlt \epsilon/[{\rm pc}]\simlt 5.6\times 10^3$).
The dashed black lines
are the same as in the upper panels, and correspond to the run with $N_{\rm p}=752^3$, 
$\epsilon=43.75\, {\rm pc}$; residuals in the lower panels
are computed with respect to these curves. Provided $\epsilon/\epsilon_{\rm fid}\simlt 8$, 
differences due to softening are most pronounced for the lowest mass haloes. Indeed, for a broad
range of softening, $4\simgt \epsilon/\epsilon_{\rm fid}\simgt 1/64$, both sets of mass functions agree 
with that of the higher-resolution run to within about $10$ per cent for haloes resolved with more than
$\sim$100 particles, although deviations depend systematically 
(and non-monotonically; see Figure~\ref{fig:NumDens} below) on $\epsilon$. 

Agreement between the halo mass functions depends on mass and force resolution in different ways depending on
how masses are defined. As discussed by \citet{Tinker2008} (see, also, \citealt{Lukic2009}), SO
masses, such as $\rm M_{200}$, are sensitive to the integrated mass profile within a fixed overdensity,
whereas FoF masses are measured within iso-density contours, and are less sensitive to the precise
distribution of mass within them. We therefore expect FoF mass functions to be more robust to changes in numerical
parameters than those based on spherical overdensities, a claim that is supported by the results
presented in Figure~\ref{fig:MF}. Indeed, \citet{Tinker2008} show that convergence in SO mass functions
worsens considerably with increasing overdensity.

\citet{Lukic2007} and \citealt{Power2016} showed that, unsurprisingly, halo mass functions are heavily suppressed on 
scales below which $\epsilon$ exceeds the halo virial radius; none of our runs are in that regime
(the largest softening we test, $\epsilon=5.6\,{\rm kpc}$, is equal to the virial radius of a halo
of mass ${\rm M_{200}}\approx 2\times 10^7\,{\rm M_\odot}$, well below the mass limit of our halo
finder). Our runs address a more subtle question: what impact does softening have on the abundance
of haloes when $\epsilon\ll r_{200}$?

Figure~\ref{fig:NumDens} shows the softening dependence of the {\em total} number density of haloes in our
$N_p=376^3$ runs based
on $\rm M_{\rm FoF}$ and $\rm M_{\rm 200}$ masses. From top to bottom, the different panels show results for
$N\geq 32$, 100, 256 and 800 particles, respectively; the corresponding mass thresholds are indicated
in each panel. The filled and open
squares correspond to $\rm M_{\rm FoF}$ and $\rm M_{\rm 200}$, respectively; symbols are colour-coded 
by $\epsilon$ as in Figure~\ref{fig:MF}. For comparison, filled and open circles show the number
densities of haloes above the same {\em mass} thresholds in the $N_{\rm p}=752^3$ runs. The grey
shaded regions centred on these circles indicate a 5 per cent change in total abundance. When
including the lowest-mass haloes ($N\leq 100$), the total number density grows gradually until
$\epsilon\approx 4\times\epsilon_{\rm fid}$, where it peaks, before decreasing rapidly for larger
$\epsilon$. For \textsc{eagle}'s fiducial softening length halo abundances are within a few per cent of the
maximum attained for any $\epsilon$. However, for both mass definitions, the total abundance of
haloes across all $\epsilon$ varies by $\simgt 5$ per cent for $N\leq 100$, suggesting that per
cent-level convergence in halo mass functions demands higher particle numbers.

What is the true abundance of haloes above each mass threshold? It is tempting to assume
that convergence is achieved when the results of low- and high-resolution runs agree.
If this were the case, then convergence in the abundance of haloes containing
a few dozen particles is clearly ruled out (in the upper panel of Figure~\ref{fig:NumDens} these
runs {\em never} overlap). It is important to note, however, that the $\epsilon-$dependence
of $n(M)$ weakens as $N$ increases, and agreement between runs of different mass resolution
improves. For $N\geq 256$ (second panel from the bottom), a weak dependence on $\epsilon$ is
still evident at the level of $\approx 5$ per cent in our $N_{\rm p}=376^3$ run, but has largely
disappeared for $N\geq 800$ (bottom panel), where the total abundances of haloes in both runs
agree (within the Poisson noise) for all values of $\epsilon$ studied (except, perhaps,
for the largest value and $N_{\rm p}=376^3$).

If $N\geq 800$ is sufficient to achieve convergence, then results from our $N_{\rm p}=752^3$ runs
represent robust measurements of $n(M)$ in all but the upper-most panel (haloes of fixed mass
are resolved with 8 times the number of particles in this run, so have $N\geq 800$ in each of the lower
three panels). In all cases, our low-resolution runs are able
to reproduce these abundances, but only for particular values of $\epsilon$, which differ
depending on how mass is defined (with FoF masses preferring small softening, $\epsilon \simlt \epsilon_{\rm fid}/8$,
and $\rm M_{200}$ preferring large values, $\epsilon \simgt 2-4\, \epsilon_{\rm fid}$). We conclude
that, regardless of mass definition, halo mass functions exhibit a small ($\approx 5$ per cent)
but measurable dependence on $\epsilon$ for haloes resolved with $\simlt 800$ particles.

\begin{figure}
  \includegraphics[width=0.47\textwidth]{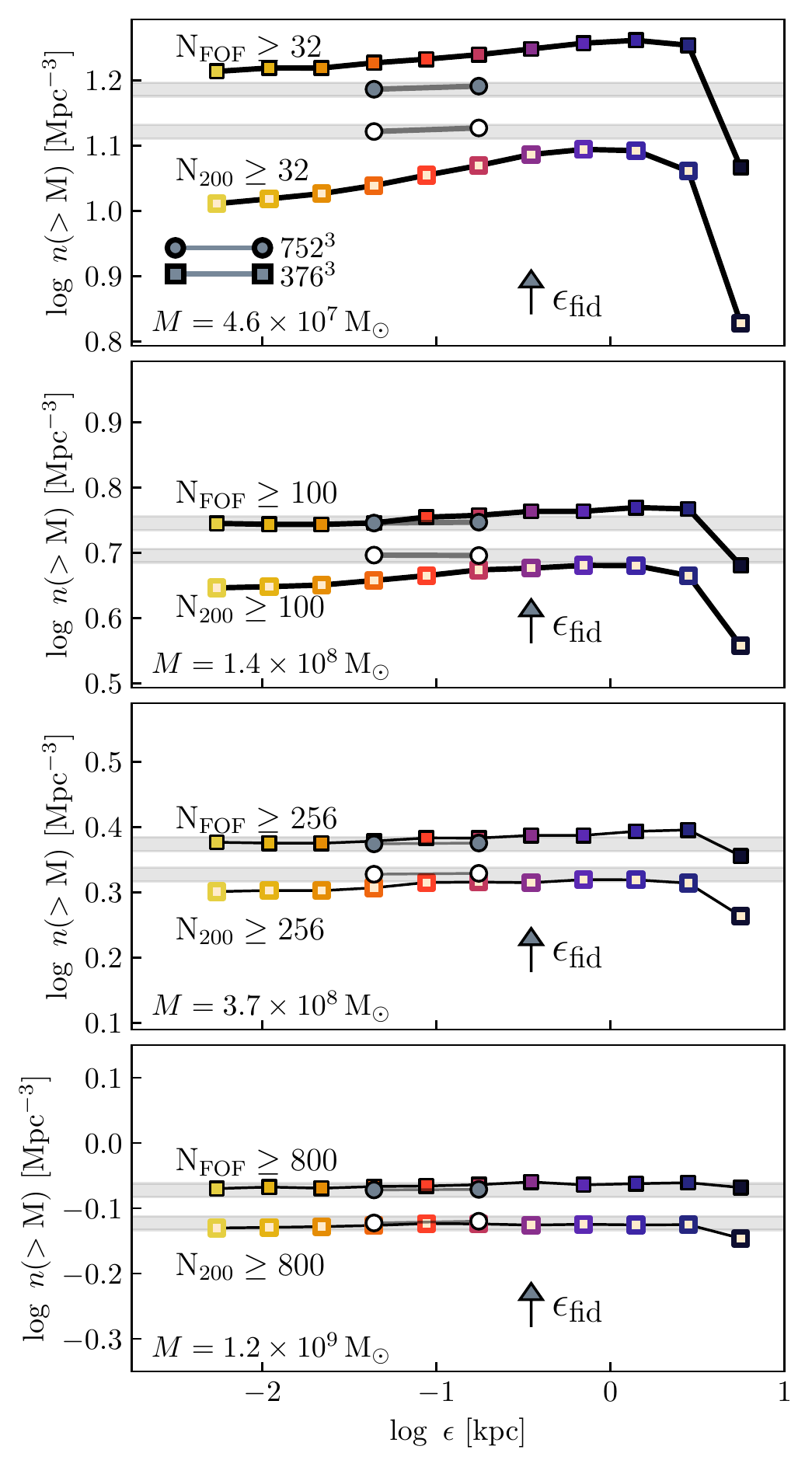}
  \caption{Total present-day number density of haloes in our $N_{\rm p}=376^3$
      run (colored squares). From top to bottom, panels corresponds to the cumulative
      abundance of haloes resolved with $\geq 32$, $100$, $256$ and $800$ particles
      and are plotted as a function of gravitational softening, $\epsilon$. Filled
      and open symbols correspond to ${\rm M_{FoF}}$ and ${\rm M_{200}}$ masses,
      respectively. Circles show the abundance of haloes of corresponding mass in our
      $N_{\rm p}=752^3$ run, which are resolved with 8 times the number of particles.
      As such, haloes in the first and second panels from the top contain, respectively,
      $N\geq 256$ and $\geq 800$ particles in the high-resolution run. Note that the
      total abundances of haloes becomes insensitive to $\epsilon$ for large $N$.}
  \label{fig:NumDens}
\end{figure}

\subsection{Dependence on redshift evolution of $\epsilon$}
\label{SSecRconvz}

\begin{figure}
  \includegraphics[width=0.47\textwidth]{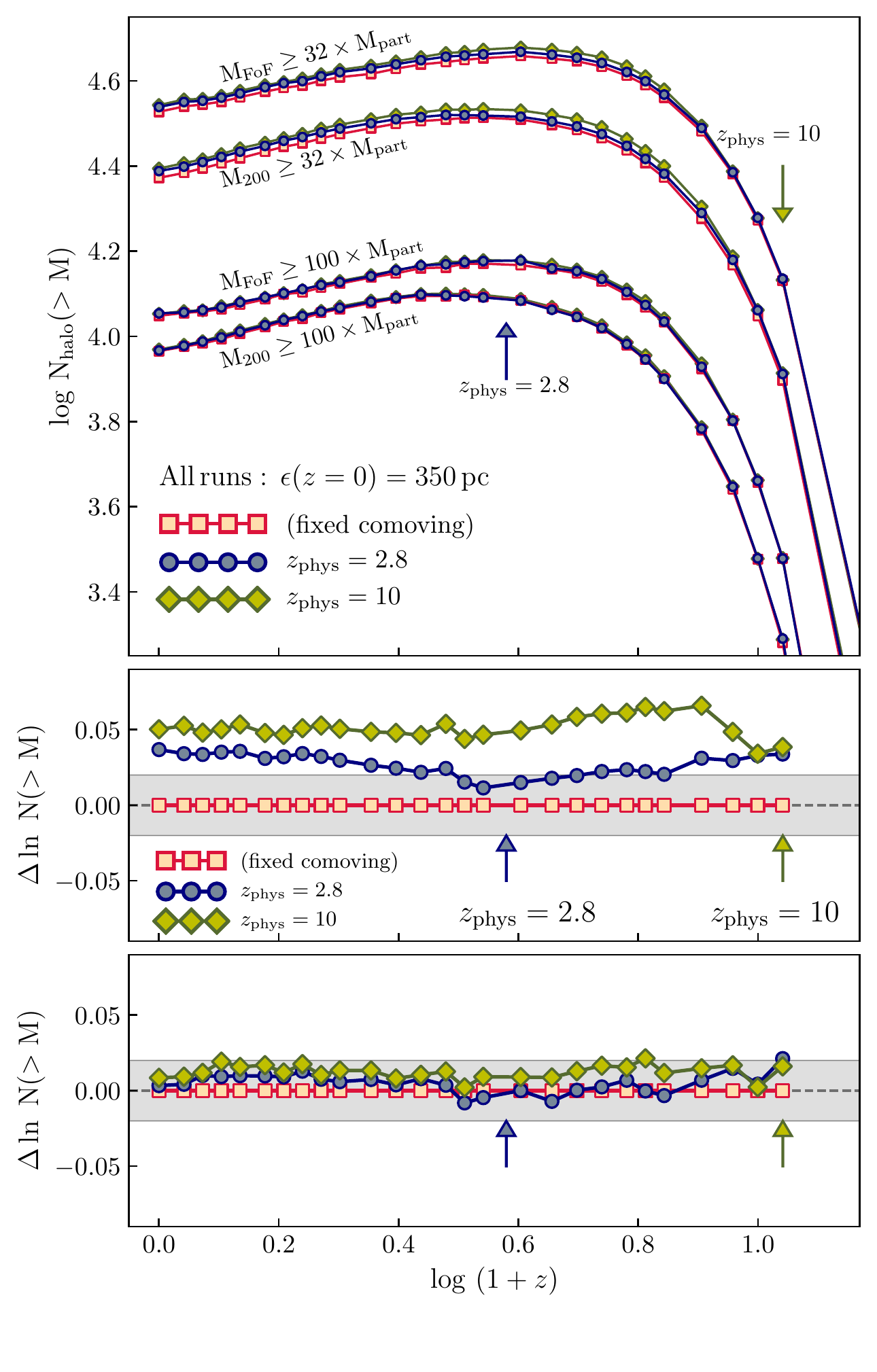}
  \caption{Redshift evolution of the total number of haloes resolved with with
    $\geq 100$ and $\geq 32$ particles (lower and upper sets of curves, respectively) for
    runs with different $z_{\rm phys}$. Each run used $N_{\rm p}=376^3$ particles and reached 
    a $z=0$ softening length of $\epsilon=350\, {\rm pc}$. Curves marked by squares, circles
    and diamonds correspond to $z_{\rm phys}=0$, $2.8$ and $10$, respectively.
    The lower two panels show the residuals with respect to 
    the $z_{\rm phys}=0$ run for the ${\rm M_{200}}$ mass definition. Grey 
    shaded regions indicate a $\pm 2$ per cent deviation from the run with fixed
    comoving softening.}
  \label{fig:MF_z_comb}
\end{figure}

The gravitational softening length, initially fixed in comoving coordinates, reaches a 
maximum {\em physical} value at $z_{\rm phys}= 2.8$, after which it remains constant in proper
coordinates. Other cosmological simulations often use a fixed comoving softening at all $z$, 
whereas some opt for fixed proper softening lengths. 
What effect, if any, does this have on the halo mass function? In order to find out,
two additional ``fiducial'' ($N_{\rm p}=376^3$) runs were carried out using
$z_{\rm phys}=10$ and 0 (i.e., fixed in comoving units at all times), each reaching
a $z=0$ softening parameter of $\epsilon_{\rm fid}=350\, {\rm pc}$. 
The upper panel of Figure~\ref{fig:MF_z_comb} shows the redshift evolution of the {\em total number} of haloes above
$32$ and $100-$particle thresholds. As above, masses are computed at all $z$ using both $M_{\rm FoF}$ 
(solid lines) and $M_{\rm 200}$ (dashed) definitions. The connected (blue) circles in this
plot show the results for $z_{\rm phys}=2.8$; other curves correspond to $z_{\rm phys}=10$ (green
diamonds) and 0 (red squares). 

Overall, the results are similar, with the 
largest differences being limited to the lowest-mass haloes. The middle and bottom panels of Figure~\ref{fig:MF_z_comb}
show the residuals with respect to the run with $z_{\rm phys}=0$ (for clarity, the comparison is limited
to ${\rm M_{200}}$ masses in this plot).
For $N_{200}\geq 100$ (lower panel), differences are never larger than $\approx 2$ per cent 
(highlighted by the grey shaded region). For $N_{200}\geq 32$ differences are still small, $\simlt 5$ per cent,
but systematic: higher $z_{\rm phys}$ corresponds to higher $n(M)$.
This is qualitatively consistent with the results in Figure~\ref{fig:NumDens}, which indicates that
increasing $\epsilon$ results in a slight boost in the numbers of low-mass haloes, at least up to a point. 
For a given $z=0$ softening length, higher $z_{\rm phys}$ implies {\em larger} physical softening lengths at 
$z>z_{\rm phys}$, and also enhances slightly the numbers of low-mass haloes. 
Note as well the slight boost in $n(M)$ (of order a couple per cent) after $z_{\rm phys}$.
These subtle changes in the abundance of low-mass haloes may impact the star formation histories
of low-mass galaxies that inhabit them, provided they are sufficiently massive to promote efficient 
gas cooling.

\subsection{Section summary}
\label{Sec1Summ}

The results of this section can be summarized as follows:
\begin{itemize}
\item Dark matter halo mass functions may differ by as much as 20 per cent
  for haloes resolved with $\geq$100 particles depending on how mass is defined.
  At fixed mass, $M_{200}$ masses yield the lowest overall number densities and $M_{\rm FoF}$ the highest.
  Comparing haloes at fixed number density,
  this suggests that, on average, FoF masses exceed those based on $M_{200}$ by
  a factor of $\approx 1.15-1.2$, though a more detailed comparison is required to
  properly assess the systematics \citep[see, e.g.,][for a thorough discussion]{Tinker2008}.

\item For haloes containing $\simgt$100 particles, FoF mass functions
  converge to better than $\approx$5 per cent for runs carried out with our ``fiducial'' softening,
  $\epsilon_{\rm fid}/l=0.011$ at $z=0$, regardless of particle number $N_{\rm p}$ (note $\epsilon_{\rm fid}=700$, 350, 
  and $175\, {\rm pc}$ for $N_{\rm p}=188^3$, 376$^3$ and 752$^3$, respectively),
  and to $\simlt$3 per cent when $\epsilon=43.75\,{\rm pc}$ was kept fixed for all $N_{\rm p}$.
  For haloes resolved with $\geq 32$ particles, corresponding to the resolution limit of \textsc{subfind},
  FoF mass functions converge to within $\approx 10$ per cent for fiducial
  softening values, and to within $\approx 5$ per cent for $\epsilon=43.75\,{\rm pc}$. The overall
  trend is such that lower mass resolution results in a systematic {\em increase} in the
  total number of FoF haloes, driven mainly by a slight increase in the number of poorly resolved
  systems. For $M_{200}$ the converse is true: runs of lower resolution produce systematically
  {\em fewer} haloes. Differences are small but systematic, reaching $\approx 5$ per cent 
  in our fiducial runs for haloes resolved with at least $N_{200}\geq 100$, and $\approx 10$
  per cent for runs with $\epsilon=43.75\,{\rm pc}$.

\item At fixed mass resolution (corresponding to $N_{\rm p}=376^3$) $M_{\rm FoF}$ mass functions converge
  to within $\approx 6$ (15) per cent for $N_{\rm FoF}\geq 100$ (32), for a range of $\epsilon$ spanning
  nearly a factor of $10^3$ ($5.46\,{\rm pc}$ to $5.6\, {\rm kpc}$). For 
  $M_{200}$, we obtain convergence at the 11 and 15 per cent level for $N_{\rm 200}\simgt 100$ and $\simgt 32$,
  respectively, although larger differences are found for the most extreme values of $\epsilon$ tested. 
  There is also a systematic trend: increasing $\epsilon$ results in a gradual increase in the abundance of
  haloes composed of $\simlt 100$ particles, provided it remains smaller than about $8\times\epsilon_{\rm fid}\approx 0.09\, l$,
  where the total abundance of haloes peaks.

\item Changing the redshift $z_{\rm phys}$ below which the softening parameter remains fixed
  in proper units (rather than comoving) can have a comparable affect on the abundance of
  low-mass $\simlt 100$-particle haloes. When $\epsilon$ is fixed at $z=0$ a
  higher $z_{\rm phys}$ implies larger {\em physical} softening lengths for all $z>z_{\rm phys}$. 
  As we have seen above, larger softening lengths tend to enhance slightly the formation of
  low-$N$ haloes, provided $\epsilon$ does not become {\rm too} large. For example, for $N_{\rm p}=376^3$
  and $\epsilon(z=0)=350\,{\rm pc}$, increasing $z_{\rm phys}$ from 0 to 10 results in a $\simlt$2 
  per cent increase in the abundance of haloes with $N_{200}\geq 100$; for $N_{200}\geq 32$
  deviations are not larger than $\approx 5$ per cent. 

\end{itemize}

Overall, our results suggest that halo mass functions are a robust result of N-body simulations
once a definition of halo mass has been specified. Variations in halo abundances with numerical
parameters tend to be restricted to poorly resolved systems containing fewer than $100$ particles.
Yet, however slight, these changes may have a noticeable impact on the first generation of star
formation if these haloes happen to have masses comparable to the threshold for efficient gas cooling.
Calibrating sub-grid models at fixed $N_{\rm p}$ and $\epsilon$ may help mitigate any non-physical
effects brought about by star formation in this first generation of poorly resolved haloes, but may
not easily adapt to increasing or decreasing resolution.
As a result, simulations that adopt numerical parameters that differ from those for which the models
were calibrated may yield noticeably different star formation histories and/or galaxy properties.
We will address these issues in a companion paper.

\section{Conclusions}
\label{SecConclusion}

We have carried out a systematic convergence study of the median and statistical
properties of DM haloes in fully cosmological, dark matter-only simulations. Unlike
previous work, which targeted {\em single} haloes, we focused our analysis on their
{\em median} spherically-averaged density profiles as a function of mass and on several mass-dependent 
structural scaling relations. After verifying the
need for fine timesteps to resolve halo centres, we tested the sensitivity of halo
profiles to total particle number, $N_{\rm p}$, and force softening length, $\epsilon$.
We also revisited the calculation of the convergence radius originally provided by
P03, and derived its explicit (but weak) dependence on the gravitational softening
(eq.~\ref{eqKapLud}).

In addition to mass profiles, we also studied convergence in the abundance
of haloes as a function of mass, focusing on runs carried out with different particle
numbers and redshift-dependent softening lengths. Our main results can be summarized as 
follows:

\begin{enumerate}
\setlength\itemsep{1em}
\item {\em Softening and 2-body relaxation:} Softening does not significantly affect
  2-body relaxation times, which are primarily driven by particle number. This is
  contrary to common belief, despite being well documented in the literature
  \citep[e.g.][]{Huang1993,Theis1998,Dehnen2001}.
  The result may at first seem puzzling because encounters between particles give
  rise to velocity perturbations that scale approximately as $\delta v\sim G\, m_{\rm DM}/(b\,v)$,
  increasing significantly for encounters with small impact parameter, $b$.
  Nevertheless, close encounters are rare and, as pointed out by \citet{Chandrasekhar1942}
  and \citet{SpitzerHart1971}, the large numbers of distant encounters dominate the cumulative effect
  of perturbations: relaxation is driven by discreteness on large
  scales, where most of the mass is \citep[see, e.g.,][]{HernquistBarnes1990}. Softening
  does, however, suppress small-scale ``collisions'' between particles: for Plummer
  softening interactions between particles become increasingly unimportant
  for separations $\simlt \epsilon/\sqrt{2}$. Softening therefore serves the purpose of
  {\em smoothing} the matter distribution on small scales, allowing force estimates
  to more faithfully represent a {\em continuous} matter field.

\item {\em Convergence of the median mass profiles of CDM haloes:}
  Provided $\epsilon$ and timestep size are appropriately chosen, 2-body relaxation imposes a 
  strict and well-defined lower limit on the spatial resolution of collisionless CDM haloes
  (Figure~\ref{fig:VcircConv}).
  Convergence in mass profiles, for example, can only be achieved at radii beyond which
  2-body relaxation times are a sizeable fraction of a Hubble time, or longer. As a result, convergence radii
  are primarily determined by the enclosed particle number, though analytic arguments suggest
  a weak (approximately logarithmic) dependence on softening. Softening appears to compromise the
  spatial resolution of simulations only if it is {\em larger} than the convergence radius
  dictated by 2-body scattering: in this case, our results suggest that
  $r_{\rm conv}\approx 2\times \epsilon$ (Figure~\ref{fig:VcircConvB}).

  Particle interactions tend to drive the local velocity distribution towards a
  Maxwellian and, occasionally, impart velocities exceeding the local escape speed.
  The net effect of 2-body relaxation is therefore to monotonically suppress the central
  densities of haloes, allowing convergence radii to be determined empirically by comparing
  runs of different (mass) resolution. A simple but accurate description of our measured convergence
  radii can be obtained using eq.~\ref{eqP03} or \ref{eqKapLud}, where $\kappa\equiv t_{\rm relax}/t_{\rm H}$;
  better convergence is achieved for higher values of $\kappa$.
  Circular velocities, for example, converge to within $\approx 3$, 10 and 20 per cent for $\kappa=0.566$, 0.177
  and 0.106, respectively (Figure~\ref{fig:rc_kappa}; values of $\kappa$ for arbitrary levels of convergence can be
  approximated using eq.~\ref{eq:poly}). 

  These results, valid at all redshifts (Figure~\ref{fig:rc_z}), are qualitatively consistent
  with those of P03 and N10, but differ in the details.
  P03, for example, find that $10$ per cent convergence in $V_c(r)$ requires $\kappa\approx 0.6$,
  while N10 find $\kappa=7.5$ yields $\Delta V_c/V_c\approx 0.025$; both are {\em larger} than
  the values of $\kappa$ we advocate for the same level of convergence ($\approx 0.2$ and 
  $\approx 0.6$, respectively).
  It is worth emphasizing, however, that both previous studies focused on {\em single}
  dark matter haloes of fixed virial mass, whereas our results apply to {\em median}
  mass profiles that span a broad range of halo masses. 
  Differences between our results and theirs may arise as a result of sampling haloes
  possessing a wide range of concentrations at fixed mass: since $r_{\rm conv}$ (and its
  relation to $\kappa$) depends explicitly on profile {\em shape}, it will necessarily
  vary from halo to halo. Another possibility is that, by averaging, we smooth-out
  idiosyncrasies of individual systems--such as asymmetric shapes, or locally dominant
  substructure--thereby minimizing the effects of collective relaxation.
  The convergence criteria advocated by P03 and N10 are therefore
  more conservative than ours, a result echoed by the recent work of \citet{Zhang2018}. 

  At fixed mass resolution, the convergence radii anticipated by eq.~\ref{eqP03} exhibit a weak 
  dependence on $N_{200}$, varying by about a factor of two for NFW haloes with $N_{200}$ ranging from 
  $\sim  10^2$ to $\sim 10^8$. Indeed, the {\em measured} median convergence radii of haloes in 
  our simulations are compatible with a much simpler approximation in which $r_{\rm conv}$ is 
  simply a fixed fraction of the mean inter-particle spacing, $l$. For example, $< 10$ per cent 
  convergence in circular velocity is obtained approximately at a radius 
  $r_{\rm conv}(z)=0.055\times l(z)$ (eq.~\ref{eqp03ApproxII}; Figure~\ref{fig:VcircConv}).
  Softening lengths for cosmological simulations should be chosen with this in mind. 

\item {\em The optimal softening for cosmological simulations:}  
  We note that the ``optimal'' softening length, $\epsilon_{\rm opt}/r_{200}=0.005\times (N_{200}/10^5)^{-1/3}$,
  advocated by \citet{vandenBosch2018a} for NFW haloes is typically a factor of 2 to 4
  {\em smaller} than $r_{\rm conv}$ at fixed $N_{200}$. It is therefore unlikely to 
  compromise the central mass profiles of simulated haloes. Furthermore, since 
  $\epsilon_{\rm opt}\propto N_{200}^{-1/3}$ the ratio $\epsilon_{\rm opt}/m_{\rm DM}^{1/3}$ is
  fixed and $\epsilon_{\rm opt}$ can be conveniently expressed in units of the mean inter-particle
  distance: $\epsilon_{\rm opt}/l\approx 0.017$ (this is a factor $\approx 1.6$ larger than the maximum
  physical softening length for our fiducial runs). We suggest that softening lengths of order 
  $\epsilon_{\rm opt}$ be employed in future large-scale simulations. This is a factor of 3
  smaller than the radius for which we find better than 10 per cent convergence in circular velocity
  (eq.~\ref{eqp03ApproxII}) and it is comparable to values
  adopted for most recent cosmological runs (Section~\ref{SSecPrelim}). It is important to note,
  however, that this recommendation is based on our systematic study of convergence in the central
  mass profiles of DM haloes; other simulation statistics may prefer different $\epsilon$. 

\item {\em Convergence of halo mass functions:} The mass functions of {\em central} CDM 
  haloes are a robust prediction of cosmological simulations once a halo mass definition has
  been specified. For our fiducial softening lengths, Friends-of-friends ($M_{\rm FoF}$) and
  spherical overdensity ($M_{200}$) mass functions,
  converge to within $\approx 10$ per cent of those obtained from a higher-resolution run,
  above a mass scale corresponding to $\approx 32$ particles (Figure~\ref{fig:MF}).
  Convergence is better at higher masses,
  reaching $\approx 5$ per cent for haloes resolved with at least 100 particles. These results are valid
  for a wide range of softening lengths, as demonstrated in Figure~\ref{fig:MF}, where we
  compared the mass functions in a suite of 11 simulations that varied
  $\epsilon$ from $\approx 5.5\,{\rm pc}$ ($1/64^{th}$ of $\epsilon_{\rm fid}$) to
  $\approx 5.6\,{\rm kpc}$ (16 times  $\epsilon_{\rm fid}$).
  For all but the largest softening length, all of these runs converge to within $\approx 5$ per
  cent for FoF mass, and to with $\approx 10$ per cent for $M_{200}$ provided haloes are resolved with at
  least 100 particles. Although small, the deviations depend {\em systematically} on $\epsilon$,
  particularly for haloes resolved with low numbers of particles ($\simlt 100$).
  For both mass definitions, the {\em total} abundance of haloes containing of order a few dozen
  to a few hundred particles increases systematically with increasing $\epsilon$, reaching a
  maximum at 4 to 8 times the fiducial softening (corresponding to roughly 4 to 8 per cent of the
  mean inter-particle separation) before rapidly declining (Figure~\ref{fig:NumDens}).
  We emphasize, however, that such
  large softening lengths are undesirable as they compromise the innermost structure of DM haloes,
  limiting the simulation's spatial resolution.

\end{enumerate}

Overall, our results confirm and extend prior work on the accuracy and reliability of
cosmological simulations of collisionless cold dark matter
\citep[e.g. P03;][]{Navarro2004,Diemand2004,Lukic2007,Springel2008b,Stadel2008,Power2016,vandenBosch2018a,vandenBosch2018b}.
The primary catalyst of this
work was a follow-up study in which we repeat a number of these simulations but including
either adiabatic hydro-dynamics, or a fully-calibrated set of sub-grid physics models for galaxy
formation. In our opinion, it is therefore necessary to first highlight and expound the level
of convergence in these pure DM runs, validating the numerical results in order to facilitate
the interpretation of more complex hydrodynamical simulations.

We close by reiterating that our convergence study focused exclusively on 
``main'' DM haloes, and that our results are unlikely to apply directly to sub-structure.
As emphasized recently by \citet{vandenBosch2018a} and \citet{vandenBosch2018b}, the evolution
of substructure in haloes is governed by several phenomena, both physical and numerical, that do not
necessarily apply to isolated systems. We do, however, encourage future studies such as ours
targeted explicitly at the structural evolution of substructure in hierarchical cosmologies. 

\section*{Acknowledgments}
We thank Lydia Heck and the cosma support team for their invaluable assistance,
Alejandro Alejandro Ben\'{i}tez-Llambay and Matthieu Schaller for helpful conversations,
and our anonymous referee for a thoughtful report.
Our work has also benefited from various public {\textsc{python}} packages, including: 
{\textsc{scipy}} \citep{scipy}, {\textsc{numpy}} \citep{numpy}, {\textsc{matplotlib}} 
\citep{matplotlib} and {\textsc{ipython}} \citep{ipython}. ADL acknowledges financial 
supported from the Australian Research Council through their Future Fellowship scheme 
(project number FT160100250). This work was supported by the Science and Technology Facilities 
Council [ST/P000541/1]. This work used the DiRAC Data Centric system at Durham University, 
operated by the Institute for Computational Cosmology on behalf of the STFC DiRAC HPC 
Facility (www.dirac.ac.uk. This equipment was funded by a BIS National E-infrastructure 
capital grant ST/K00042X/1, STFC capital grant ST/K00087X/1, DiRAC Operations grant 
ST/K003267/1 and Durham University. DiRAC is part of the National E-Infrastructure.

\bibliographystyle{mn2e}
\bibliography{paper}

\end{document}